\documentclass[twocolumn,aps,prd,superscriptaddress,nofootinbib,floatfix,preprintnumbers]{revtex4-1}

\usepackage{bm,amsmath,amsfonts,amssymb,graphicx,xspace,placeins,multirow}

\usepackage{hyperref}
\usepackage{breakurl}
\usepackage{array}
\usepackage{slashed}
\usepackage{xspace}
\usepackage[italic]{hepnames}
\usepackage{booktabs}
\usepackage{hhline}
\usepackage{multirow}
\usepackage{header3}
\usepackage{lineno}
\usepackage{orcidlink}

\allowdisplaybreaks
\newcommand{\GeV}{\text{GeV}}

\def\d{{\rm d}}

\newcommand{\Bbar}{\,\overline{\!B}{}}
\newcommand{\Dbar}{\,\overline{\!D}{}}

\newcommand{\Vub}{V_{ub}}

\newcommand{\GF}{\ensuremath{G_{F}}}

\newcommand{\Vcba}{\ensuremath{\left| V_{cb} \right|}}

\newcommand{\bdlnu}{\ensuremath{B \to D \, \ell\,\bar{\nu}_{\ell}}\xspace}
\newcommand{\bdslnu}{\ensuremath{B \to D^* \, \ell\,\bar{\nu}_{\ell}}\xspace}

\newcommand{\dd}{\mathrm{d}}

\makeatletter
\g@addto@macro\bfseries{\boldmath}
\makeatother

\definecolor{red}{rgb}{0.9, 0,0}

\begin{document}


\title{Measurement of Differential Distributions of $B \to D^* \ell \bar \nu_\ell$ and Implications on $|V_{cb}|$}

\noaffiliation
  \author{M.~T.~Prim\,\orcidlink{0000-0002-1407-7450}} 
  \author{F.~Bernlochner\,\orcidlink{0000-0001-8153-2719}} 
  \author{F.~Metzner\,\orcidlink{0000-0002-0128-264X}} 
  \author{K.~Lieret\,\orcidlink{0000-0003-2792-7511}} 
  \author{T.~Kuhr\,\orcidlink{0000-0001-6251-8049}} 
  
  \author{I.~Adachi\,\orcidlink{0000-0003-2287-0173}} 
  \author{H.~Aihara\,\orcidlink{0000-0002-1907-5964}} 
  \author{S.~Al~Said\,\orcidlink{0000-0002-4895-3869}} 
  \author{D.~M.~Asner\,\orcidlink{0000-0002-1586-5790}} 
  \author{H.~Atmacan\,\orcidlink{0000-0003-2435-501X}} 
% ...
\collaboration{The Belle Collaboration}

\begin{abstract}
We present a measurement of the differential shapes of exclusive $B\to D^* \ell \bar{\nu}_\ell$ ($B = B^-, \bar{B}^0 $ and $\ell = e, \mu$) decays with hadronic tag-side reconstruction for the full Belle data set of $711\,\mathrm{fb}^{-1}$ integrated luminosity. 
We extract the Caprini-Lellouch-Neubert (CLN) and Boyd-Grinstein-Lebed (BGL) form factor parameters and use an external input for the absolute branching fractions to determine the Cabibbo-Kobayashi-Maskawa matrix element and find $|V_{cb}|_\mathrm{CLN} = (40.1\pm0.9)\times 10^{-3}$ and $|V_{cb}|_\mathrm{BGL} = (40.6\pm 0.9)\times 10^{-3}$ with the zero-recoil lattice QCD point $\mathcal{F}(1) = 0.906 \pm 0.013$. We also perform a study of the impact of preliminary beyond zero-recoil lattice QCD calculations on the $|V_{cb}|$ determinations.
Additionally, we present the lepton flavor universality ratio $R_{e\mu} = \mathcal{B}(B \to D^* e \bar{\nu}_e) / \mathcal{B}(B \to D^* \mu \bar{\nu}_\mu) = 0.990 \pm 0.021 \pm 0.023$, the electron and muon forward-backward asymmetry and their difference $\Delta A_{FB}=0.022\pm0.026\pm 0.007$, and the electron and muon $D^*$ longitudinal polarization fraction and their difference $\Delta F_L^{D^*} = 0.034 \pm 0.024 \pm 0.007$. The uncertainties quoted correspond to the statistical and systematic uncertainties, respectively.
\end{abstract}


\preprint{
Belle Preprint           2022-34,
KEK Preprint             2022-47
}

\maketitle

\section{Introduction}

The precise determination of the absolute value of the Cabibbo-Kobayashi-Maskawa (CKM) matrix element $V_{cb}$ is important to test the validity of the Standard Model of particle physics~\cite{PhysRevLett.10.531,km_paper}: its value constrains the amount of charge-parity ($CP$)-violating effects in the quark sector~\cite{pdg:2022} and is needed to predict branching fractions of rare decay processes~\cite{Beneke:2019slt,Bobeth:2013uxa}. Semileptonic decays into charmed hadrons offer a clean avenue to determine \Vcba: the decay rate of such processes is theoretically better understood than purely hadronic decays, and measurements of fully leptonic $B_c$ decays will only be possible at future experimental facilities~\cite{Zheng:2021xuq}. Indirect determinations with reasonable precision via loop processes are also possible~\cite{Altmannshofer:2021uub}. Existing determinations of \Vcba\ with semileptonic decays focus either on inclusive decays~\cite{Bordone:2021oof,Bernlochner:2022ucr} or on exclusive final states, with $B \to D^* \ell \bar \nu_\ell$  being the exclusive channel with the most precise results~\cite{Belle:2018ezy}. The obtained values of \Vcba\ are, however, only marginally compatible between inclusive and exclusive determinations, exhibiting a tension of about $3 \sigma$~\cite{Amhis:2022mac}.

In this paper, measurements of normalized differential distributions of $\Bbar{}^0 \to D^{*+} \ell \overline{\nu}_\ell$  and $B^- \to D^{*0} \ell \overline{\nu}_\ell$ are presented.\footnote{Charge conjugation is implied and $\ell = e,\mu$.} These distributions provide the necessary experimental input to determine the non-perturbative form factors governing the strong decay dynamics of the process. Knowledge of the functional form of the form factors in combination with information from Lattice QCD or other non-perturbative methods on their absolute normalization, allow the determination of \Vcba \ using
\begin{align}
  |V_{cb}| & = \sqrt{  \frac{ \mathcal{B}(B\to D^* \ell \bar{\nu}_\ell)  }{ \tau_B \, \Gamma(B\to D^* \ell \bar{\nu}_\ell) }  } \, .
  \label{eq:vcb}
\end{align}
Here $\mathcal{B}$ denotes an externally measured branching fraction of the process, $\Gamma$ is the predicted decay rate omitting the CKM factor $\Vcba^2$\,, and $\tau_B$ is the $B$ meson lifetime. 

To retain a high resolution in the kinematic quantities of interest and a high signal purity, we make use of the improved hadronic tagging algorithm of Ref.~\cite{Keck:2018lcd}. This algorithm hierarchically reconstructs the accompanying $B_\mathrm{tag}$ meson in the $\Upsilon(4S)\to B_\mathrm{sig} B_\mathrm{tag}$ decay in $\mathcal{O}(10000)$ exclusive hadronic decay channels and selects candidates based on a multivariate method. With this the signal $B_\mathrm{sig}$ kinematic properties are accessible, allowing for the direct calculation of the four-momentum transfer squared, $q^2 = (p_B - p_{D^*})^2$, with the $B$ $(D^*)$ meson momentum $p_B$ ($p_{D^*}$), and the three angular relations necessary to describe the full $B \to D^* \ell \bar \nu_\ell$ decay cascade (illustrated in Fig.~\ref{fig:helicity_angle_defition}). Due to the challenges of understanding absolute efficiencies when using algorithms such as that of Ref.~\cite{Keck:2018lcd}, we only focus on measuring normalized differential shapes. To determine \Vcba\ we make use of external inputs for the branching fraction. We report 1D projections of the decay angles and hadronic recoil parameter $w$, which are fully corrected for detector effects and efficiencies, and we provide the correlations to allow for a simultaneous analysis of the decay angles and $w$ in all considered decay modes.

This paper is organized as follows: Section~\ref{sec:theo} provides a brief overview on the theory of $\bdslnu$ decays, including definitions for the measured angular relations and the hadronic recoil parameter. Sections~\ref{sec:dataset} and \ref{sec:reco} summarize the analyzed data set, event reconstruction, and selection. Section~\ref{sec:fit} describes the background subtraction fit and Section~\ref{sec:unfold} the unfolding of detector resolution effects. In Section~\ref{sec:syst} an overview of the evaluated systematic uncertainties is given. Section~\ref{sec:results} presents our results and our conclusions are presented in Section~\ref{sec:conclusions}.

\vspace{0.5em}

\section{Theory of $B \to D^* \ell \bar \nu_\ell$ Decays}\label{sec:theo}

In the SM, semileptonic $B \to D^* \ell \bar \nu_\ell$  decays are mediated by a weak charged current interaction. The dominant theory uncertainty in predicting the semileptonic decay rate arises in the description of the hadronic matrix elements. These matrix elements can be represented in terms of four independent form factors $h_{A_{1-3}, V}$ in the heavy quark symmetry basis \cite{Manohar:2000dt}:
\begin{align}
\frac{ \langle D^* | \bar c \, \gamma^\mu b |  B \rangle }{\sqrt{m_B m_{D^*}} } & = i \,\, h_V \, \varepsilon^{\mu\nu\alpha\beta} \, \epsilon_\nu^* \, v_\alpha' \, v_\beta  \\
\frac{ \langle D^* | \bar c \, \gamma^\mu \, \gamma^5 b |  B \rangle }{\sqrt{m_B m_{D^*}} } & = h_{A_1} (w+1) \, \epsilon^{*\,\mu} - h_{A_2} (\epsilon^* \cdot v) \, v^\mu \nonumber \\ 
&  \phantom{=} - h_{A_3} ( \epsilon^* \cdot v) \, v'^\mu \, .
\end{align}
\begin{figure}
    \centering
    \includegraphics[width=\linewidth]{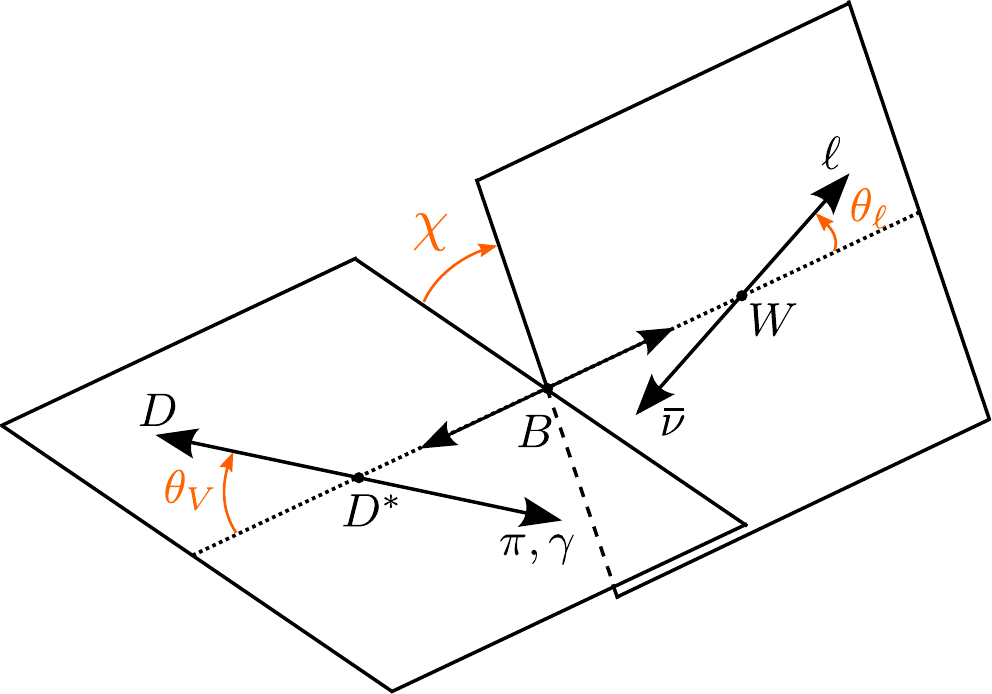}
    \caption{Visualization of the decay angles in $\bdslnu$. For definitions see text.}
    \label{fig:helicity_angle_defition}
\end{figure}
Here $w = v \cdot v' = (m_B^2 + m_{D^*}^2 - q^2)/( 2 m_B m_{D^*})$ is the hadronic recoil parameter, which can be expressed as the product of the two four-velocities $v = p_B / m_B$ and $v' = p_{D^*} / m_{D^*}$. Further, $\epsilon^*$ denotes the $D^*$ polarization vector and $\varepsilon^{\mu\nu\alpha\beta}$ is the Levi-Civita tensor. The form factors are functions of $q^2$, or equivalently $w$. For $\ell = e, \mu$ the $B \to D^*$ transition can be fully described by the form factor $h_{A1}$ and the two form factor ratios,
\begin{align}
  R_1(w) = \frac{h_V}{h_{A_1}} \, , \qquad R_2(w) = \frac{ h_{A_3} + r^* h_{A_2} }{h_{A_1}} \, ,
\end{align}
with $r^* = m_{D^*}/m_B$. 

An alternative common choice to describe the $B \to D^*$ decay transition is to represent the decay with form factors $g, f, F_1$~\cite{Boyd:1995sq,Boyd:1997kz}, which are related to the form factors of the heavy quark symmetry basis as
\begin{align}
 h_{A_1}  = \frac{f}{ m_B \sqrt{r^*} (w+1) } \, , \quad h_V = g \, m_B \, \sqrt{r^*} \, , \\
  h_{A_1} \, (w - r^* - (w-1) \, R_2)  = \frac{F_1}{m_B^2 \, \sqrt{r^*} (w+1)} \, .
\end{align}

The functional forms of the form factors have to be obtained using fits to differential distributions and/or to input from non-perturbative methods such as Lattice QCD~\cite{FermilabLattice:2021cdg,FermilabLattice:2014ysv}. There are various theoretical approaches used to parameterize the $B \to D^*$ form factors.

The BGL parameterization~\cite{Boyd:1995sq,Boyd:1997kz} makes use of dispersive bounds and applies a conformal transformation to approximate the form factors as a series expansion. The conformal transformation maximizes the statistical power of the data by ensuring a fast convergence of the expansion. Following Ref.~\cite{Grinstein:2017nlq} we introduce the conformal variable
\begin{align}
  z = \frac{ \sqrt{w+1} - \sqrt{2 } }{ \sqrt{w+1} + \sqrt{2} } \, ,
\end{align}
 and parameterize the form factors in terms of $\{ a_n, b_n, c_n \}$ expansion coefficients
\begin{align}
 g(z) & = \frac{1}{ P_g(z) \, \phi_g(z)} \, \sum_{n=0}^{n_a} \, a_n \, z^n \, ,  \\
  f(z) & = \frac{1}{ P_f(z) \, \phi_f(z)} \, \sum_{n=0}^{n_b} \, b_n \, z^n \, ,  \\
  F_1(z) & =  \frac{1}{ P_{F_1}(z) \, \phi_{F_1}(z)} \, \sum_{n=0}^{n_c} \, c_n \, z^n \, .  \\
\end{align}
Here $n_{a/b/c}$ denotes the truncation order of the expansion. Note that $c_0$ and $b_0$ are not independent, but are related via
\begin{align}
 c_0 = \left(  \frac{ (m_B - m_{D^*} ) \, \phi_{F_1}(0) }{ \phi_f(0)}  \right) b_0 \, .
\end{align}
Further, $P_j(z)$ $(j=g, f, F_1)$ are Blaschke factors, which remove poles for the region $q^2/c^2 < (m_B^2 + m_{D^*}^2)$, and $\phi_j(z)$ are the outer functions~\cite{Grinstein:2017nlq}. 

The CLN parameterization~\cite{Caprini:1997mu} applies dispersive bounds and incorporates quark model inputs from QCD sum rules to obtain a prediction for a $z$ expansion of $h_{A_1}$, with coefficients depending only on a slope parameter $\rho^2$, and normalizations $R_{1/2}(1)$. The parametrization incorporates corrections to $R_{1/2}(w)$ up to second order in $(w-1)$:
\begin{align}
 h_{A_1}(z)  & = h_{A_1}(w=1) \bigg( 1 - 8 \rho^2 \, z + (53 \, \rho^2 - 15) \, z^2  \nonumber \\
 & \phantom{h_{A_1}(1) \bigg(\qquad\qquad } - (231 \rho^2 - 91)\, z^3 \bigg) \, , \\
 R_1(w) &  = R_1(1) - 0.12 \, (w-1) \, + 0.05 \, (w-1)^2 \, , \\ 
 R_2(w) &  = R_2(1) + 0.11 \, (w-1) \, - 0.06 \, (w-1)^2 \, . 
\end{align}

In the following both of these parameterization are used to determine \Vcba\ from our measurements of the one-dimensional hadronic recoil and decay angle projections of the $B \to D^* \ell \bar \nu_\ell$ decay rate.  The decay rate is fully parameterized in terms of $w$ and the three angles introduced in Fig.~\ref{fig:helicity_angle_defition}:
\begin{itemize}
 \item $\cos \theta_\ell$: The angle between the lepton and the direction opposite the $B$ meson in the virtual $W$-boson rest frame.
 \item $\cos \theta_V$: The angle between the $D$ meson and the direction opposite the $B$ meson in the $D^*$ rest frame.
 \item $\chi$: The azimuthal angle between the two decay planes spanned by the $W-\ell$ and $D^*-D$ systems in the $B$ meson rest frame.
\end{itemize}

\section{The Belle detector and data set}\label{sec:dataset}
We analyze the full Belle data set of \mbox{$(772 \pm 10) \times 10^6$} \PB meson pairs, produced at the KEKB accelerator complex~\cite{KEKB} with a center-of-mass energy of $\sqrt{s} = \SI{10.58}{GeV}$ at the $\Upsilon(4S)$ resonance. In addition, we use $\SI{79}{fb^{-1}}$ of collision data recorded $\SI{60}{MeV}$ below the $\Upsilon(4S)$ resonance peak to derive corrections and carry out cross-checks.

The Belle detector is a large-solid-angle magnetic spectrometer that consists of a silicon vertex detector (SVD), a 50-layer central drift chamber (CDC), an array of aerogel threshold  \v{C}erenkov counters (ACC),  a barrel-like arrangement of time-of-flight scintillation counters (TOF), and an electromagnetic calorimeter comprised of CsI(Tl) crystals (ECL) located inside a superconducting solenoid coil that provides a \SI{1.5}{T} magnetic field. An iron flux return located outside of the coil is instrumented to detect $K^0_L$ mesons and to identify muons (KLM). A more detailed description of the detector, its layout and performance can be found in Ref.~\citep{Abashian:2000cg} and in references therein.

Charged tracks are identified as electron or muon candidates by combining information from multiple subdetectors into a lepton identification likelihood ratio, $\mathcal{L}_\mathrm{LID}$. For electrons the identifying features are the ratio of the energy deposition in the ECL with respect to the reconstructed track momentum, the energy loss in the CDC, the shower shape in the ECL, the quality of the geometrical matching of the track to the shower position in the ECL, and the photon yield in the ACC~\citep{HANAGAKI2002490}. Muon candidates are identified from charged track trajectories extrapolated to the outer detector. The identifying features are the difference between expected and measured penetration depth as well as the transverse deviation of KLM hits from the extrapolated trajectory~\citep{ABASHIAN200269}. Charged tracks are identified as pions or kaons using a likelihood classifier, which combines information from the CDC, ACC, and TOF subdetectors. In order to avoid the difficulties understanding the efficiencies of reconstructing $K^0_L$ mesons, they are not explicitly reconstructed in what follows.
Photons are identified as energy depositions in the ECL without an associated track.

We carry out the entire analysis in the Belle~II analysis software framework~\cite{basf2}. The recorded Belle collision data and simulated Monte Carlo (MC) samples are converted using the software described in Ref.~\cite{b2b2}. MC samples of \PB meson decays and non-resonant processes are simulated using the \texttt{EvtGen} generator~\citep{EvtGen}. The MC sample sizes correspond to approximately ten and six times the Belle collision data for \PB meson and continuum decays, respectively. The interactions of particles traversing the detector are simulated using \texttt{Geant3}~\citep{Geant3}. Electromagnetic final-state radiation (FSR) is simulated using the \texttt{PHOTOS}~\citep{Photos} package. The efficiencies from the MC simulation are corrected using data-driven methods. In particular, the slow pion efficiency, which impacts the slope of the form factor and the determination of \Vcba\, has been determined in differential bins of the slow pion momentum, using $B \to D^*\pi$ data.
We update the branching fractions for the $B\to D^{(*,**)} \ell \bar{\nu}_\ell$ decay modes and the consecutive $D^{(*)}$ decays to the latest values in Ref.~\cite{pdg:2022}. The branching fraction gap between the inclusive $B\to X_c \ell \bar{\nu}_\ell$ decays and the sum-of-exclusive decays is filled with $B\to D^{(*)}\eta \ell \bar{\nu}_\ell$ and $B\to D^{(*)}\pi \pi \ell \bar{\nu}_\ell$ decays.
The differential distributions of the $B\to D \ell \bar{\nu}_\ell$ decays are updated by reweighting the simulated data to the BGL form factor parametrization obtained from fits provided in Ref.~\citep{glattauer_BGL_params},
and for the $B\to D^* \ell \bar{\nu}_\ell$ decays to the form factor parameters given in Ref.~\citep{Ferlewicz:2020lxm}.
The decay model for the $B\to D^{**} \ell \bar{\nu}_\ell$ decays is updated to Ref.~\citep{Bernlochner:2016bci}.

\section{Event reconstruction and selection}\label{sec:reco}
We select a sample of $\bdslnu$ events with which we determine the distributions of the kinematic variables $w$, $\cos\theta_\ell$, $\cos\theta_V$, and $\chi$. In the following, $\bdslnu$ refers to all the decay channels considered. When we refer to any specific decay, the charge of the $B$ or $D^{(*)}$ meson is explicitly stated. We consider both charged and neutral $B$ mesons with the decay chains $\Bbar{}^0 \to D^{*+} \ell \overline{\nu}_\ell$, $D^{*+} \to D^0 \pi^+$ and $D^{*+} \to D^+ \pi^0$, and $B^- \to D^{*0} \ell \overline{\nu}_\ell$ with $D^{*0} \to D^0\pi^0$ respectively. The decay $D^{*0} \to D^0\gamma$ has a different Lorentz structure resulting in different angular distributions, requiring a dedicated analysis, and is therefore omitted. We reconstruct the following decays of the $D$ mesons:
$D^+ \to K^- \pi^+ \pi^+$,
$D^+ \to K^- \pi^+ \pi^+ \pi^0$,
$D^+ \to K^- \pi^+ \pi^+ \pi^+ \pi^-$,
$D^+ \to K_\mathrm{S}^0 \pi^+$,
$D^+ \to K_\mathrm{S}^0 \pi^+ \pi^0$,
$D^+ \to K_\mathrm{S}^0 \pi^+ \pi^+ \pi^-$,
$D^+ \to K_\mathrm{S}^0 K^+$,
$D^+ \to K^+ K^- \pi^+$,
$D^0 \to K^- \pi^+$,
$D^0 \to K^- \pi^+ \pi^0$,
$D^0 \to K^- \pi^+ \pi^+ \pi^-$,
$D^0 \to K^- \pi^+ \pi^+ \pi^- \pi^0$,
$D^0 \to K_\mathrm{S}^0 \pi^0$,
$D^0 \to K_\mathrm{S}^0 \pi^+ \pi^-$,
$D^0 \to K_\mathrm{S}^0 \pi^+ \pi^- \pi^0$, and
$D^0 \to K^- K^+$.

Primary charged tracks are required to have impact parameters $dr < \SI{2}{cm}$ and $|dz| < \SI{4}{cm}$, which are defined perpendicular to, and along the beam-axis, respectively. In addition to selecting the primary charged tracks to be consistent with the interaction point (IP), a transverse momentum of $p_T > \SI{0.1}{GeV/c}$ is required for these tracks. Muons, electrons, charged pions, kaons and protons are identified using information from the particle identification subsystems. Electron (Muon) tracks are further required to have a momentum in the lab frame of $p^\mathrm{Lab} > \SI{0.3}{GeV/c}$ ($p^\mathrm{Lab} > \SI{0.6}{GeV/c}$). The momenta of particles identified as electrons are corrected for bremsstrahlung by adding photons within a $\SI{2}{\degree}$ cone defined around the electron track at the point of closest approach to the IP.

Photons are selected with an energy of  $E_\gamma > \SI{100}{MeV}$, $\SI{150}{MeV}$, and $\SI{50}{MeV}$ in the forward endcap (covering the polar angle ($12^\circ < \theta <31^\circ$), backward endcap ($132^\circ < \theta < 157^\circ$) and barrel ($32^\circ < \theta < 129^\circ$) part of the calorimeter, respectively. 
The $\pi^0$ candidates are reconstructed from photon pairs and selected if their reconstructed invariant mass is within $\SI{104}{MeV/c^2}$ and $\SI{165}{MeV/c^2}$. Additionally, the difference of the reconstructed  $\pi^0$ mass from the nominal mass of $m_{\pi^0} = 135 \, \mathrm{MeV/c^2}$ has to be smaller than $3\sigma$ of the estimated mass resolution.

$K_\mathrm{S}^0$ mesons are reconstructed from two oppositely charged tracks and selected with a multivariate method and within a reconstructed invariant mass window of $\SI{398}{MeV/c^2}$ and $\SI{598}{MeV/c^2}$. The difference of the reconstructed $K_\mathrm{S}^0$ mass from the nominal value of $m_{K_\mathrm{S}^0} = 498 \, \mathrm{MeV/c^2}$ has to be smaller than $3\sigma$ of the estimated mass resolution. A description of the multivariate method can be found in Ref.~\cite{Belle:2018xst}.

$D$ meson candidates are reconstructed in the sixteen decays listed above, with mass window selection criteria depending on the final state particles involved. 
The $\pi^0$ daughter particles from these $D$ meson candidates must have a center-of-mass momentum $p_{\pi^0}^\mathrm{CMS} > \SI{0.2}{GeV/c}$, except for the final state $D^0 \to K^- \pi^+ \pi^+ \pi^- \pi^0$, where this selection is not applied. To reduce the combinatorial background, we rank the reconstructed $D$ mesons by the absolute difference of the reconstructed mass to the nominal mass ($m_{D^+} = 1.87 \, \mathrm{GeV/c^2}$, $m_{D^0} = 1.86 \, \mathrm{GeV/c^2}$) and select up to ten candidates with the lowest mass difference.

$D^*$ mesons are reconstructed in three different decay channels: $D^{*0}\to D^0\pi^0_\mathrm{slow}$, $D^{*+}\to D^+\pi^0_\mathrm{slow}$, and $D^{*+}\to D^0\pi^+_\mathrm{slow}$. We require charged slow pions to have a center-of-mass momentum smaller than $\SI{0.4}{GeV/c}$, and a mass difference $\Delta M(D, D^*) = M_{D^*}- M_D$ to be smaller than $\SI{0.155}{GeV/c^2}$ ($\SI{0.160}{GeV/c^2}$) for $D^{*+}$ ($D^{*0}$) mesons.

We reconstruct $B_\mathrm{sig}$ candidates with the selected $D^*$ candidates and lepton candidate and only impose a loose selection at this stage by requiring that the reconstructed invariant mass lies in the interval [$1.0$, $6.0$] $\mathrm{GeV/c^2}$ to reduce combinatorial background. 
We perform a global decay chain vertex fix using the \texttt{TreeFitter} \cite{Belle-IIanalysissoftwareGroup:2019dlq} implementation, to retrieve a quality indicator for our candidate particles in the form of the $p$-value of the vertex fit, which is used at a later stage. Events that cannot be fitted successfully are rejected.

$B_\mathrm{tag}$ mesons candidates are reconstructed using the Full Event Interpretation (FEI)~\cite{Keck:2018lcd}. We select candidates with a beam-constrained mass
\begin{align}
    M_\mathrm{bc}^{\mathrm{tag}} = \sqrt{s/2 - \vec p_{\mathrm{tag}}^{\,2}} > \SI{5.27}{GeV/c^2} \, ,
\end{align}
 and energy difference  $\Delta E_{\mathrm{tag}} = E_{\mathrm{tag}} - \sqrt{s}/2$ within the interval 
 \begin{align}
     \SI{-0.15}{GeV} < \Delta E_{\mathrm{tag}} < \SI{0.10}{GeV} \, ,
 \end{align}
 with $p_{\mathrm{tag}} = (E_{\mathrm{tag}} , \vec p_{\mathrm{tag}})$ denoting the 4-momentum of the $B_\mathrm{tag}$ in the center-of-mass frame.
 Exploiting the clean environment provided by the $e^+e^-$ collisions, we impose a completeness constraint on the event by recombining the $\Upsilon(4S)$ candidate from a tag- and signal $B$ meson and require that no additional charged particles are present in the event. The $\Upsilon(4S)$ candidates are reconstructed in combinations of $B_\mathrm{sig}^+ B_\mathrm{tag}^-$,
$B_\mathrm{sig}^0\bar{B}_\mathrm{tag}^0$, 
$B_\mathrm{sig}^\pm B_\mathrm{tag}^0$,
$B_\mathrm{sig}^0 B_\mathrm{tag}^\pm$,
$B_\mathrm{sig}^0 B_\mathrm{tag}^0$,
and their charge conjugates.
The reconstructed invariant mass of the $\Upsilon(4S)$ candidate must be in the range $M^{\Upsilon(4S)} \in [7.0, 13.0]\, \mathrm{GeV/c^2}$.

Continuum events are suppressed using event shape variables, such as the magnitude of the thrust of final-state particles from both $B$ mesons, the reduced Fox-Wolfram moment $R_2$, the modified FoxWolfram moments~\cite{SFW}, and CLEO Cones~\cite{cleocones}. These variables are combined using a multivariate classifier with an optimized implementation of gradient boosted decision trees~\cite{Keck:2017gsv}.

We apply a final best candidate selection on all candidates, to reduce the number of candidates per event to a single one. We select the candidate with the lowest $E_\mathrm{ECL}$, which is the sum of unassigned photon clusters in the full event reconstruction. Subsequently, if there is more than one such candidate, we select the candidate with the smallest $|\Delta E_\mathrm{tag}|$. If this selection remains inconclusive, we select a random candidate.

\section{Reconstruction of kinematic quantities and signal extraction}\label{sec:fit}
\label{sec:signalextraction}

\begin{figure*}
    \centering
    \includegraphics[width=0.24\linewidth]{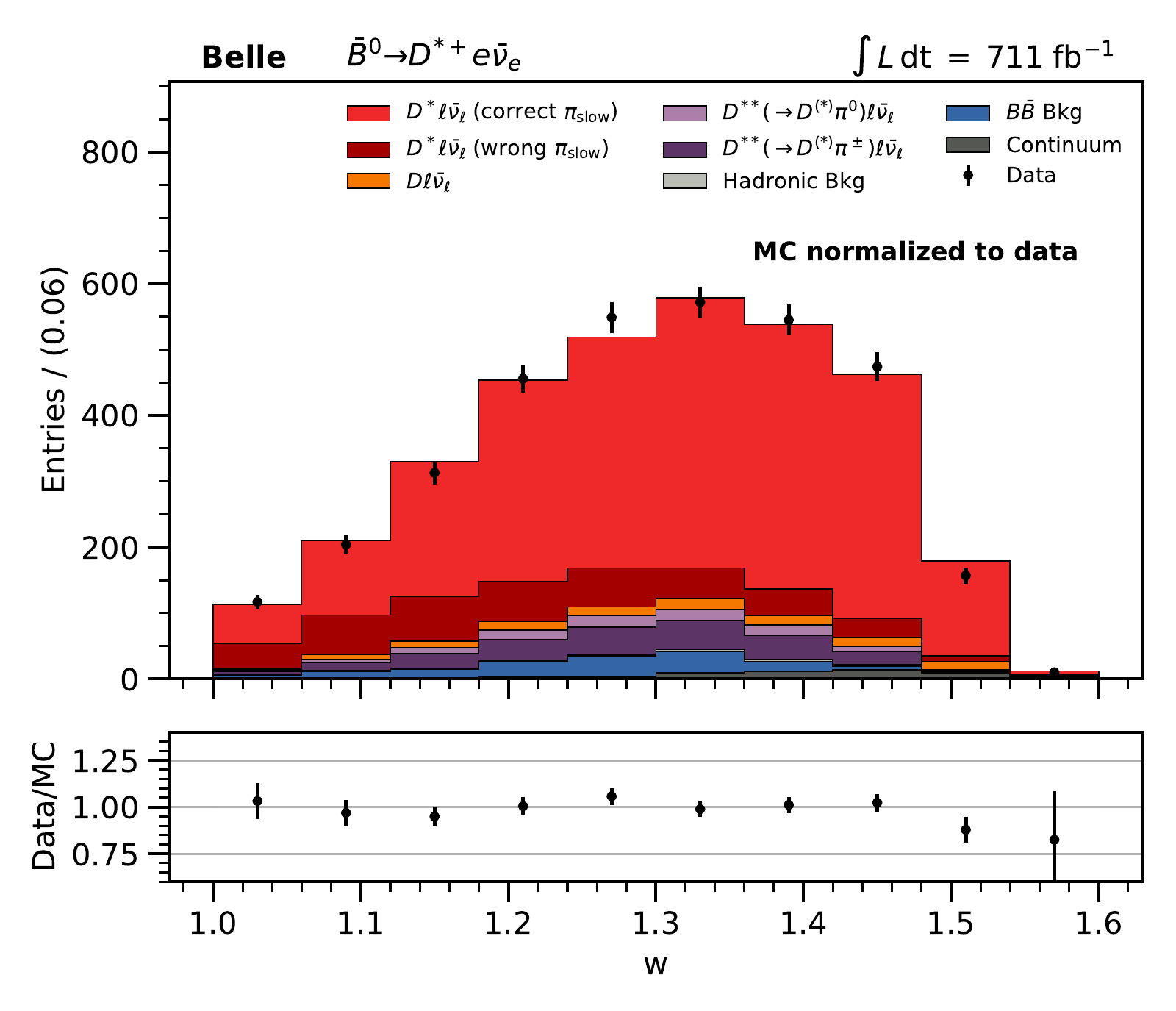}
    \includegraphics[width=0.24\linewidth]{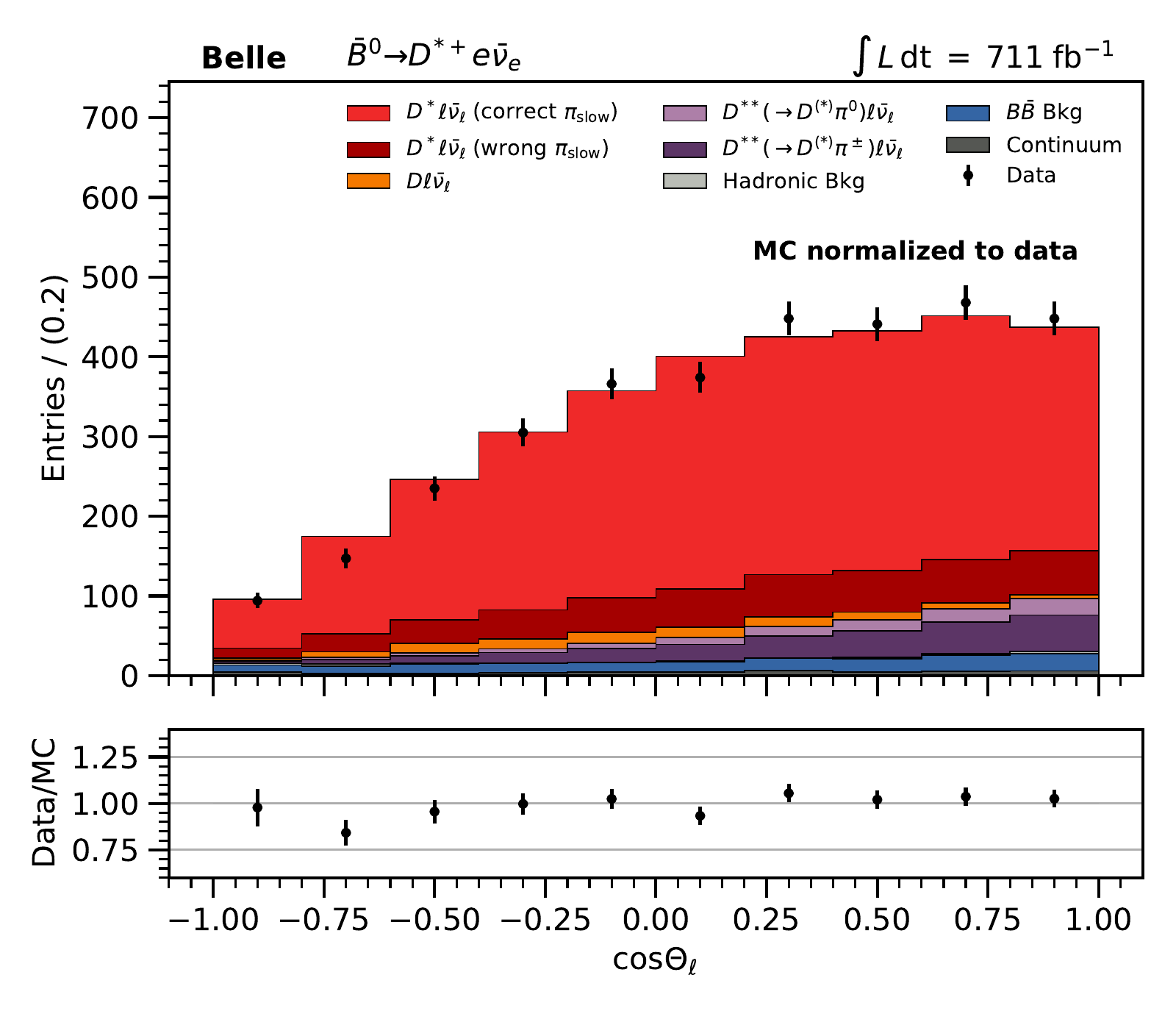}
    \includegraphics[width=0.24\linewidth]{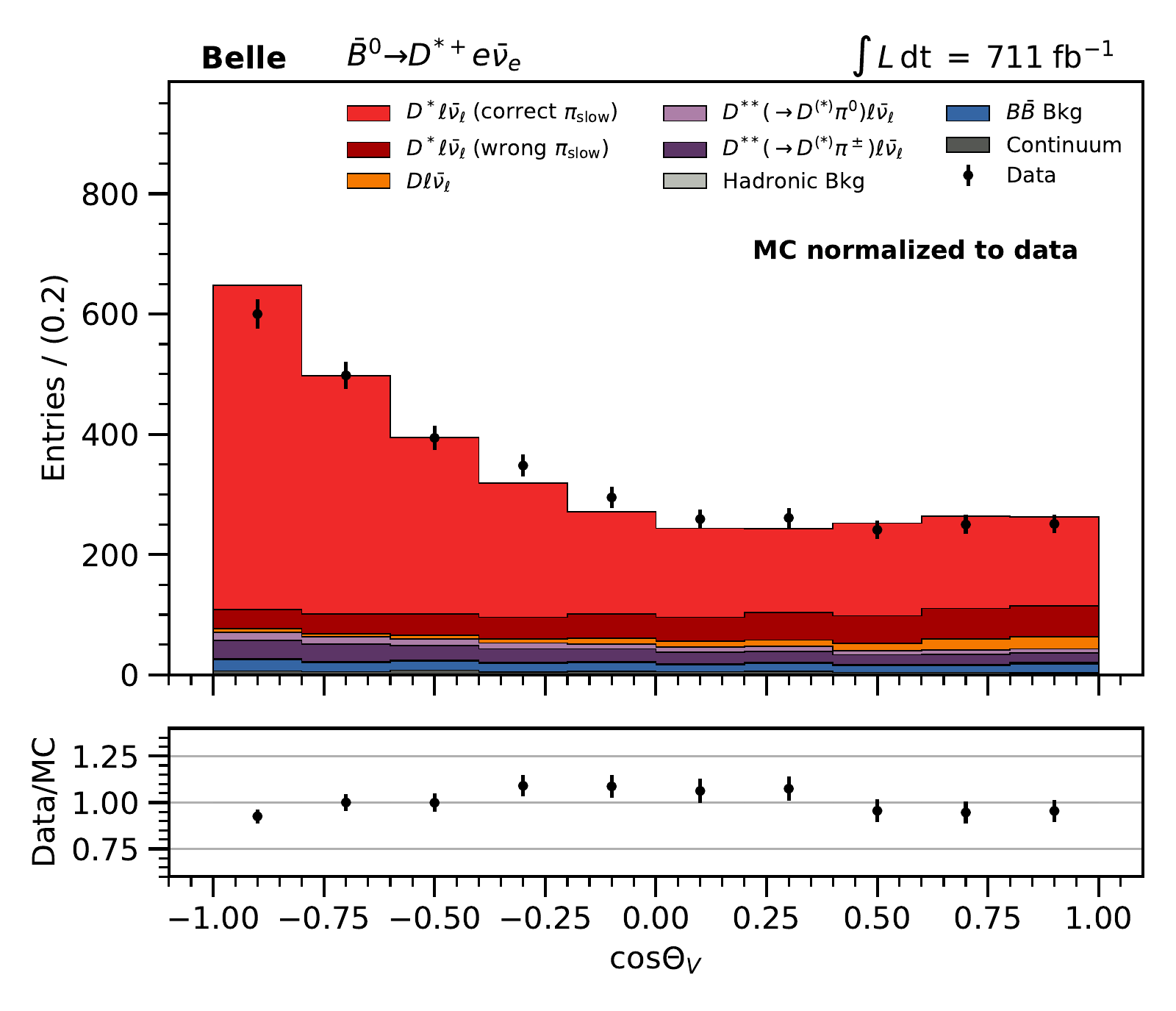}
    \includegraphics[width=0.24\linewidth]{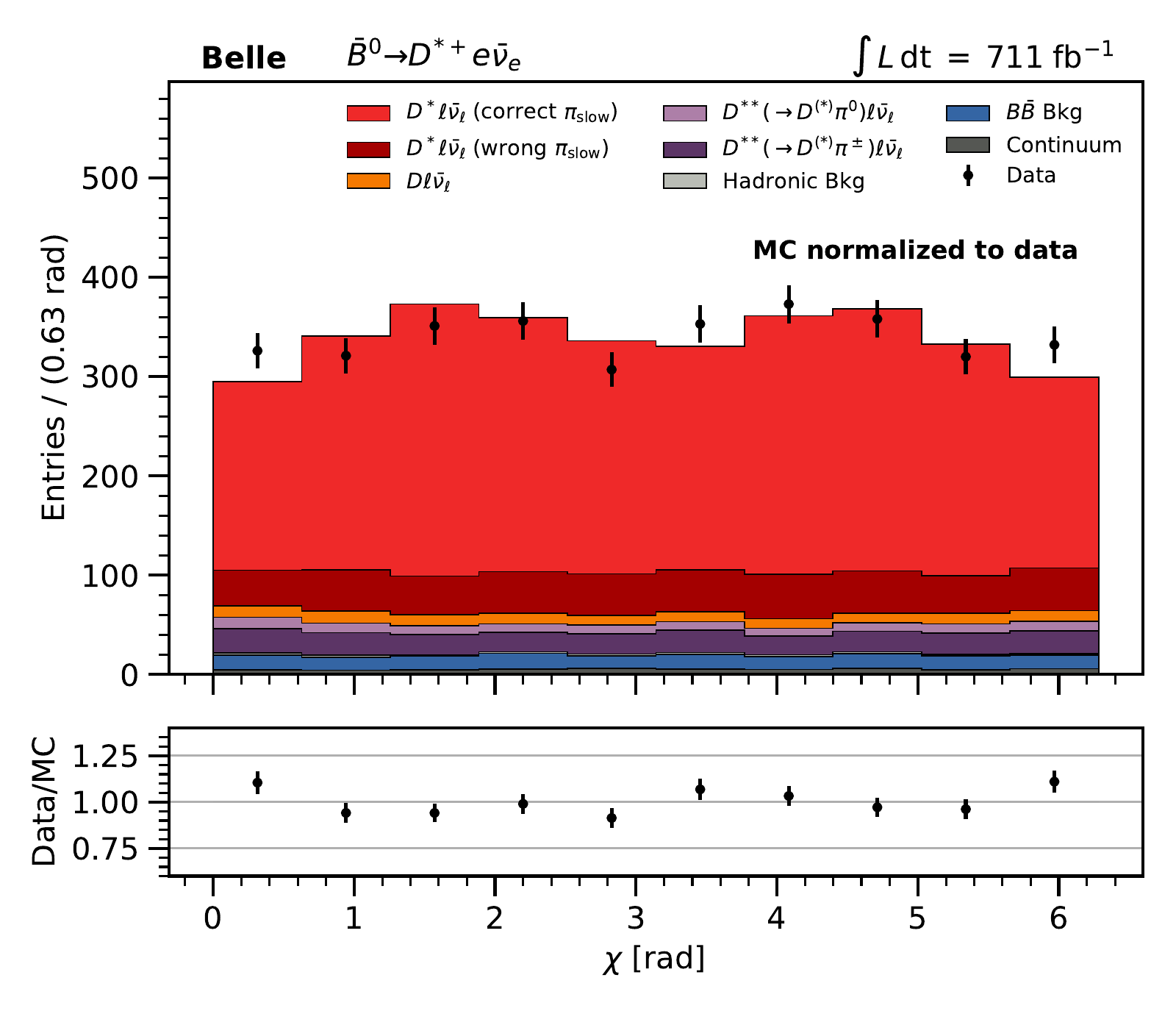}\\
    \includegraphics[width=0.24\linewidth]{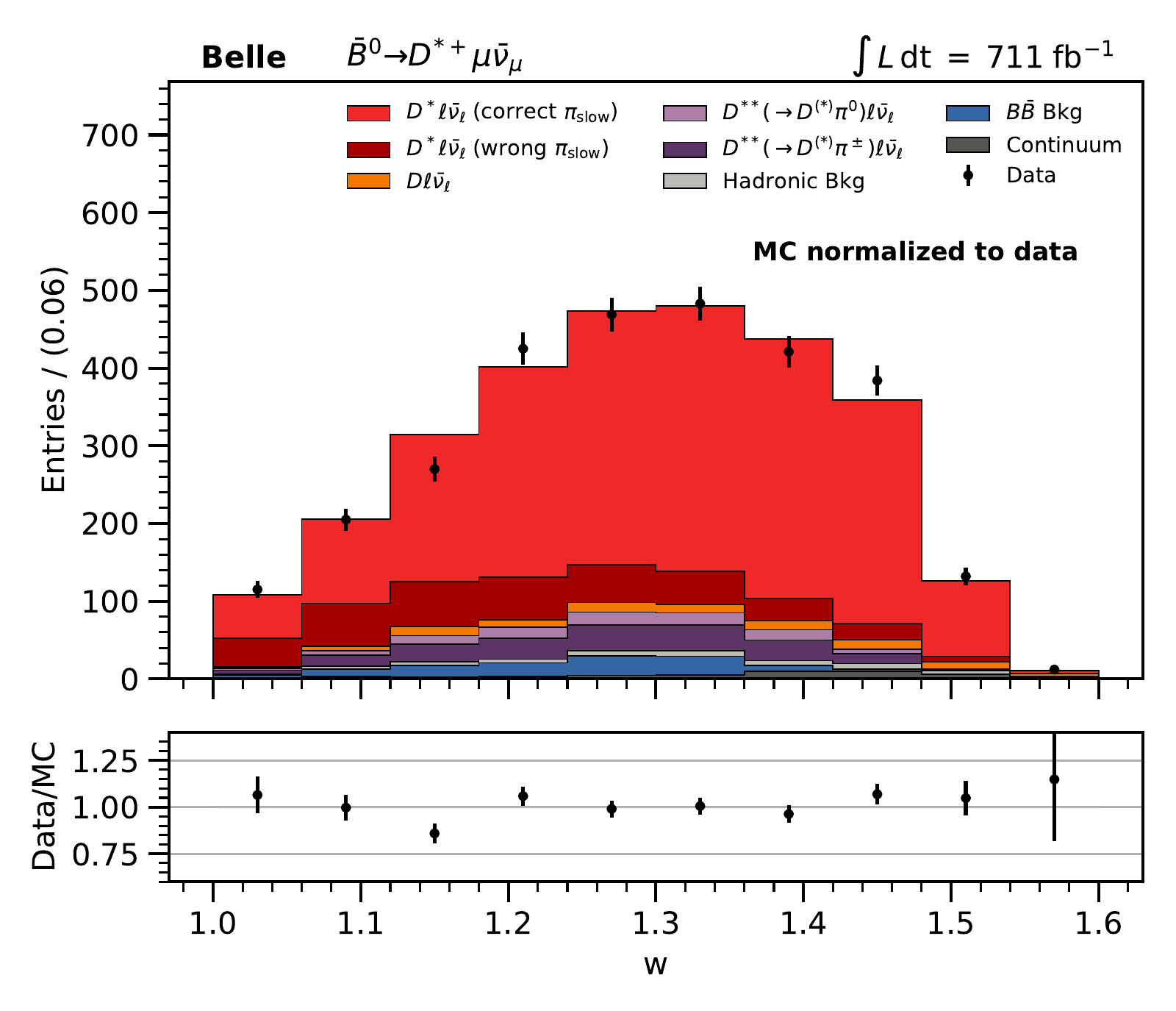}
    \includegraphics[width=0.24\linewidth]{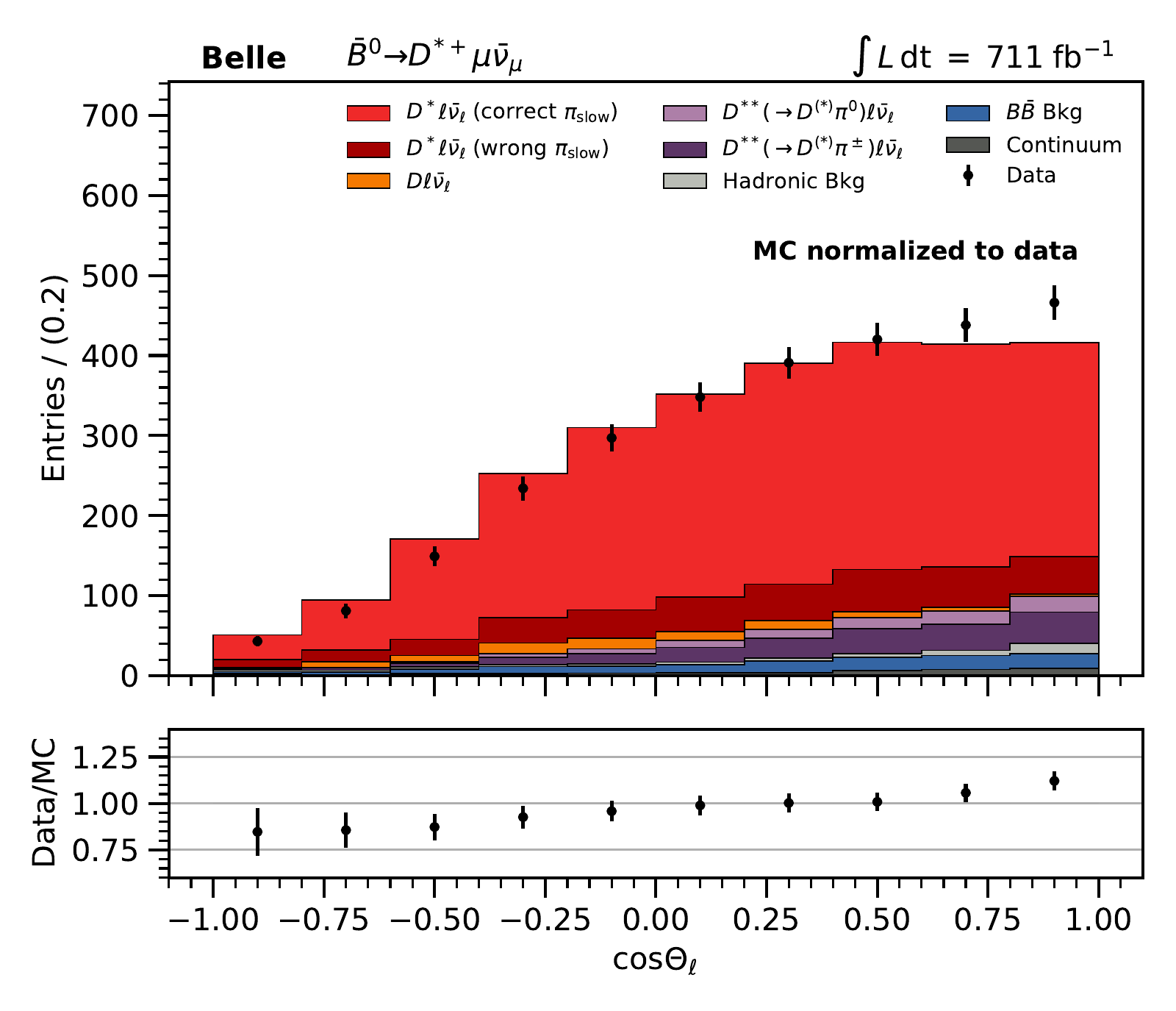}
    \includegraphics[width=0.24\linewidth]{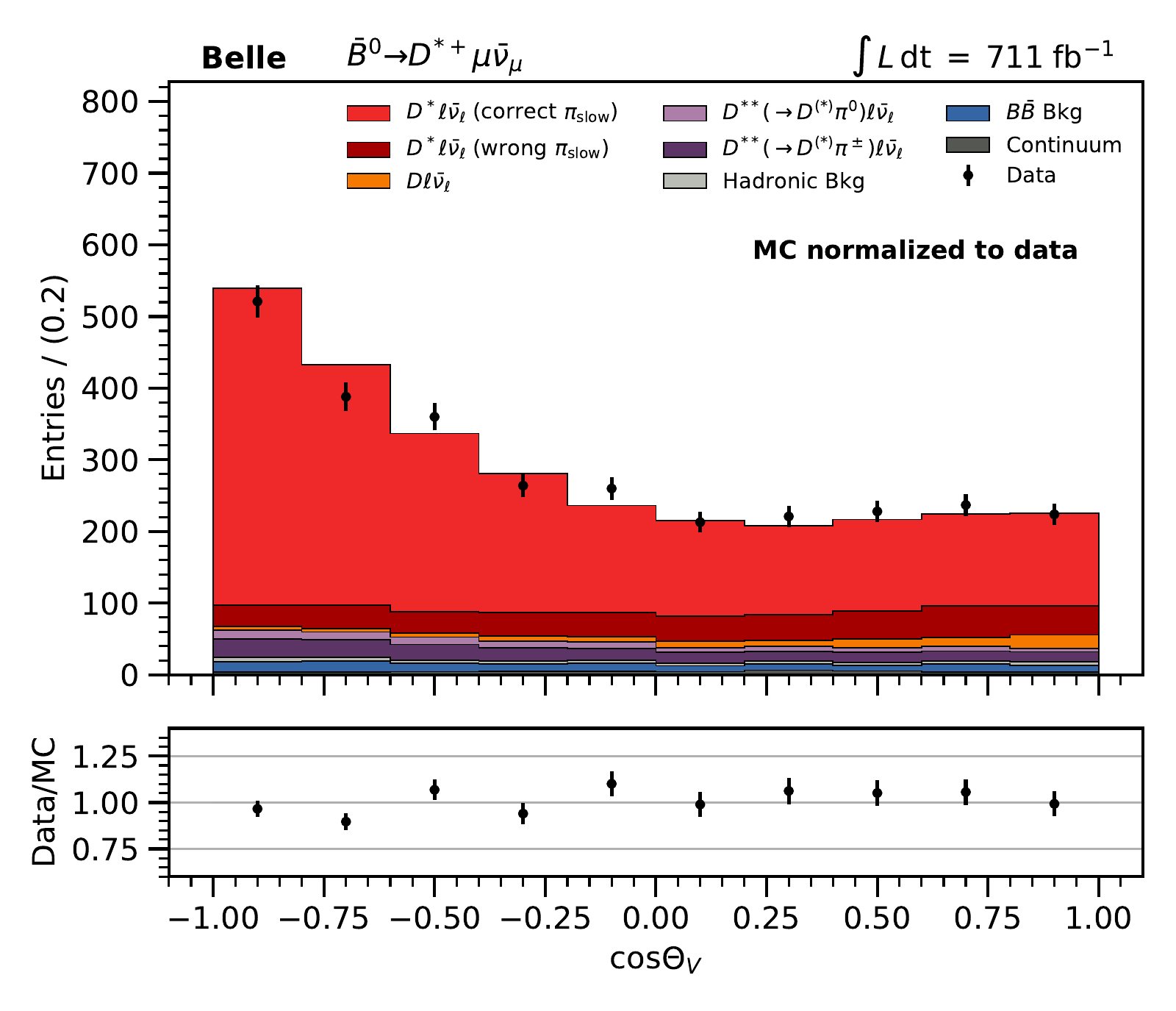}
    \includegraphics[width=0.24\linewidth]{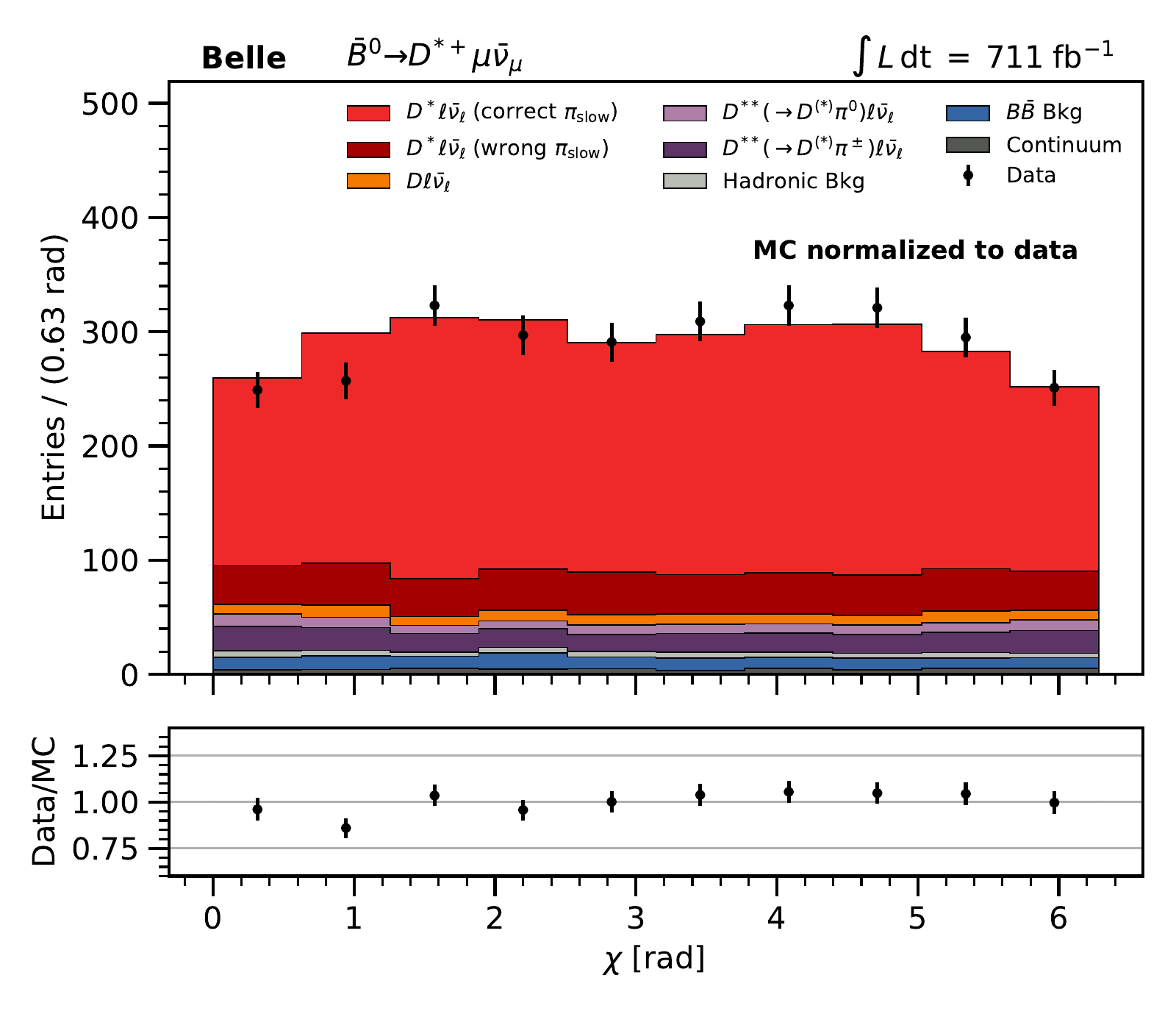}\\
    \includegraphics[width=0.24\linewidth]{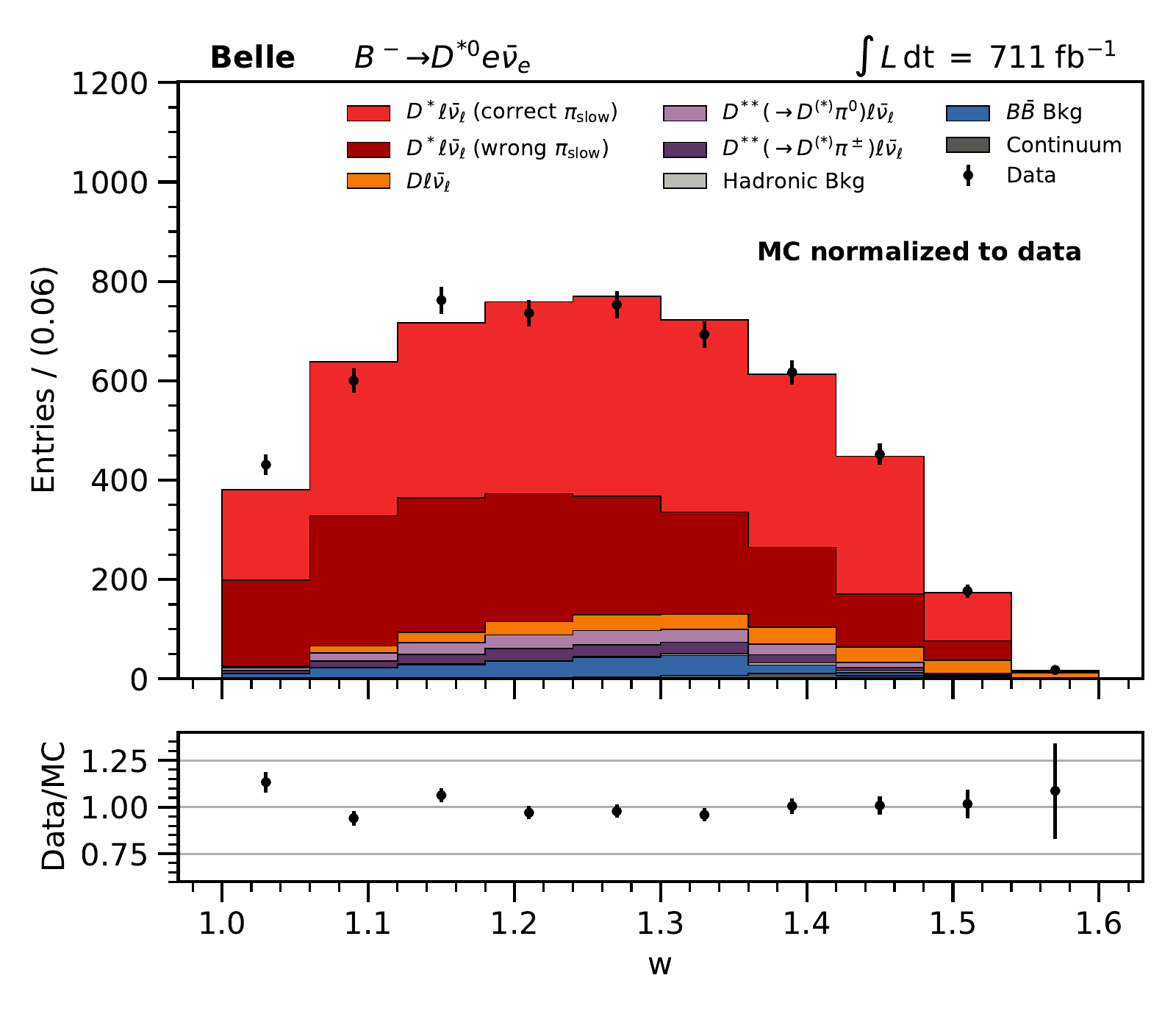}
    \includegraphics[width=0.24\linewidth]{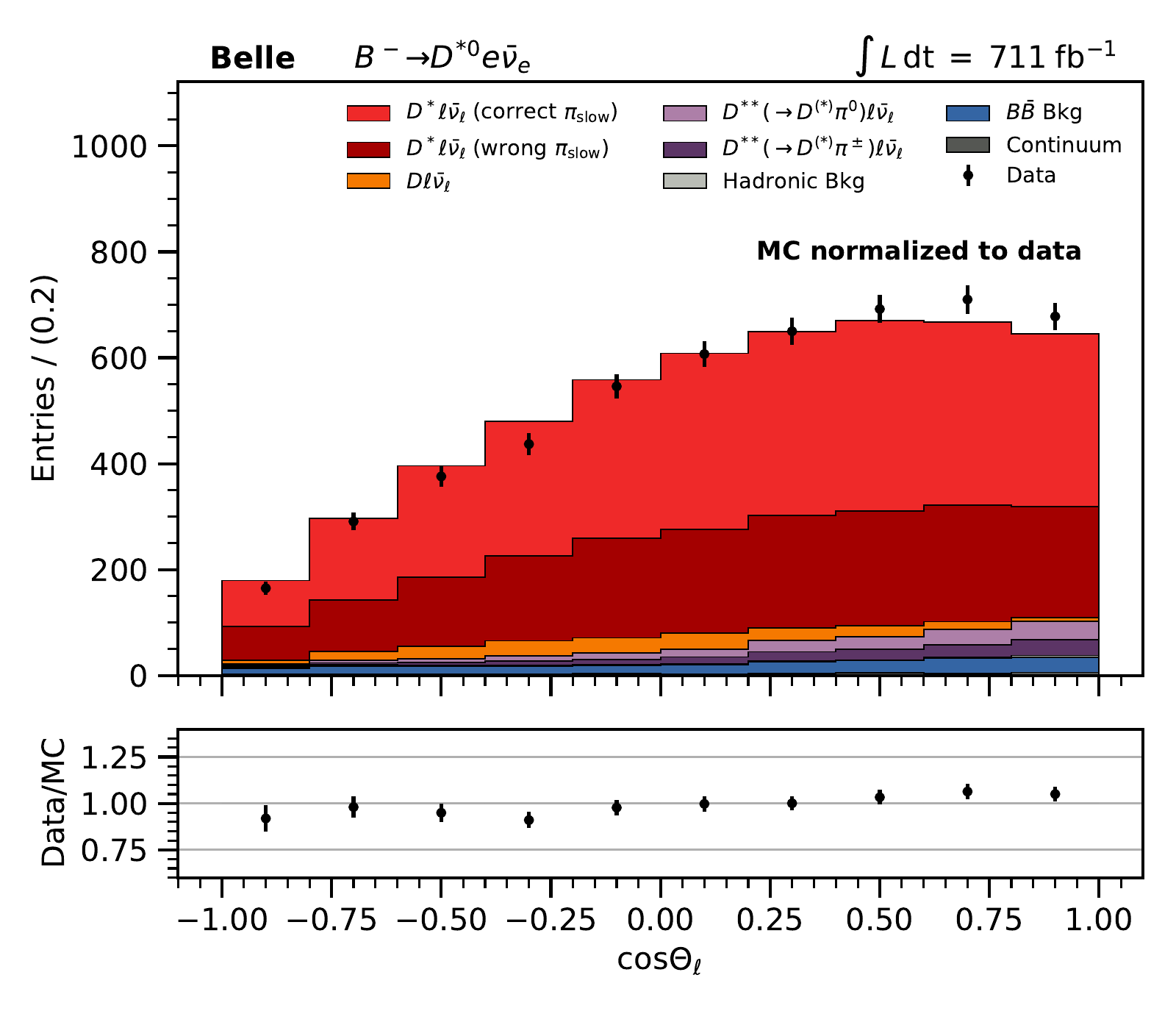}
    \includegraphics[width=0.24\linewidth]{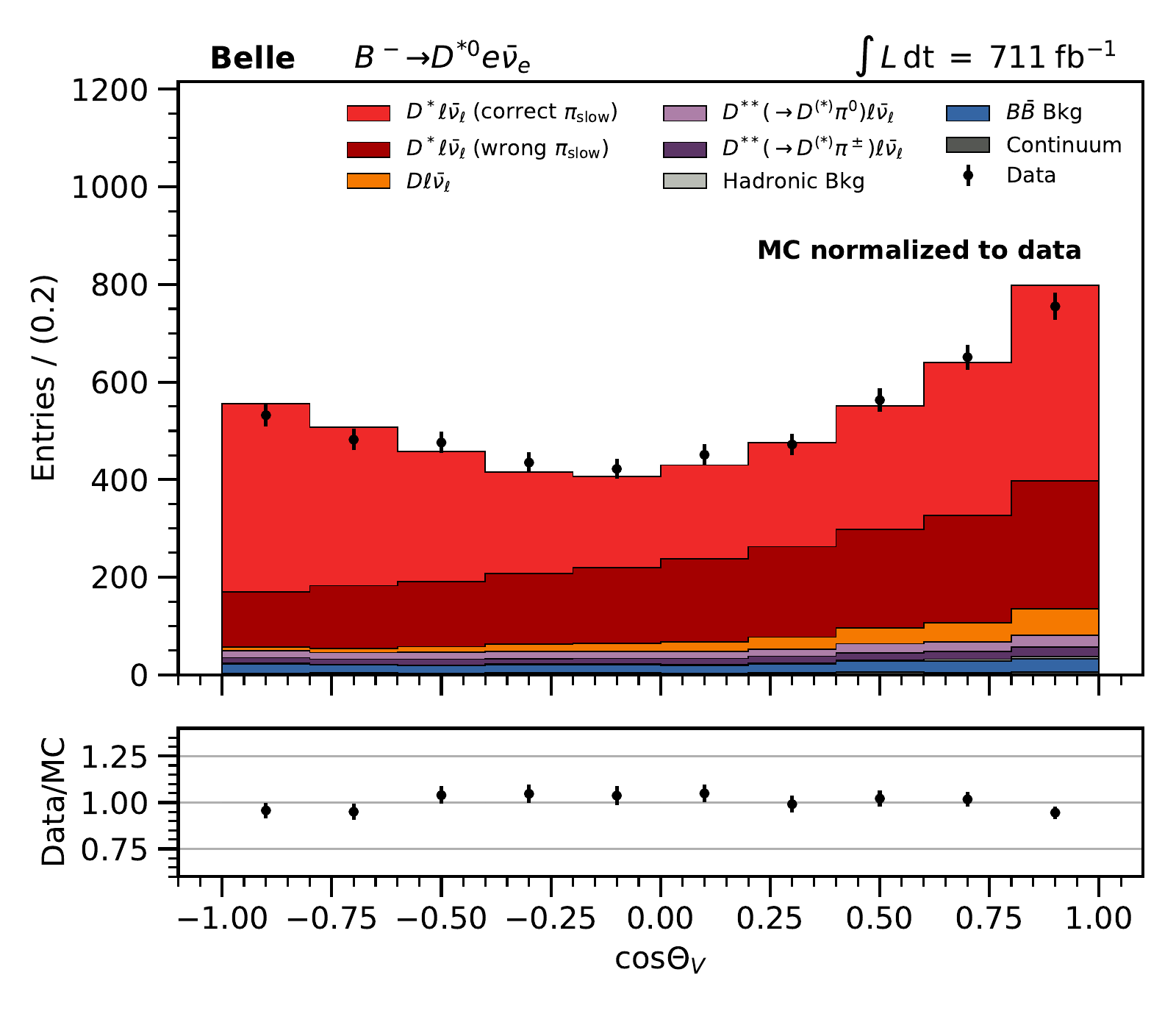}
    \includegraphics[width=0.24\linewidth]{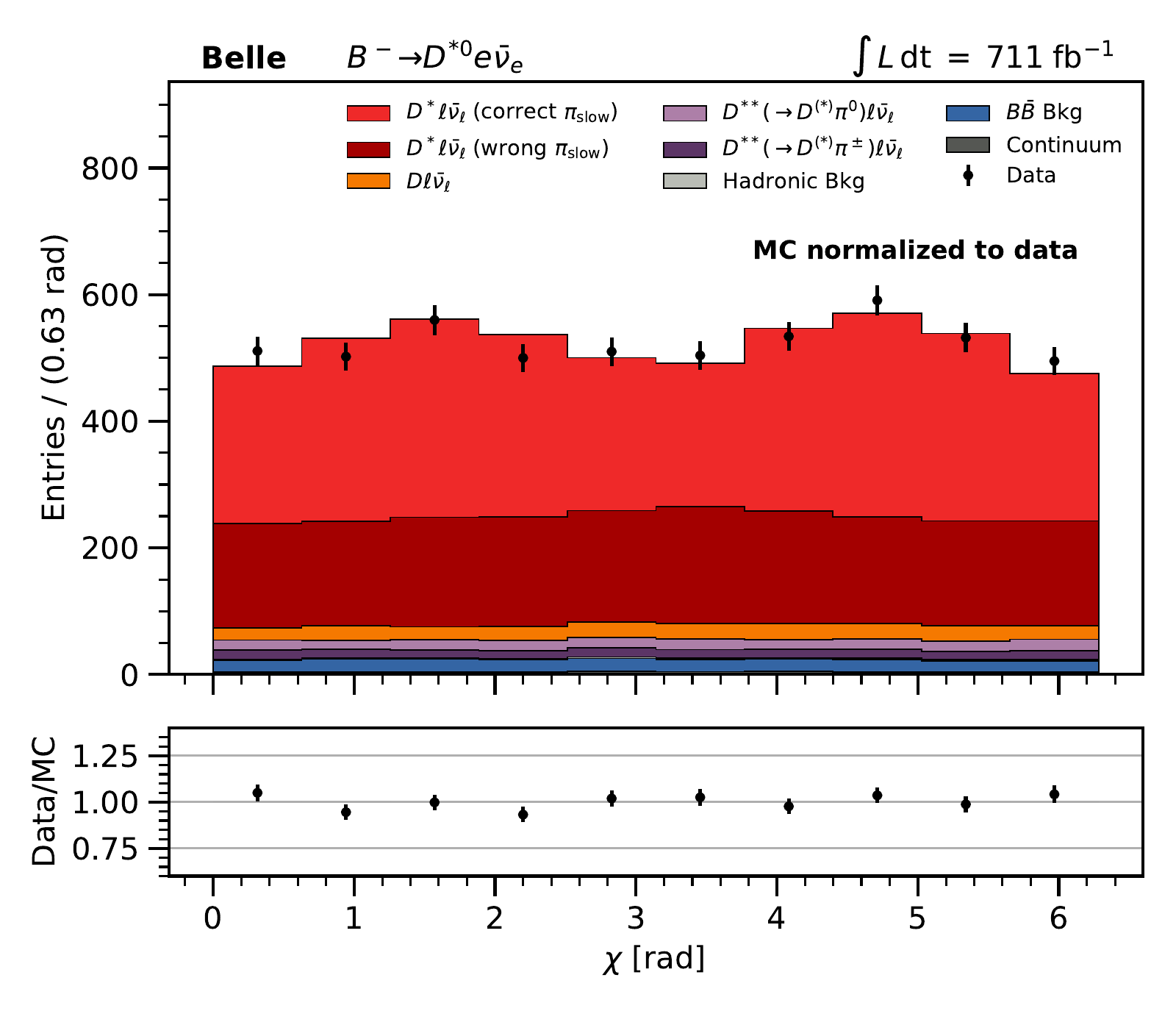}\\
    \includegraphics[width=0.24\linewidth]{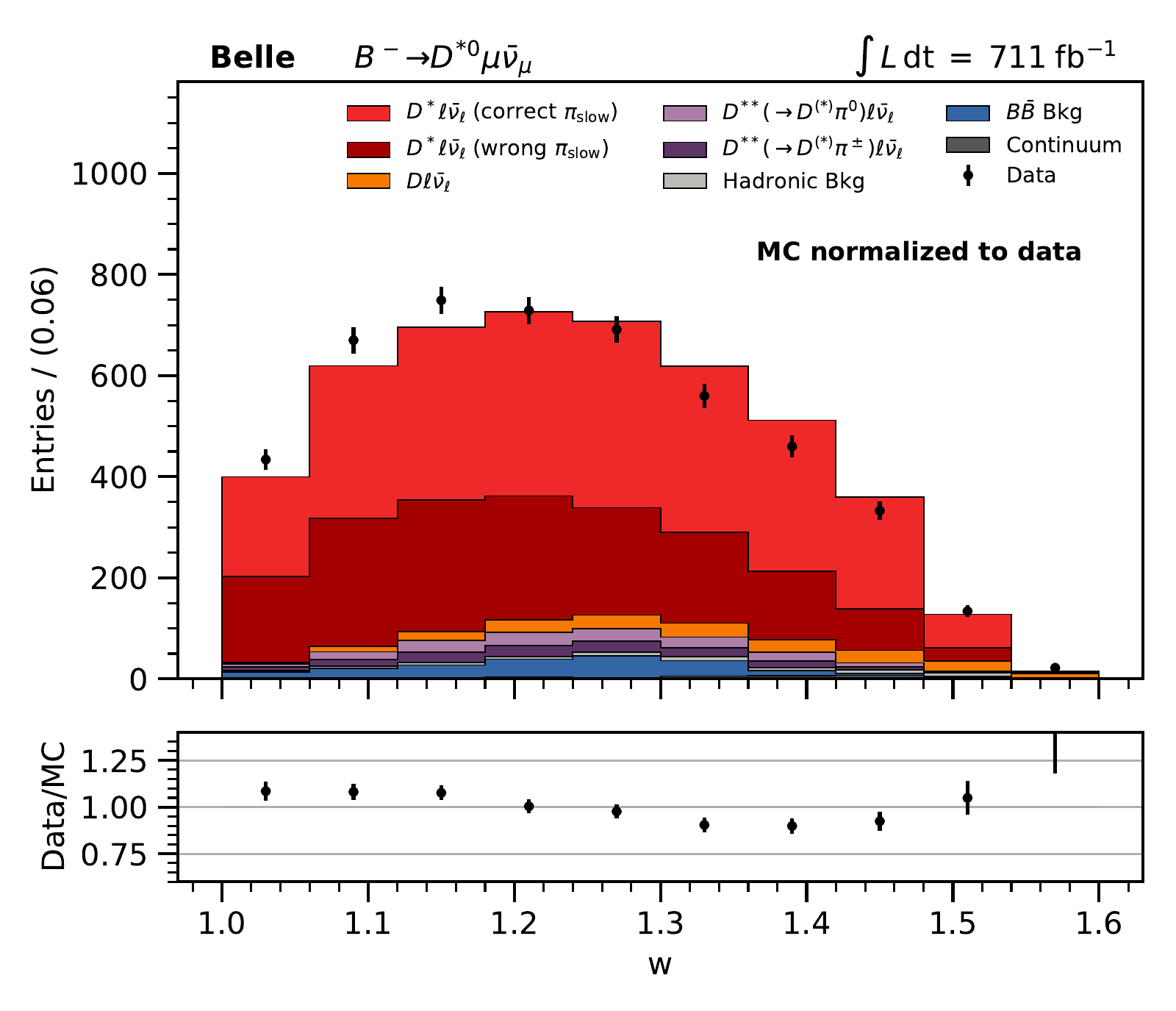}
    \includegraphics[width=0.24\linewidth]{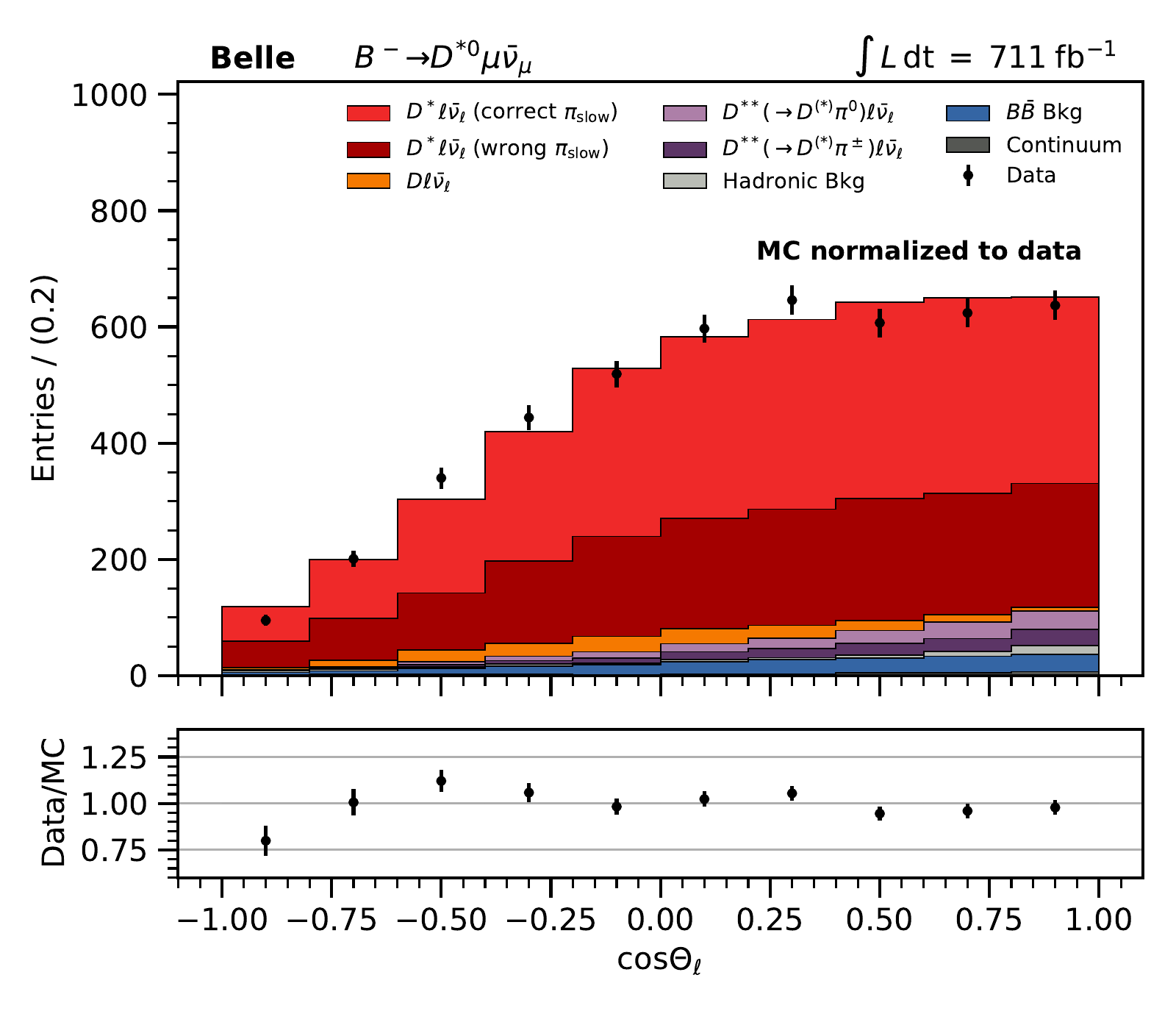}
    \includegraphics[width=0.24\linewidth]{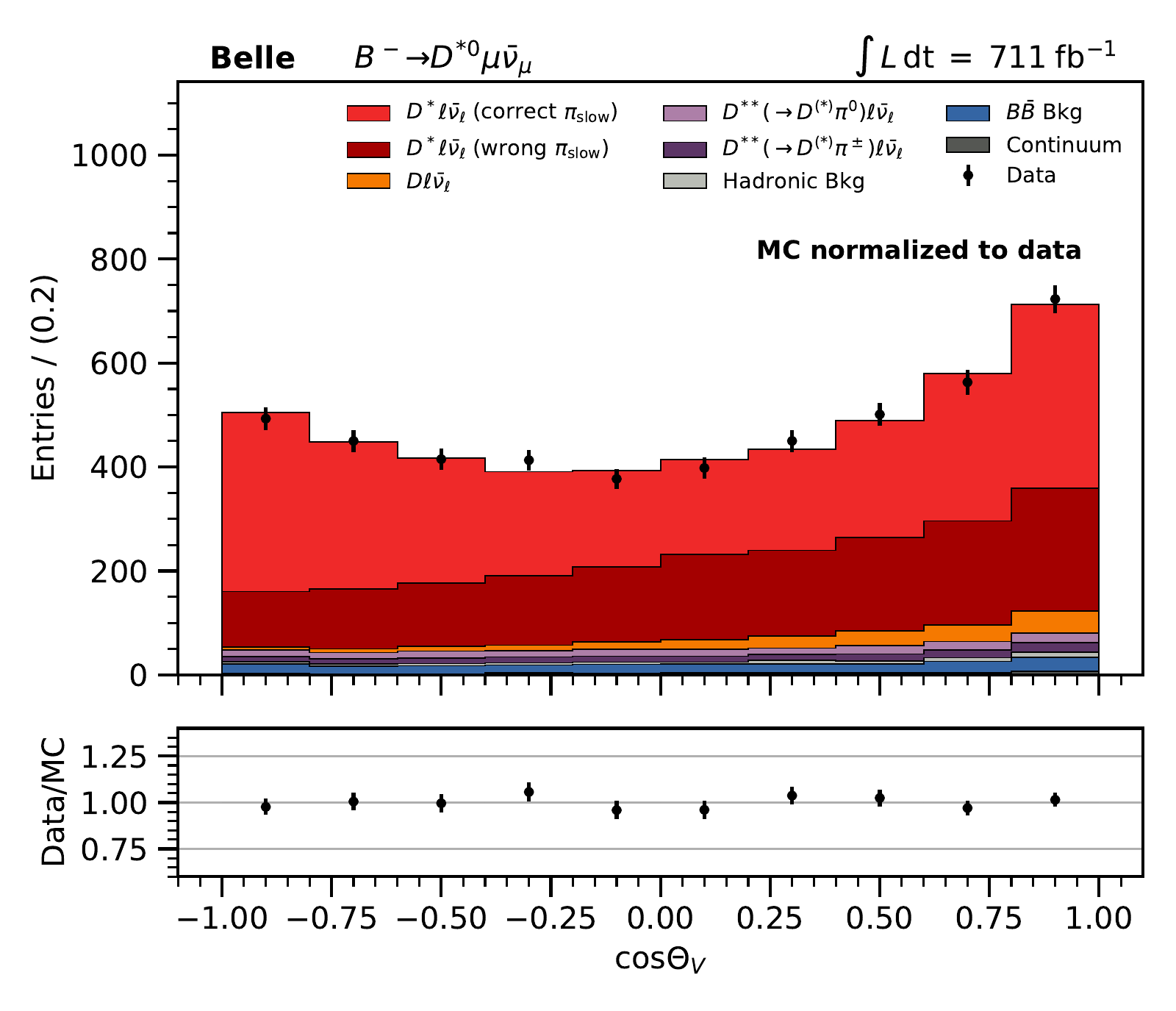}
    \includegraphics[width=0.24\linewidth]{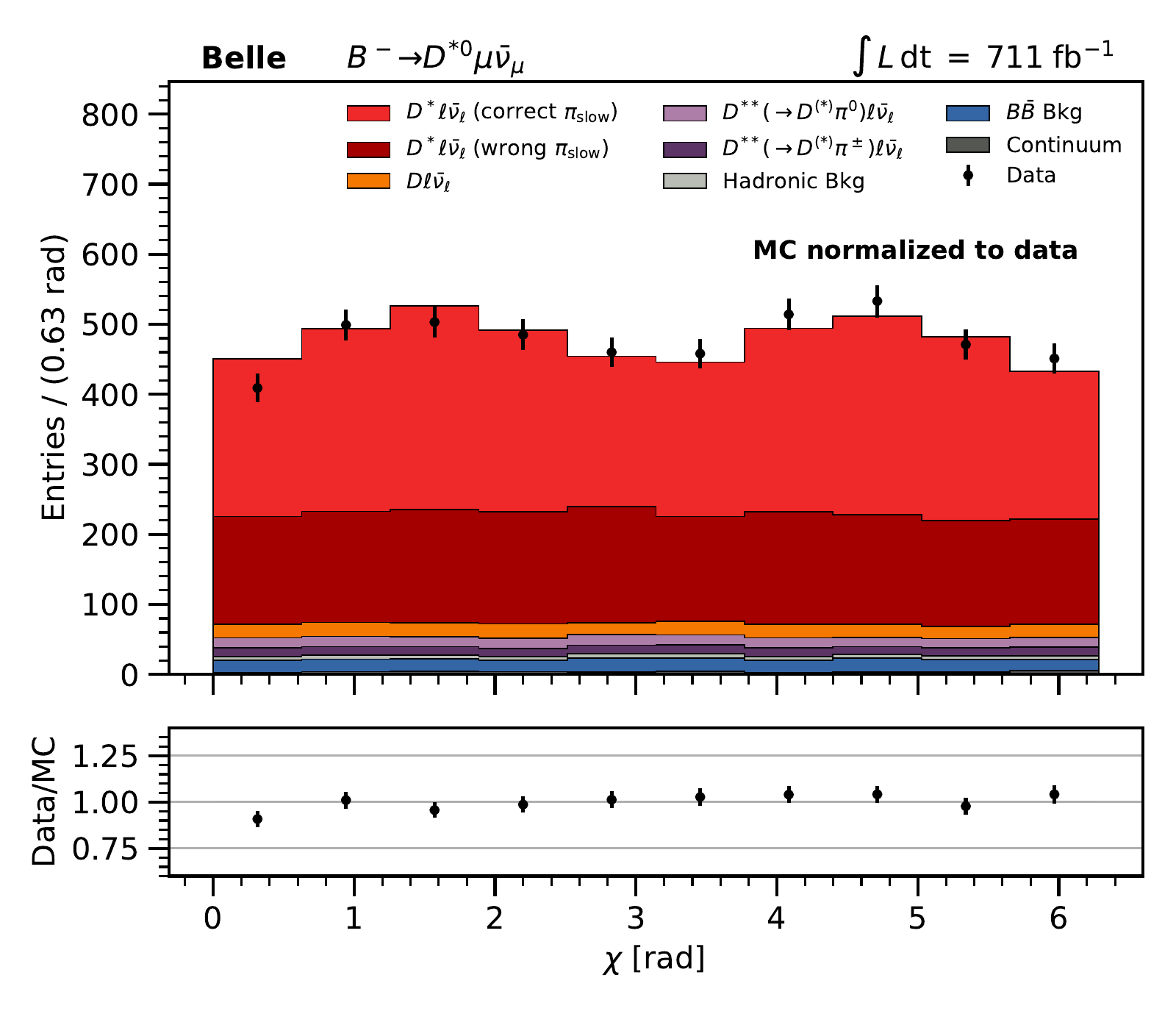}\\
    \caption{The differential distributions of the kinematic variables describing the differential decay rate of $\bdslnu$ are shown in our four considered decay modes. The MC expectation was normalized to the number of observed events in data.}
    \label{fig:kinematicquantities}
\end{figure*}

After our selection is applied, the distributions of the kinematic variables describing the decay are shown in Fig.~\ref{fig:kinematicquantities}. 

We perform background subtraction with binned likelihood fits to the squared missing mass distribution defined as
\begin{equation}
    M_\mathrm{miss}^2 = p_\mathrm{miss}^2 = \left( p_{e^+e^-} - p_\mathrm{tag} - p_{D^*} - p_\ell \right)^2 \,, 
\end{equation}
where the momenta of the colliding $e^+e^-$-pair, the reconstructed tag side $B$, the reconstructed signal side $D^*$, and the signal lepton are denoted as $p_{e^+e^-}$, $p_\mathrm{tag}$, $p_{D^*}$, and $p_\ell$, respectively.

\begin{figure}
    \centering
    \includegraphics[width=\linewidth]{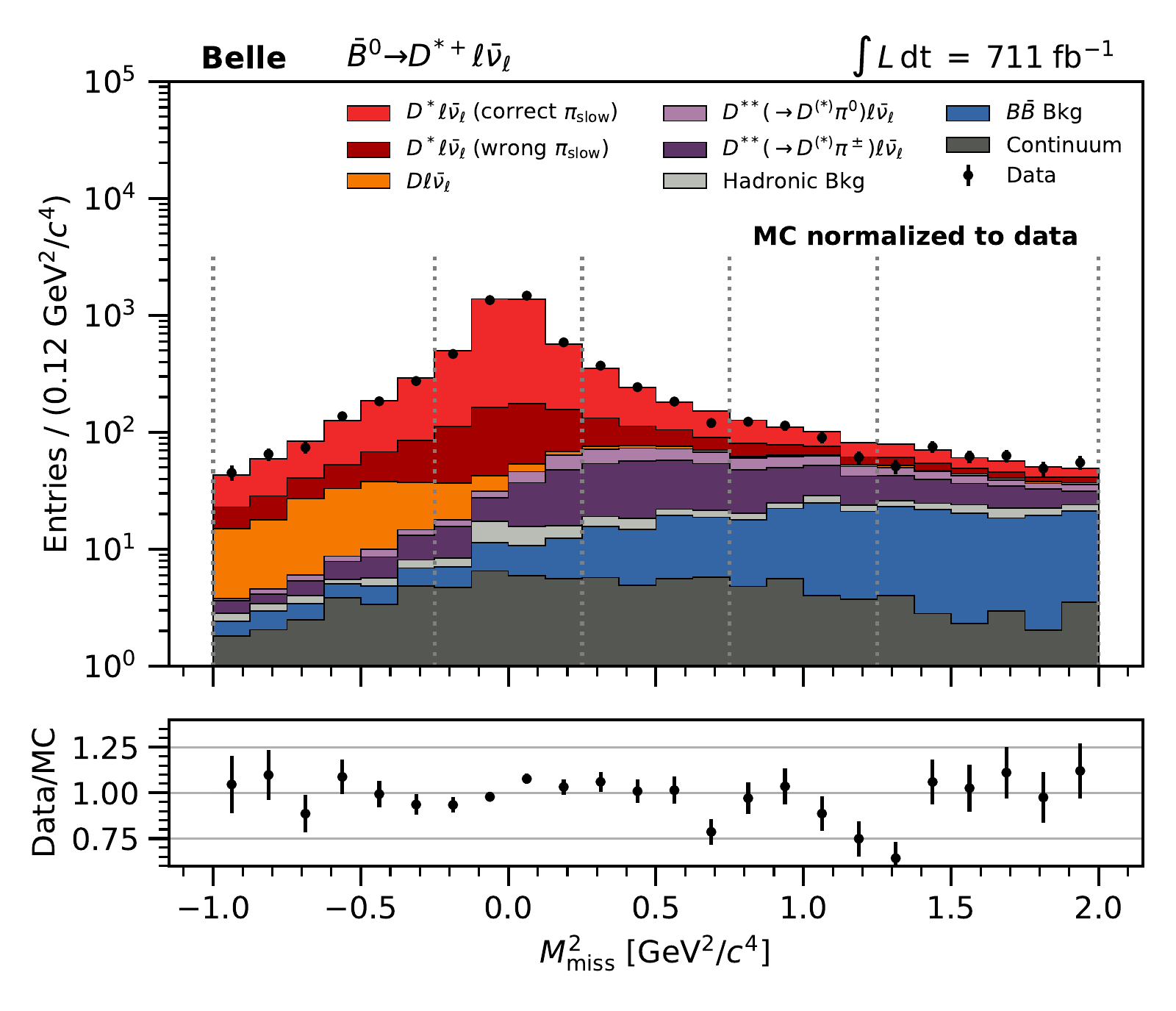}
    \includegraphics[width=\linewidth]{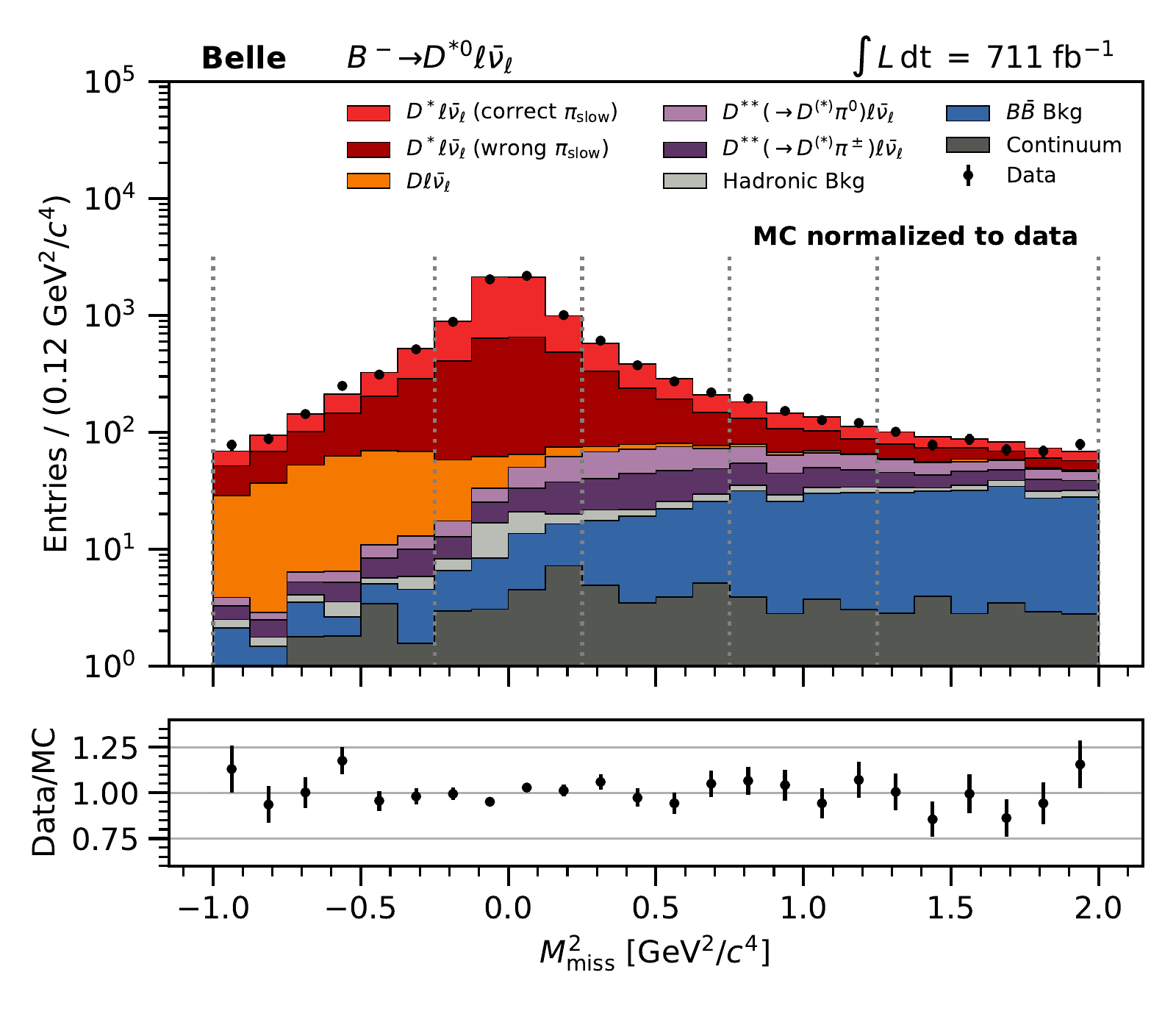}
    \caption{The reconstructed $M_\mathrm{miss}^2$ distribution after our final selection for the $\bar{B}^0 \to D^{*+} \ell \bar{\nu}_\ell$ (top) channel and the $B^- \to D^{*0} \ell \bar{\nu}_\ell$ (bottom) channel. In this plot we average over the electron and muon mode. The grey dotted lines indicate the binning used for the signal extraction described in the text.}
    \label{fig:mm2selection}
\end{figure}

The resolution of the signal events is underestimated in the MC and corrected for by smearing the $M_\mathrm{miss}^2$ distribution in the vicinity of the peak close to zero \si{\GeV\squared/c^2}. 
This is achieved by convolving the MC $M_\mathrm{miss}^2$ distribution with an asymmetric Laplace distribution, whose parameters are optimized to minimize data and MC disagreements.
This procedure increases the root-mean-square of the MC distribution in the vicinity of the peak ($-0.5\, \mathrm{GeV}^2/c^4 < M_\mathrm{miss}^2 < 0.5\, \mathrm{GeV}^2/c^4$) by approximately 2.8\% from $0.197\,\mathrm{GeV}^2/c^4$ to $0.203\,\mathrm{GeV}^2/c^4$ and behaves similarly for $\bar{B}^0$ and $B^-$ decays. The inclusive distribution after the correction for the $\bar{B}^0$ and $B^-$ modes is shown in Fig.~\ref{fig:mm2selection}.

Likelihood fits are carried out separately for $\bar{B}^0$ and $B^-$ candidates, and $e$ and $\mu$ modes, in bins of $w$ and the decay angles in order to determine the \bdslnu signal yields in the chosen bins by subtracting the background yields. We choose ten equidistant bins with a width of $0.05$ for $w \in [1, 1.5]$ and widen the last bin to recover all events outside of the physical region due to resolution effects and to be insensitive to lepton mass effects. The ten bins in each of the three angular variables $\cos\theta_\ell$, $\cos\theta_V$, and $\chi$ are also chosen with equidistant binning. In total $4 \times 40 = 160$ separate fits are carried out.

The likelihood function for a given fit is constructed from the product of individual Poisson distributions $\mathcal{P}$ and nuisance-parameter (NP) constraints $\mathcal{G}_k$,
\begin{equation} \label{eq:likelihood}
 \mathcal{L} = \prod_i^{\rm bins} \, \mathcal{P}\left( n_i ; \nu_i \right)  \times \prod_j^{\rm systematics} \, \mathcal{G}_j \, ,
\end{equation}
with $n_i$ denoting the number of observed data events and $\nu_i$ the total number of expected events in a given bin $i$. We divide the $M_\mathrm{miss}^2$ spectrum into 5 bins with the bin edges at $[-1.0, -0.25, 0.25, 0.75, 1.25, 2.0]\,\mathrm{GeV}^2/c^4$. This coarse binning is chosen to reduce the sensitivity to resolution effects in the peak region to negligible level. The number of expected events in a given bin, $\nu_i$, is estimated using MC simulation,
\begin{equation}\label{eq:nui}
 \nu_i = \sum_k^{\rm processes} \, f_{ik} \, \eta_k  \, ,
\end{equation}
with $\eta_k$ the total number of events from a given process $k$ with the fraction $f_{ik}$ of such events being reconstructed in the bin $i$.
The likelihood Eq.~\ref{eq:likelihood} is numerically maximized using \texttt{iminuit}~\cite{James:1975dr,iminuit} to fit the values of two different categories, $\eta_k$ ($k:$ Signal, Background), using the observed events. We split the simulated data into two categories to define the templates used in the fit: 
\begin{itemize}
\item[-] \emph{Signal} is defined to be a MC truth matched lepton originating from a semileptonic $B \to D^* \ell \bar{\nu}_\ell\,, \ell=e,\mu$ decay. The $D^*$ meson does not have to be correctly reconstructed for the $B$ to be considered a signal candidate. 
\item[-] \emph{Background}, concretely: $B \to D l \bar{\nu}_l$ with $ l=e, \mu, \tau$ decays, $B \to D^* \tau \bar{\nu}_\tau$ decays, $B \to D^{**} \ell \bar{\nu}_\ell$ decays, hadronic background where the reconstructed lepton is a misidentified kaon or pion, other processes from $B$ decays, and continuum.
\end{itemize}

\begin{figure}
    \centering
    \includegraphics[width=\linewidth]{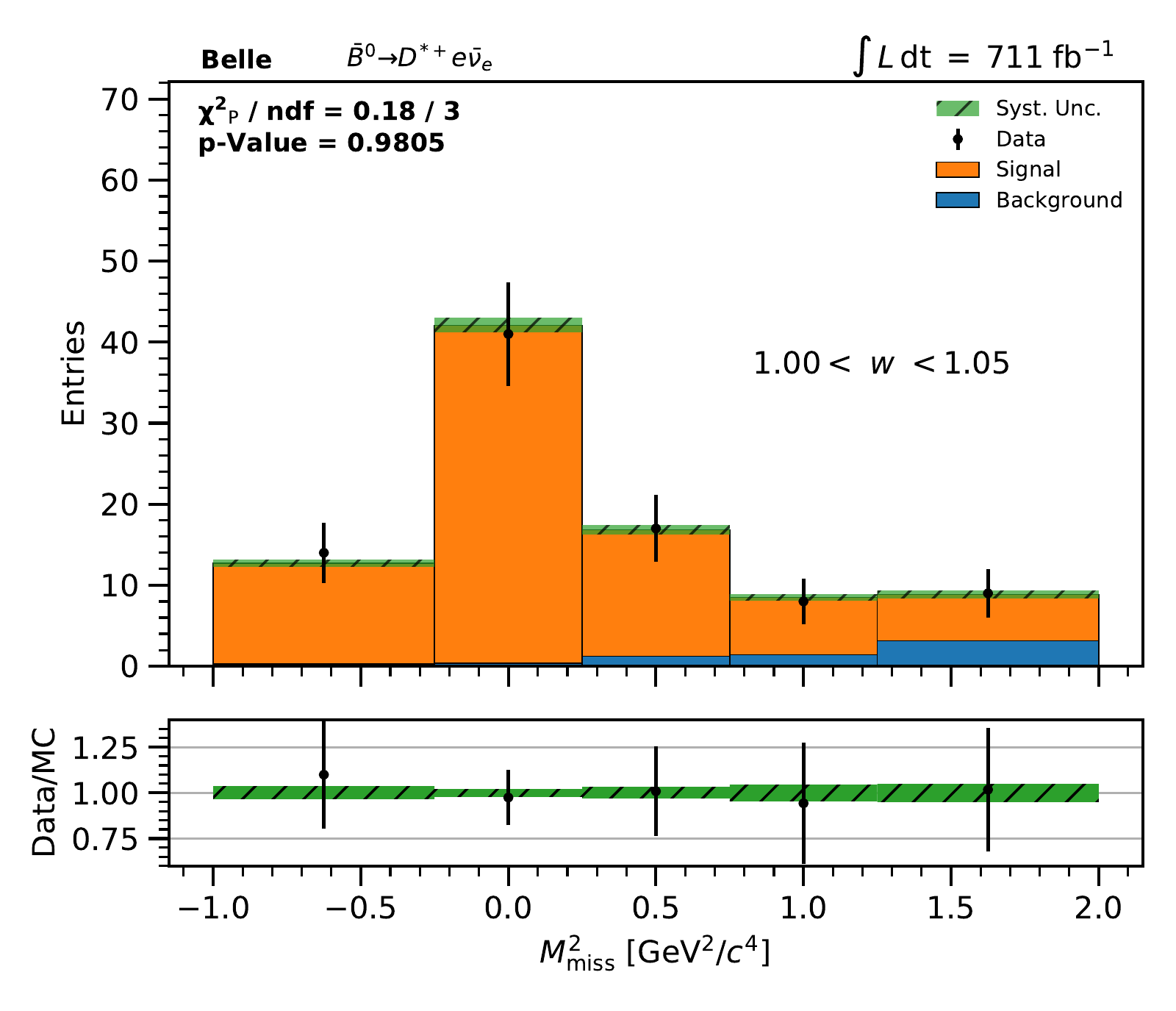}
    \caption{The post-fit $M_\mathrm{miss}^2$ distribution in the $\bar{B}^0 \to D^* e \bar{\nu}_e$ mode, in the $1 < w < 1.05$ bin. 
    }
    \label{fig:mm2fitexample}
\end{figure}

\begin{figure}
    \centering
    \includegraphics[width=\linewidth]{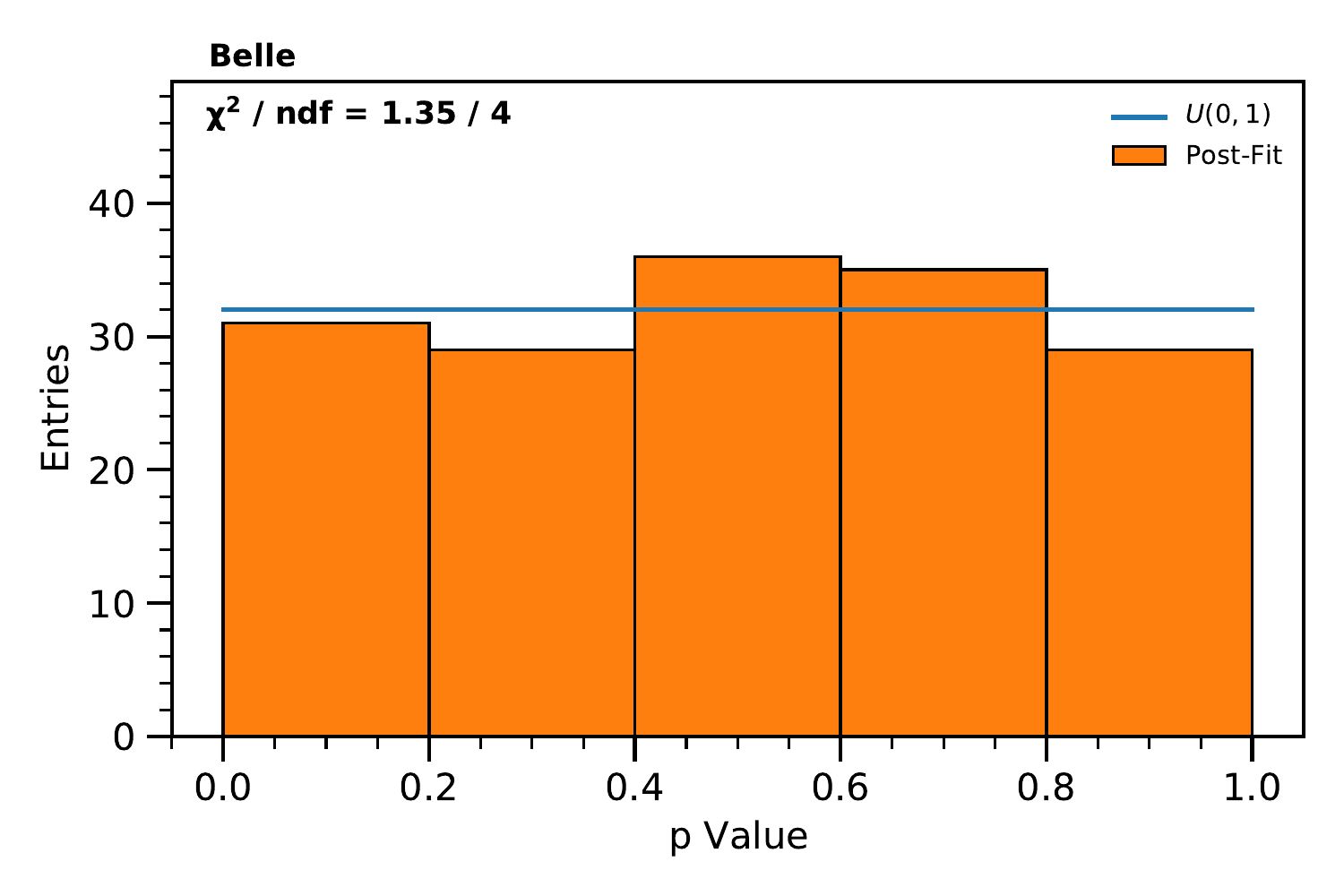}
    \caption{The $p$-value distribution for the 160 fits performed in different decay channels and kinematic regions. The distribution is compatible with the expected uniform behavior.}
    \label{fig:chi2fitdistro}
\end{figure}

An example fit is shown in Fig.~\ref{fig:mm2fitexample} for the $1 < w < 1.05$ bin. The goodness-of-fit of likelihoods can be calculated in the large sample limit \cite{BAKER1984437} with
\begin{equation}
    \chi_P^2 = 2 \sum_{i=0}^N \left( n_i \log \frac{n_i}{\hat{\nu}_i} + \hat{\nu}_i - n_i \right)\,,
\end{equation}
where $\hat{\nu}_i$ is the estimated number of events in bin $i$. The $p$-value is calculated as
\begin{equation}
     \int_{\chi^2_P}^\infty f_{\chi^2}(x|k=3) \mathrm{d}x\,,
\end{equation}
with $k=3$ degrees of freedom and $f_{\chi^2}$ denoting the $\chi^2$ distribution. The corresponding $p$-value distribution for all 160 fits is shown in Fig.~\ref{fig:chi2fitdistro} and is compatible with the expected uniform behavior.

We determine the statistical correlation between the marginalized distributions of the full four-dimensional rate by considering: 
\begin{enumerate}
\item The statistical correlation of the data.
\item The sample overlap in the MC distributions and the systematic uncertainties on the signal and background shapes on $M_\mathrm{miss}^2$. This is used to correlate the fit shape uncertainties  between measured bins associated with the finite sample size of the MC simulation.
\item The other systematic shape uncertainties, discussed further in Sec.~\ref{sec:syst}, are negligibly small and we treat them as fully correlated between individually measured bins.
\end{enumerate} 

The statistical correlation of the data between different bins of different observables is determined by sampling with replacement from the selected recorded data and repeated fits to resolve Pearson correlation coefficients as small as $r_{\rm data} \approx 0.01$. For cases without statistical overlap, e.g.\ neighbouring bins in the same marginal distribution, we set the correlation to zero.

We further determine the expected correlation in the MC distributions by using the sample overlap
\begin{equation}
    r_{\rm MC} = \frac{n_{xy}}{\sqrt{n_x} \sqrt{n_y}}
\end{equation}
in the peak region $\SI{-0.25}{GeV^2/c^4} < M_\mathrm{miss}^2 < \SI{0.25}{GeV^2/c^4}$. Here, $n_{x/y}$ refers to the number of events in a given bin of an observable $x, y = w, \cos\theta_\ell, \cos\theta_V, \chi$ and $n_{xy}$ refers to the events that are in both bins of both observables under consideration.

\section{Unfolding of differential yields}\label{sec:unfold}
The resolution caused by detector effects and mis-reconstructed $D^{*}$ mesons causes migrations of events into neighbouring bins in the kinematic distributions. These effects must be corrected for in order to compare the measured distribution with a theoretical distribution. We proceed by unfolding our measured spectrum, but also provide all components necessary to forward fold a theoretical distribution.

\begin{figure}
    \centering
    \includegraphics[width=0.75\linewidth]{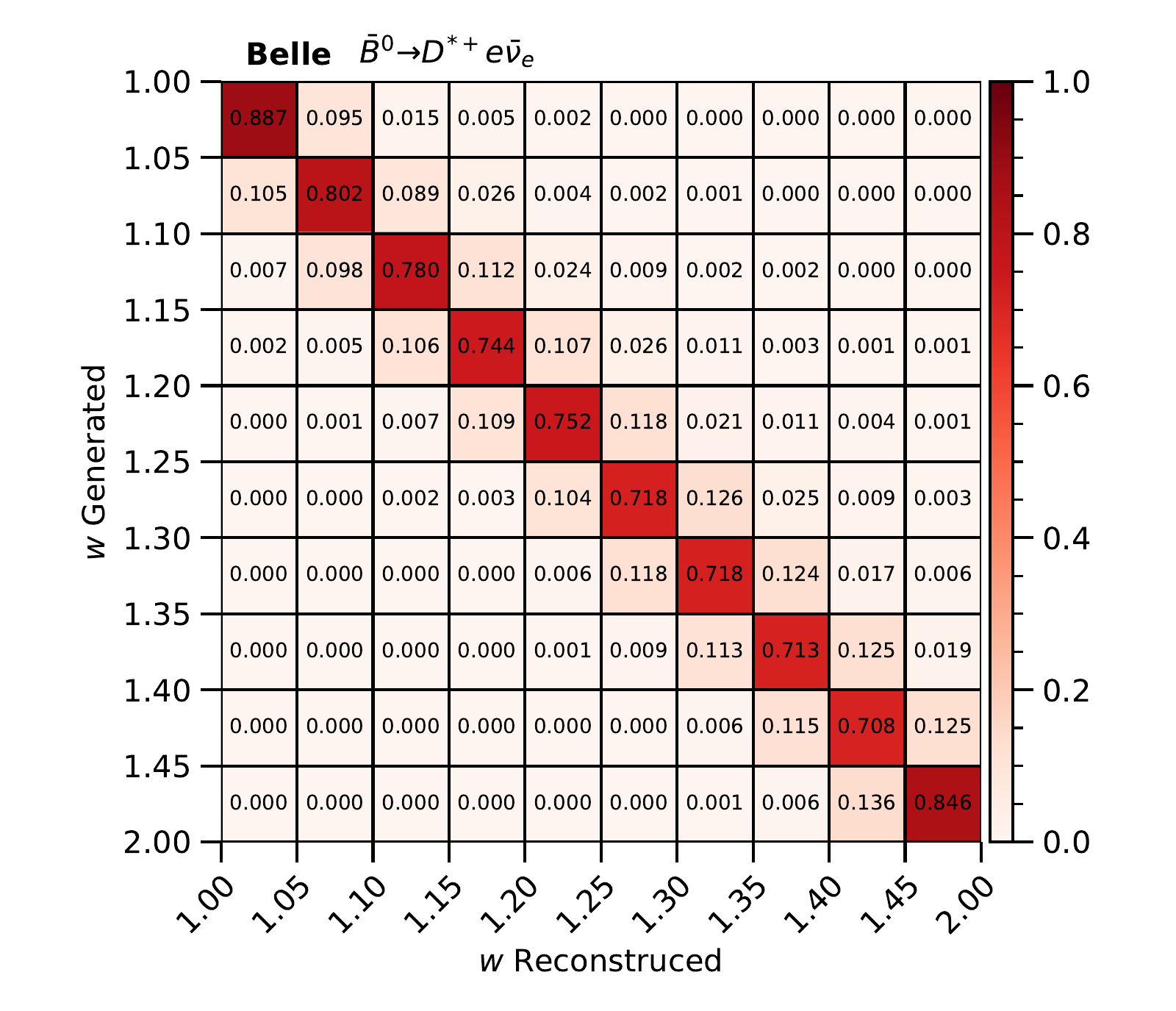}\\
    \includegraphics[width=0.75\linewidth]{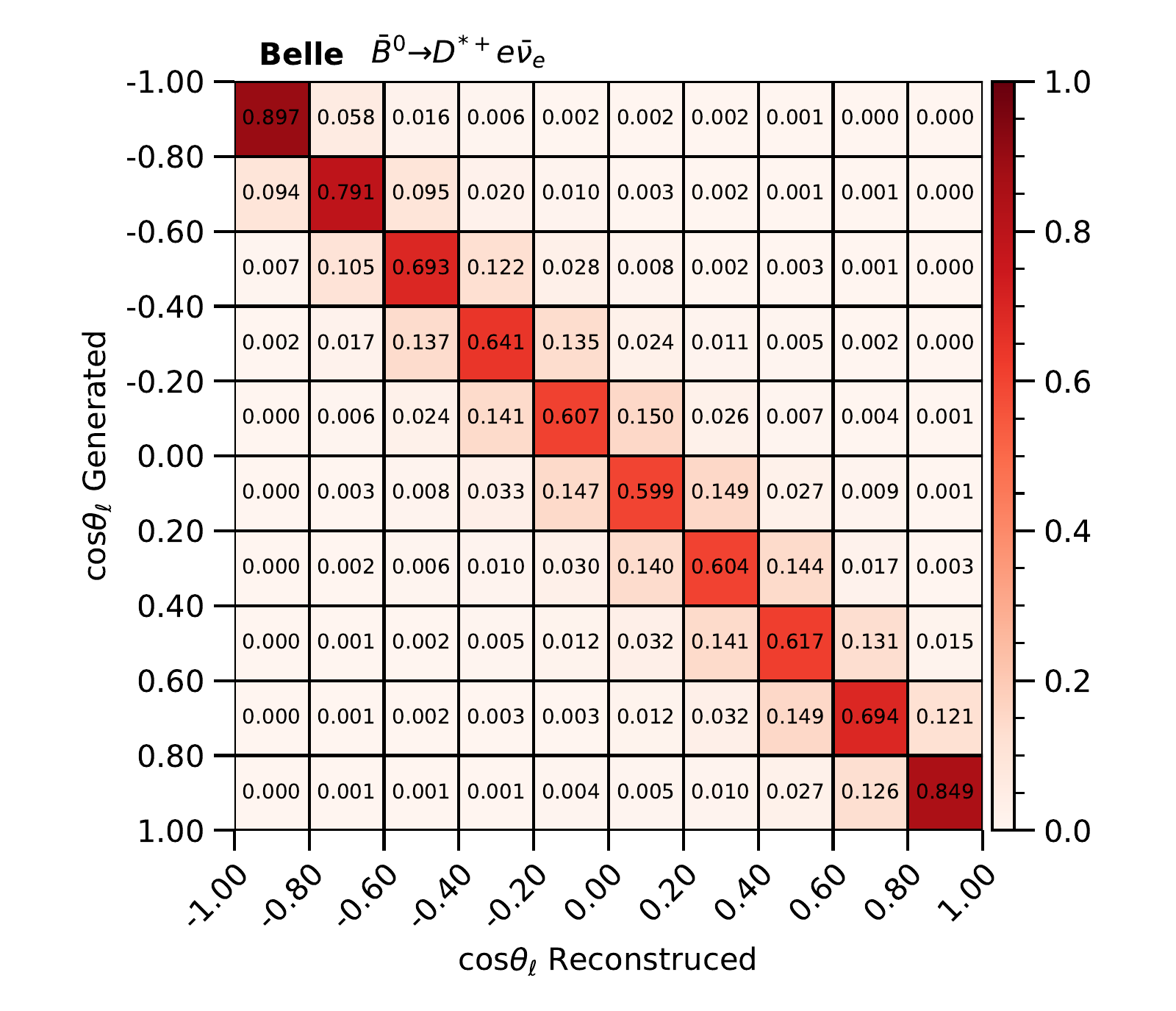}\\
    \includegraphics[width=0.75\linewidth]{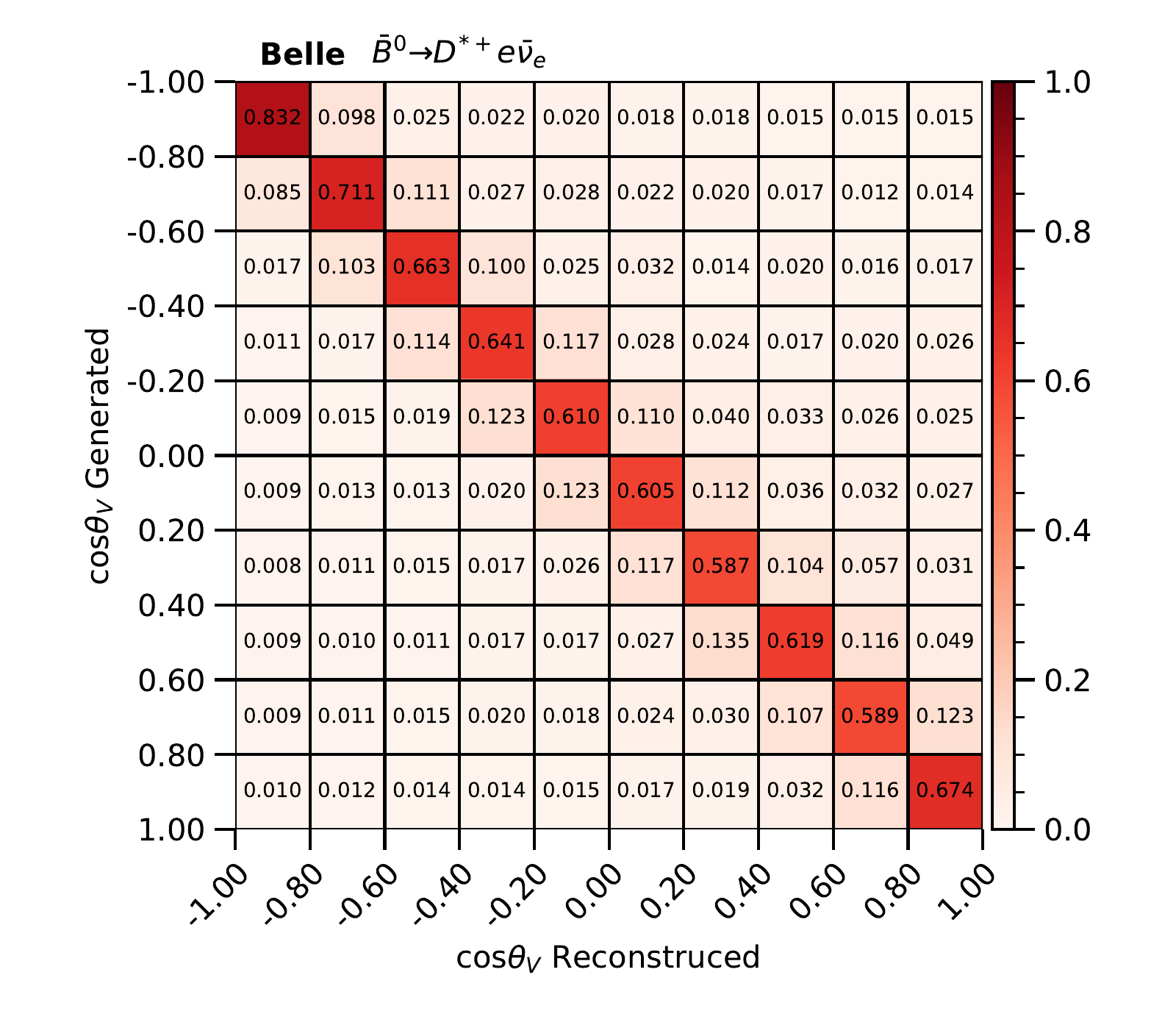}\\
    \includegraphics[width=0.75\linewidth]{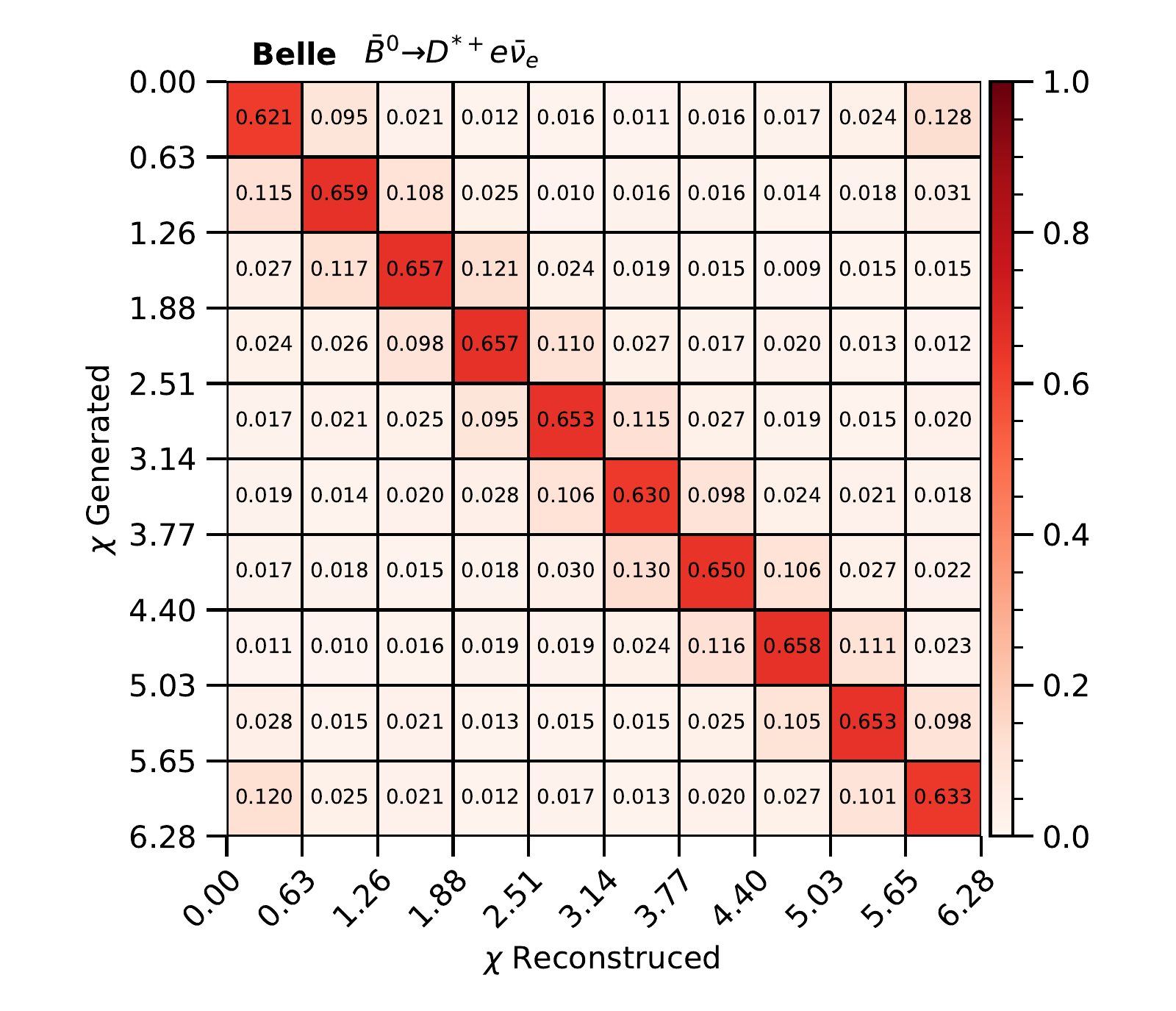}
    \caption{Migration matrices for the $\bar{B}^0 \to D^* e \bar{\nu}_e$ mode, for the four marginal distributions: $w, \cos\theta_\ell, \cos\theta_V, \chi$. These matrices transform the reconstructed to the generated quantity.
    }
    \label{fig:migration}
\end{figure}

The migrations can be quantified by determining a detector response matrix $R$, which encodes the probability $P$ of an event within a true bin to migrate into a reconstructed bin:
\begin{equation}
    R_{ij} = P(\mathrm{reco \, bin} \, i \, | \,  \mathrm{true \, bin } \, j)\,.
\end{equation}
These matrices are determined for each of the four decay modes individually using simulated events, and illustrated for the $\bar{B}^0 \to D^* e \bar{\nu}_e$ decay mode in Fig.~\ref{fig:migration}. The response matrices are dominated by diagonal entries and exhibit a similar structure in each of the four modes.

We unfold the signal yields determined in Sec.~\ref{sec:signalextraction} using matrix inversion. This produces the best linear unbiased maximum likelihood estimator given by 
\begin{equation}
    \hat{\vec{\mu}} = R^{-1} \hat{\vec{n}}\,,
\end{equation}
with $\hat{\vec n}$ being our estimated background subtracted yields. 

\begin{figure}
    \centering
    \includegraphics[width=\linewidth]{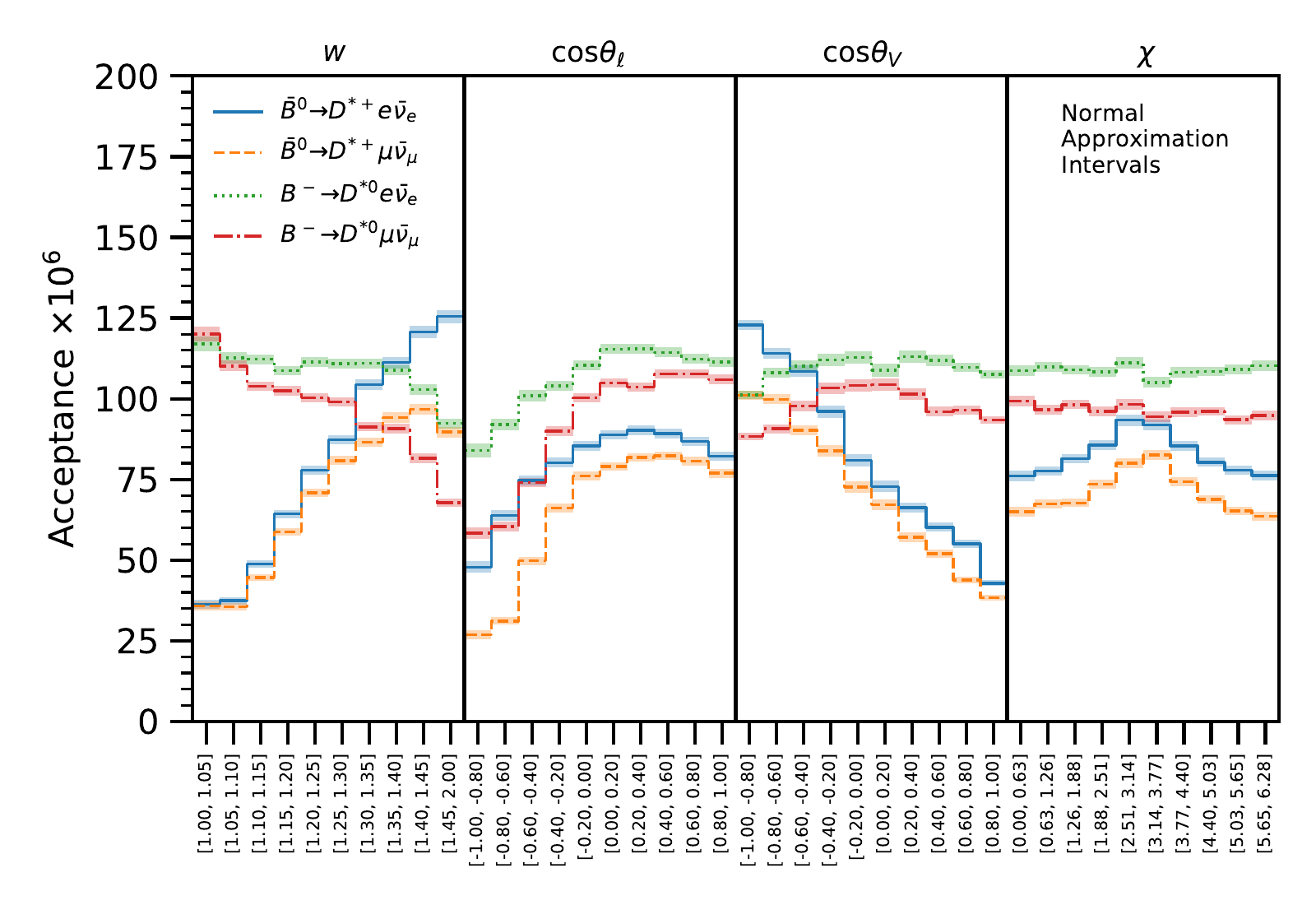}
    \caption{Acceptance functions for the four decay modes considered. As expected they behave differently for charged and neutral $B$ mesons, due to the charged and neutral slow pion reconstruction. The uncertainty on the acceptance is statistical only and calculated using normal approximation intervals. Additional systematic uncertainties are considered, for details see the text. }
    \label{fig:acceptance}
\end{figure}

We correct for acceptance effects, and reverse the impact of FSR photons from \texttt{PHOTOS} on the measured distributions. The acceptance functions for all modes are shown in Fig.~\ref{fig:acceptance}.

\begin{figure}
    \centering
    \includegraphics[width=\linewidth]{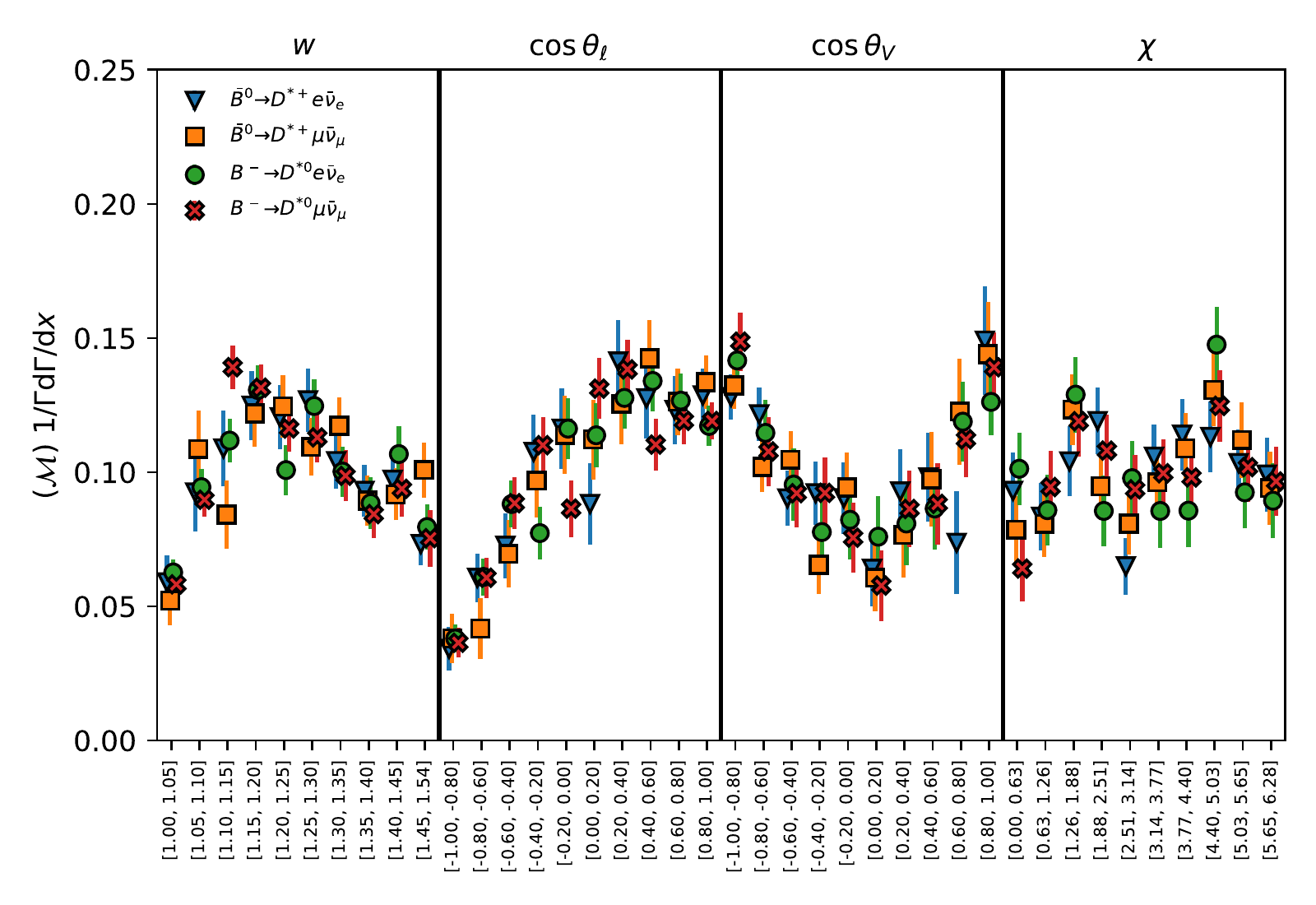}
    \caption{Our determined shapes for the four decay modes using matrix inversion to correct for the migrations and applying the acceptance correction. 
    }
    \label{fig:shapes}
\end{figure}

We find the shapes in the kinematic quantities, shown in Fig.~\ref{fig:shapes} and tabulated in Table~\ref{tab:yields}, after correcting our background subtracted yields for the migration and acceptance. 

\begin{table*}
    \centering
    \caption{Normalized partial branching ratios $\Delta \Gamma$ in the observed bin $\Delta x$ and the corresponding uncertainties for all channels and projections.}
    \label{tab:yields}
    \begin{tabular}{llrrrrrrrr}
\hline
\hline
& & \multicolumn{2}{c}{$\bar{B}^0 \to D^{*+} e \bar{\nu}_e$} & \multicolumn{2}{c}{$\bar{B}^0 \to D^{*+} \mu \bar{\nu}_\mu$} & \multicolumn{2}{c}{$B^- \to D^{*0} e \bar{\nu}_e$} & \multicolumn{2}{c}{$B^- \to D^{*0} \mu \bar{\nu}_\mu$} \\
       &              & $\Delta \Gamma / \Delta x$ & $\sigma$ & $\Delta \Gamma / \Delta x$ & $\sigma$ & $\Delta \Gamma / \Delta x$  & $\sigma$ & $\Delta \Gamma / \Delta x$ & $\sigma$ \\
Projection & Bin &                                                     &                                   &                                                         &                                       &                                                     &                                   &                                                         &                                       \\
\hline
$w$ & [1.00, 1.05) &                                              0.059 &                             0.010 &                                              0.052 &                                 0.009 &                                              0.063 &                             0.005 &                                              0.058 &                                 0.004 \\
       & [1.05, 1.10) &                                              0.092 &                             0.015 &                                              0.109 &                                 0.014 &                                              0.094 &                             0.007 &                                              0.090 &                                 0.006 \\
       & [1.10, 1.15) &                                              0.109 &                             0.014 &                                              0.084 &                                 0.013 &                                              0.112 &                             0.008 &                                              0.139 &                                 0.008 \\
       & [1.15, 1.20) &                                              0.125 &                             0.013 &                                              0.122 &                                 0.012 &                                              0.131 &                             0.009 &                                              0.131 &                                 0.009 \\
       & [1.20, 1.25) &                                              0.120 &                             0.012 &                                              0.124 &                                 0.012 &                                              0.101 &                             0.009 &                                              0.116 &                                 0.009 \\
       & [1.25, 1.30) &                                              0.127 &                             0.012 &                                              0.109 &                                 0.011 &                                              0.125 &                             0.010 &                                              0.113 &                                 0.009 \\
       & [1.30, 1.35) &                                              0.104 &                             0.010 &                                              0.117 &                                 0.010 &                                              0.100 &                             0.009 &                                              0.099 &                                 0.010 \\
       & [1.35, 1.40) &                                              0.093 &                             0.010 &                                              0.089 &                                 0.009 &                                              0.088 &                             0.010 &                                              0.084 &                                 0.009 \\
       & [1.40, 1.45) &                                              0.097 &                             0.009 &                                              0.092 &                                 0.010 &                                              0.107 &                             0.011 &                                              0.094 &                                 0.010 \\
       & [1.45, 1.51) &                                              0.073 &                             0.008 &                                              0.101 &                                 0.010 &                                              0.080 &                             0.008 &                                              0.075 &                                 0.011 \\
\hline
$\cos \theta_\ell$ & [-1.00, -0.80) &                                              0.034 &                             0.008 &                                              0.038 &                                 0.009 &                                              0.038 &                             0.005 &                                              0.036 &                                 0.006 \\
       & [-0.80, -0.60) &                                              0.061 &                             0.009 &                                              0.042 &                                 0.011 &                                              0.061 &                             0.007 &                                              0.061 &                                 0.008 \\
       & [-0.60, -0.40) &                                              0.073 &                             0.012 &                                              0.070 &                                 0.013 &                                              0.088 &                             0.009 &                                              0.088 &                                 0.010 \\
       & [-0.40, -0.20) &                                              0.108 &                             0.014 &                                              0.097 &                                 0.014 &                                              0.077 &                             0.010 &                                              0.110 &                                 0.011 \\
       & [-0.20, 0.00) &                                              0.116 &                             0.015 &                                              0.114 &                                 0.015 &                                              0.116 &                             0.011 &                                              0.086 &                                 0.011 \\
       & [0.00, 0.20) &                                              0.088 &                             0.015 &                                              0.112 &                                 0.015 &                                              0.114 &                             0.012 &                                              0.131 &                                 0.011 \\
       & [0.20, 0.40) &                                              0.141 &                             0.015 &                                              0.126 &                                 0.015 &                                              0.128 &                             0.012 &                                              0.138 &                                 0.011 \\
       & [0.40, 0.60) &                                              0.128 &                             0.015 &                                              0.142 &                                 0.014 &                                              0.134 &                             0.011 &                                              0.110 &                                 0.010 \\
       & [0.60, 0.80) &                                              0.123 &                             0.013 &                                              0.126 &                                 0.012 &                                              0.127 &                             0.010 &                                              0.119 &                                 0.009 \\
       & [0.80, 1.00) &                                              0.129 &                             0.010 &                                              0.134 &                                 0.010 &                                              0.117 &                             0.007 &                                              0.119 &                                 0.007 \\
\hline
$\cos \theta_V$ & [-1.00, -0.80) &                                              0.128 &                             0.008 &                                              0.132 &                                 0.009 &                                              0.142 &                             0.011 &                                              0.149 &                                 0.011 \\
       & [-0.80, -0.60) &                                              0.122 &                             0.010 &                                              0.102 &                                 0.009 &                                              0.115 &                             0.012 &                                              0.108 &                                 0.013 \\
       & [-0.60, -0.40) &                                              0.090 &                             0.010 &                                              0.105 &                                 0.011 &                                              0.095 &                             0.013 &                                              0.092 &                                 0.013 \\
       & [-0.40, -0.20) &                                              0.092 &                             0.012 &                                              0.065 &                                 0.011 &                                              0.078 &                             0.014 &                                              0.092 &                                 0.013 \\
       & [-0.20, 0.00) &                                              0.090 &                             0.013 &                                              0.094 &                                 0.013 &                                              0.082 &                             0.015 &                                              0.076 &                                 0.013 \\
       & [0.00, 0.20) &                                              0.064 &                             0.014 &                                              0.061 &                                 0.013 &                                              0.076 &                             0.015 &                                              0.058 &                                 0.013 \\
       & [0.20, 0.40) &                                              0.093 &                             0.016 &                                              0.077 &                                 0.016 &                                              0.081 &                             0.016 &                                              0.086 &                                 0.014 \\
       & [0.40, 0.60) &                                              0.098 &                             0.017 &                                              0.097 &                                 0.018 &                                              0.086 &                             0.015 &                                              0.088 &                                 0.015 \\
       & [0.60, 0.80) &                                              0.074 &                             0.019 &                                              0.123 &                                 0.020 &                                              0.119 &                             0.015 &                                              0.112 &                                 0.014 \\
       & [0.80, 1.00) &                                              0.149 &                             0.020 &                                              0.144 &                                 0.020 &                                              0.126 &                             0.012 &                                              0.139 &                                 0.013 \\
\hline
$\chi$ & [0.00, 0.63) &                                              0.093 &                             0.014 &                                              0.079 &                                 0.012 &                                              0.101 &                             0.013 &                                              0.064 &                                 0.012 \\
       & [0.63, 1.26) &                                              0.083 &                             0.013 &                                              0.081 &                                 0.012 &                                              0.086 &                             0.013 &                                              0.094 &                                 0.013 \\
       & [1.26, 1.88) &                                              0.104 &                             0.013 &                                              0.123 &                                 0.013 &                                              0.129 &                             0.014 &                                              0.119 &                                 0.013 \\
       & [1.88, 2.51) &                                              0.119 &                             0.012 &                                              0.095 &                                 0.012 &                                              0.086 &                             0.013 &                                              0.108 &                                 0.013 \\
       & [2.51, 3.14) &                                              0.065 &                             0.011 &                                              0.081 &                                 0.011 &                                              0.098 &                             0.014 &                                              0.094 &                                 0.013 \\
       & [3.14, 3.77) &                                              0.106 &                             0.012 &                                              0.096 &                                 0.011 &                                              0.086 &                             0.014 &                                              0.100 &                                 0.013 \\
       & [3.77, 4.40) &                                              0.114 &                             0.013 &                                              0.109 &                                 0.013 &                                              0.086 &                             0.014 &                                              0.098 &                                 0.013 \\
       & [4.40, 5.03) &                                              0.113 &                             0.013 &                                              0.131 &                                 0.014 &                                              0.148 &                             0.014 &                                              0.125 &                                 0.013 \\
       & [5.03, 5.65) &                                              0.103 &                             0.013 &                                              0.112 &                                 0.014 &                                              0.092 &                             0.013 &                                              0.102 &                                 0.013 \\
       & [5.65, 6.28) &                                              0.099 &                             0.014 &                                              0.094 &                                 0.014 &                                              0.089 &                             0.014 &                                              0.097 &                                 0.013 \\
\hline
\hline
\end{tabular}

\end{table*}

The self-consistency of the measurement is checked by comparing pairs of distributions, and by comparing all four distributions, taking their covariance matrices into account. We ignore the effects of different masses between $\bar{B}^0$ and $B^-$, which are significantly smaller than the measured uncertainties on our shapes. Details ($\chi^2$~/~ndf and $p$-values) are listed in Table~\ref{tab:averages}. 

\begin{table}
    \centering
    \caption{The compatibility of the measurements from the different decay modes determined with the statistical and systematic covariance matrix and the statistical covariance matrix only. All modes agree well with each other.}
    \label{tab:averages}
    \resizebox{\linewidth}{!}{
    \begin{tabular}{lrrrr}
\hline
\hline
{} & $\chi^2$ / dof &     $p$ & $\chi^2_\mathrm{stat}$ / ndf & $p_\mathrm{stat}$ \\
\hline
$B \to D^{*} \ell \bar{\nu}_\ell$            &     94.7 / 108 &  0.82 &                  102.0 / 108 &              0.65 \\
$\bar{B}^0 \to D^{*+} \ell \bar{\nu}_\ell$         &      26.3 / 36 &  0.88 &                    27.7 / 36 &              0.84 \\
$B^- \to D^{*0} \ell \bar{\nu}_\ell$         &      31.6 / 36 &  0.68 &                    33.8 / 36 &              0.57 \\
$B^{(0,-)} \to D^{*(+,0)} e \bar{\nu}_e$     &      27.4 / 36 &  0.85 &                    29.2 / 36 &              0.78 \\
$B^{(0,-)} \to D^{*(+,0)} \mu \bar{\nu}_\mu$ &      42.5 / 36 &  0.21 &                    45.7 / 36 &              0.13 \\
\hline
\hline
\end{tabular}

    }
\end{table}

\section{Systematic Uncertainties}
\label{sec:syst}

For the $M_\mathrm{miss}^2$ fits we studied uncertainties originating from the branching fractions and form factor parameterizations of the \bdslnu and \bdlnu decays in our simulated events, the uncertainty from the overall limited MC statistics, the lepton identification efficiency, the efficiencies for reconstruction of tracks, neutral pions, slow pions, and $K_\mathrm{S}^0$ mesons, and the uncertainties of the parameters describing the resolution smearing function.

The effect of systematic uncertainties is directly incorporated into the likelihood function in Eq.~\ref{eq:likelihood}. For this we introduce a vector of nuisance parameters, $\boldsymbol{\theta}_k$, for each fit template $k$. Each vector element represents one bin. The nuisance parameters are constrained in the likelihood using multivariate Gaussian distributions $\mathcal{G}_k = \mathcal{G}_k( \boldsymbol{0}; \boldsymbol{\theta}_k, \Sigma_k ) $, with $\Sigma_k$ denoting the systematic covariance matrix for a given template $k$. The systematic covariance is constructed from the sum over all possible uncertainty sources affecting a template $k$, i.e.
\begin{equation}
 \Sigma_k = \sum_{s}^{\text{error sources}} \Sigma_{ks} \, ,
\end{equation}
with $\Sigma_{ks} $ the covariance matrix of error source $s$. 

The impact of nuisance parameters is included in Eq.~\ref{eq:nui} as follows. The fractions $f_{ik}$ for all templates are rewritten as
\begin{equation}
 f_{ik} = \frac{ \eta_{ik}^{\rm MC} }{ \sum_j \eta_{jk}^{\rm MC} } \to  \frac{ \eta_{ik}^{\rm MC} \left( 1 + \theta_{ik} \right) }{ \sum_j \eta_{jk}^{\rm MC} \left( 1 + \theta_{jk} \right)  },
\end{equation}
to take into account shape uncertainties. Here $\theta_{ik}$ represents the nuisance parameter vector element of bin $i$ and $\eta_{ik}^{\rm MC}$ the expected number of events in the same bin for event type $k$ as estimated from the simulation. The systematic effects on the shape of $M_\mathrm{miss}^2$ have a small impact on the yields in $M_\mathrm{miss}^2$ with the largest uncertainty from the finite sample size of the simulated MC templates.

For the unfolding and acceptance correction procedure we consider uncertainties originating from the $D$ decay branching fractions, the \bdslnu form factors, the limited MC statistics, the lepton identification efficiency, and the efficiencies for reconstruction of tracks, neutral pions, slow pions, and $K_\mathrm{S}^0$ mesons.
The impact of these systematic effects on the unfolding and acceptance correction is determined by varying the MC sample used to determine the migration matrices and acceptance function within the uncertainty of the given systematic effect, and repeating the unfolding and acceptance correction procedure.

The calibration factors for the FEI are determined from a study of hadronically tagged inclusive $B\to X_c \ell \bar{\nu}_\ell$ decays. The study is performed in bins of the FEI signal probability and the tag-side channels. The calibration factors are defined as the ratio of expected and measured number of events in each bin.
The absolute efficiency of the FEI cancels in the measurement of the shapes. The impact of the FEI on the measured shapes is determined by weighting the events after removing FEI calibration factors and determining the difference after applying unfolding and acceptance correction. We treat this uncertainty as fully correlated.

The individual contributions of the uncertainties to the normalized shapes are listed in Appendix~\ref{app:systematics}.

\section{Determination of the form factors and Implications on $|V_{cb}|$}\label{sec:results}
We use the averaged $B\to D^* \ell \bar{\nu}_\ell$ shapes to fit the BGL and CLN form factor parameterizations to the data. We minimize the $\chi^2$ defined by
\begin{align}
 \chi^2 = & \left(\frac{\Delta\vec{\Gamma}^{\rm m}}{\Gamma^{\rm m}} - \frac{\Delta\ \vec{\Gamma}^{\rm p}({\vec{x}})}{\Gamma^{\rm p}({\vec{x}})}\right) C^{-1}_\mathrm{exp} \left(\frac{\Delta \vec \Gamma^{\rm m}}{\Gamma^{\rm m}} - \frac{\Delta\vec \Gamma^{\rm p}({\vec{x}})}{\Gamma^{\rm p}({\vec{x}})}\right)^T \nonumber \\
   & + (\Gamma^\mathrm{ext} - \Gamma^{\rm p}(\vec{x}))^2 / \sigma(\Gamma^\mathrm{ext})^2 \nonumber \\
  & + (h_{X} - h_{X}^{\rm LQCD}) C^{-1}_\mathrm{LQCD} (h_{X} - h_{X}^{\rm LQCD}) \,,
\end{align}
with the measured (predicted) differential rate $\Delta \vec \Gamma^\mathrm{m (p)} / \Gamma^\mathrm{m (p)}$, where the predicted rate is a function of the form factor coefficients $\vec{x}$ and \Vcba. The rate is calculated assuming the meson masses of $m_B = \SI{5.28}{GeV}$ and $m_{D^*} = \SI{2.01}{GeV}$, and the lepton as massless. $C_\mathrm{exp}$ ($C_\mathrm{LQCD}$) is the covariance matrix of the experimental (lattice) data. 

We rely on external branching fractions provided by HFLAV~\cite{Amhis:2022mac} to determine  \Vcba \ : 
\begin{align}
\mathcal{B}(B^-\to D^{*0} \ell \bar{\nu}_\ell) = (5.58 \pm 0.22)\% \, , \\
\mathcal{B}(\bar{B}^0\to D^{*+} \ell \bar{\nu}_\ell) = (4.97 \pm 0.12)\% \, .
\end{align}
We combine these branching fractions assuming isospin and by using the $B^{+/0}$ lifetimes $\tau_{\bar{B}^0} = 1.520 \, \mathrm{ps}$ and $\tau_{B^-} = 1.638 \, \mathrm{ps}$ from Ref.~\cite{pdg:2022}. Expressing this average as a $\bar{B}^0$ branching fraction we find:
\begin{align}
\mathcal{B}(\bar{B}^0\to D^{*+} \ell \bar{\nu}_\ell) = (5.03 \pm 0.10)\% \, .
\end{align} 

The form factor normalization is constrained at zero-recoil with $h_X = h_{A_1}(1) = 0.906 \pm 0.013$ from Ref.~\cite{FermilabLattice:2014ysv} for our nominal fit scenario. 
For the BGL form factor fit, we truncate the series based on the result of a nested hypothesis test (NHT) \cite{Bernlochner:2019ldg} with the additional constraint that the inclusion of additional coefficients do not result in correlations of larger than $r=0.95$. This leads to the choice of $n_a=1$, $n_b=2$, $n_c=1$ free parameters, with the constraint for $c_0$ defined in Eq.~(12). More details about the NHT can be found in Appendix~\ref{app:nht}. For the CLN type parameterization we determine three coefficients: $\rho^2$ , $R_1(1)$, and $R_2(1)$. 

\begin{figure}
    \centering
    \includegraphics[width=\linewidth]{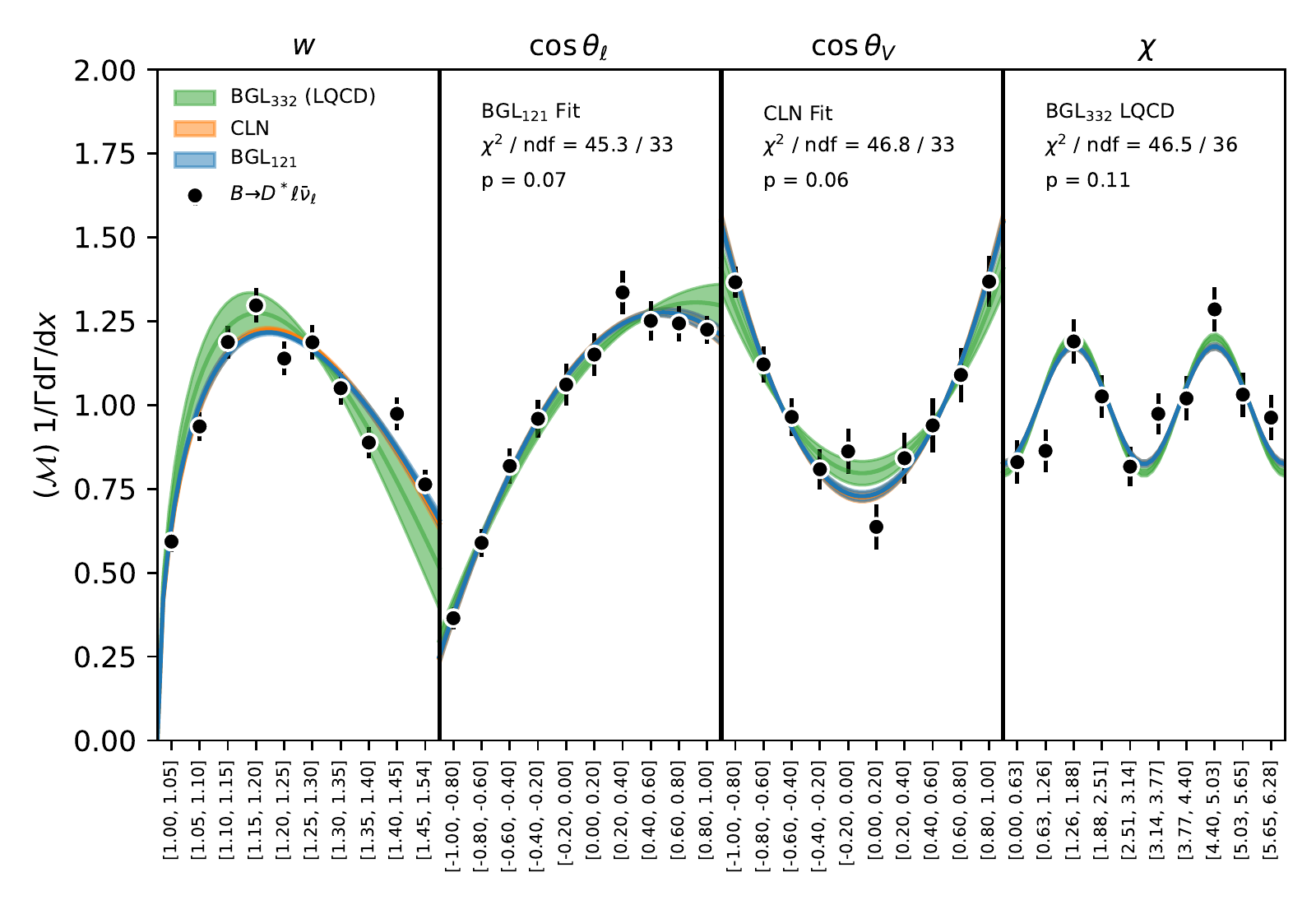}
\caption{The fitted shapes for both BGL (blue) and CLN (orange) parametrization. Both parametrizations are able to explain the data, and are compatible with each other. Note that the BGL (blue) band almost completely overlays the CLN (orange) band. The green band is the prediction using BGL coefficients from lattice QCD calculations in \citep{FermilabLattice:2021cdg}.
    }
    \label{fig:BGL121vsCLN}
\end{figure}

Both form factor parameterizations are able to describe the data with $p$-values of $7\%$ and $6\%$ for BGL and CLN, respectively, and the extracted \Vcba\ values of both determinations are compatible. The fitted shapes are shown in Fig.~\ref{fig:BGL121vsCLN} (red and blue bands) and the numerical values for the coefficients and \Vcba \ are listed in Table~\ref{tab:BGL121} and Table~\ref{tab:CLN} for BGL and CLN, respectively. In the figure we also show the recent beyond zero-recoil prediction of Ref.~\cite{FermilabLattice:2021cdg} as a green band. Its agreement with the measured spectra has a $p$-value of 11\%. We also perform fits to our measured $\bar{B}^0$ and $B^-$ shapes separately, with the corresponding external branching fraction input. The results are compatible with each other, and the individual extracted \Vcba\ values are listed in Table~\ref{tab:vcbIsospinScenarios}. We observe a discrepancy between the \Vcba\, values from the charged- and neutral-only fits ($p=5\%$). Correcting for the existing disagreement between the charged and neutral input branching fractions from HFLAV~\cite{Amhis:2022mac} and comparing the full set of BGL coefficients and \Vcba\, we recover a $p$-value of 20\%.

Additionally, we tested explicitly the impact of the d'Agostini bias~\cite{DAgostini:1993arp} on the reported results. The impact of this bias on our quoted values of \Vcba\ and the form factor parameters is approximately a factor of 30 smaller than the quoted uncertainties and we thus do not apply an additional correction. 

We also test the impact of the preliminary lattice results that constrain the $B \to D^*$ form factors beyond zero recoil of Ref.~\citep{FermilabLattice:2021cdg} using two scenarios: 
\begin{enumerate}
    \item Inclusion of $h_{A_1}$ beyond zero recoil:\\ $h_X \equiv h_{A_1}(w)$ , 
    \item Inclusion of the full lattice information: \\ $h_X \equiv h_X(w) = \{h_{A_1}(w), R_1(w), R_2(w)\}$,
\end{enumerate}
where we consider the points at $w=\{1.03,1.10,1.17\}$ and use the provided correlations between the lattice data points. We translate the lattice data points and propagate their uncertainty and correlation into predictions of $R_1(w) = (w+1) m_B m_{D^*} g(w)/f(w)$ and $R_2(w) = (w-r) / (w-1) - F_1(w) / (m_B (w-1) f(w))$ with $r=m_{D^*} / m_B$. 

Including lattice points for $h_{A_1}$ beyond zero-recoil results in a good fit $(p_{\rm BGL} = 11\%, p_{\rm CLN} = 9\%)$ compatible with our nominal scenario. Including the full lattice information results in a poor fit $(p_{\rm BGL} = 2\%, p_{\rm CLN} = 2\%)$, where the disagreement is predominantly generated in $R_2(w)$. The extracted \Vcba\, values in the different lattice scenarios are compatible with each other, as shown in Table~\ref{tab:vcbLatticeScenarios}. We also investigate the beyond zero-recoil lattice data for an equivalent number of BGL coefficients $N_a=3$, $N_b=3$, $N_c=2$ as used in Ref.~\cite{FermilabLattice:2021cdg}. We find a much higher value of $|V_{cb}|= (42.67 \pm 0.98) \times 10^{-3}$ with a $p$-value of 5\%. The full details of the fit can be found in Appendix~\ref{app:bgl332}.

\begin{table}
    \centering
    \caption{Fitted BGL$_{121}$ coefficients and correlations.}
    \label{tab:BGL121}
    \resizebox{\linewidth}{!}{
    \begin{tabular}{lrrrrrr}
\hline
\hline
{} &          Value & \multicolumn{5}{l}{Correlation} \\
\hline
$a_0 \times 10^3$      & $  25.98\pm1.40 $ &    $    1.00$ & $  0.26$ & $ -0.23$ & $  0.28$ & $ -0.31$ \\
$b_0 \times 10^3$      & $  13.11\pm0.18 $ &    $    0.26$ & $  1.00$ & $ -0.01$ & $ -0.01$ & $ -0.62$ \\
$b_1 \times 10^3$      & $ -7.86\pm12.51 $ &    $   -0.23$ & $ -0.01$ & $  1.00$ & $  0.26$ & $ -0.47$ \\
$c_1 \times 10^3$      & $  -0.92\pm0.97 $ &    $    0.28$ & $ -0.01$ & $  0.26$ & $  1.00$ & $ -0.49$ \\
$|V_{cb}| \times 10^3$ & $  40.55\pm0.91 $ &    $   -0.31$ & $ -0.62$ & $ -0.47$ & $ -0.49$ & $  1.00$ \\
\hline
\hline
\end{tabular}

    }
\end{table}

\begin{table}
    \centering
    \caption{Fitted CLN coefficients and correlations.}
    \label{tab:CLN}
    \begin{tabular}{lrrrrr}
\hline
\hline
{} &         Value & \multicolumn{4}{l}{Correlation} \\
\hline
$\rho^2$               & $  1.22\pm0.09$ &       $ 1.00$ & $  0.58$ &  $-0.88 $ & $ 0.37  $ \\
$R_1(1)$               & $  1.37\pm0.08$ &       $ 0.58$ & $  1.00$ &  $-0.66 $ & $ -0.03 $ \\
$R_2(1)$               & $  0.88\pm0.07$ &       $-0.88$ & $ -0.66$ &  $ 1.00 $ & $ -0.14 $ \\
$|V_{cb}| \times 10^3$ & $ 40.11\pm0.85$ &       $ 0.37$ & $ -0.03$ &  $-0.14 $ & $ 1.00  $ \\
\hline
\hline
\end{tabular}

\end{table}

\begin{table}
    \centering
    \caption{Extracted $|V_\mathrm{cb}| \times 10^{3}$ values with our fitted form factor coefficients to the averaged $B^- \to D^* \ell \nu$, $\bar{B}^0 \to D^* \ell \nu$, and $\bdslnu$ shapes, with the external input for the absolute branching fractions described in the text, and our nominal scenario for the lattice input: $h_{A_1}(1) = 0.906 \pm 0.013$ from~\cite{FermilabLattice:2014ysv}.}
    \label{tab:vcbIsospinScenarios}
    \begin{tabular}{lll}
\hline
\hline
{} & BGL$_{121}$ &         CLN \\
\hline
$B^+ \to D^{*0} \ell \bar{\nu}_\ell$ & $ 42.0\pm1.2$ & $ 41.4\pm1.2$ \\
$\bar{B}^0 \to D^{*+} \ell \bar{\nu}_\ell$ & $ 38.5\pm1.3$ & $ 38.3\pm1.1$ \\
$B \to D^{*} \ell \bar{\nu}_\ell$    & $ 40.6\pm0.9$ & $ 40.1\pm0.9$ \\
\hline
\hline
\end{tabular}

\end{table}

\begin{table}
    \centering
    \caption{Extracted $|V_\mathrm{cb}| \times 10^{3}$ values with our fitted form factor coefficients to the averaged $\bdslnu$ shape, with the external input for the absolute branching fractions described in the text, and different scenarios for the lattice input.}
    \label{tab:vcbLatticeScenarios}
    \begin{tabular}{lll}
\hline
\hline
{} & BGL$_{121}$ &         CLN \\
\hline
$h_{A_1}(1)$                     & $ 40.6\pm0.9$ & $ 40.1\pm0.9$ \\
$h_{A_1}(w)$                     & $ 40.2\pm0.9$ & $ 40.0\pm0.9$ \\
$h_{A_1}(w)$, $R_1(w)$, $R_2(w)$ & $ 39.3\pm0.8$ & $ 39.4\pm0.9$ \\
\hline
\hline
\end{tabular}

\end{table}

Using on our measured $\cos\theta_\ell$ shapes we determine the forward-backward asymmetry over the full $w$ phase-space,
\begin{equation}
    A_\mathrm{FB} = \frac{\int_0^1 \d \cos_\ell \dd \Gamma / \dd \cos_\ell - \int_{-1}^0 \d \cos_\ell \dd \Gamma / \dd \cos_\ell}{\int_0^1 \d \cos_\ell \dd \Gamma / \dd \cos_\ell + \int_{-1}^0 \d \cos_\ell \dd \Gamma / \dd \cos_\ell}\,,
\end{equation}
by summing the last five and first five bins in the measured shape of $\cos\theta_\ell$ considering the correlations of the uncertainties.
We also determine the differences
\begin{equation}
    \Delta A_\mathrm{FB} = A_\mathrm{FB}^\mu - A_\mathrm{FB}^e\,.
\end{equation}
The numerical values are tabulated in Table~\ref{tab:AFB} and Table~\ref{tab:DeltaAFB} for $A_\mathrm{FB}$ and $\Delta A_\mathrm{FB}$ respectively. 

\begin{table}
    \centering
    \caption{The forward-backward asymmetries for the four decay modes and $\bar{B}^0B^-$ averages. The first uncertainty is statistical and the second uncertainty is systematic.}
    \label{tab:AFB}    
    \begin{tabular}{lr}
\hline
\hline
{} &              $A_\mathrm{FB}$ \\
\hline
$\bar{B}^0 \to D^{*+} e \bar{\nu}_e$                &  $0.218 \pm 0.030 \pm 0.008$ \\
$\bar{B}^0 \to D^{*+} \mu \bar{\nu}_\mu$            &  $0.280 \pm 0.032 \pm 0.009$ \\
$B^- \to D^{*0} e \bar{\nu}_e$                &  $0.239 \pm 0.023 \pm 0.007$ \\
$B^- \to D^{*0} \mu \bar{\nu}_\mu$            &  $0.236 \pm 0.023 \pm 0.006$ \\
$B^{(0,-)} \to D^{*(+,0)} e \bar{\nu}_e$     &  $0.230 \pm 0.018 \pm 0.005$ \\
$B^{(0,-)} \to D^{*(+,0)} \mu \bar{\nu}_\mu$ &  $0.252 \pm 0.019 \pm 0.005$ \\
\hline
\hline
\end{tabular}

\end{table}

\begin{table}
    \centering
    \caption{The difference of the forward-backward asymmetries for the $\bar{B}^0$ and $B^-$ modes, and for the $\bar{B}^0B^-$ averages. The first uncertainty is statistical and the second uncertainty is systematic.}
    \label{tab:DeltaAFB}
    \begin{tabular}{lr}
\hline
\hline
{} &        $\Delta A_\mathrm{FB}$ \\
\hline
$\bar{B}^0 \to D^{*+} \ell \bar{\nu}_\ell$ &   $0.062 \pm 0.044 \pm 0.011$ \\
$B^- \to D^{*0} \ell \bar{\nu}_\ell$ &  $-0.003 \pm 0.033 \pm 0.009$ \\
$B \to D^{*} \ell \bar{\nu}_\ell$    &   $0.022 \pm 0.026 \pm 0.007$ \\
\hline
\hline
\end{tabular}

\end{table}

\begin{figure}
    \centering
    \includegraphics[width=\linewidth]{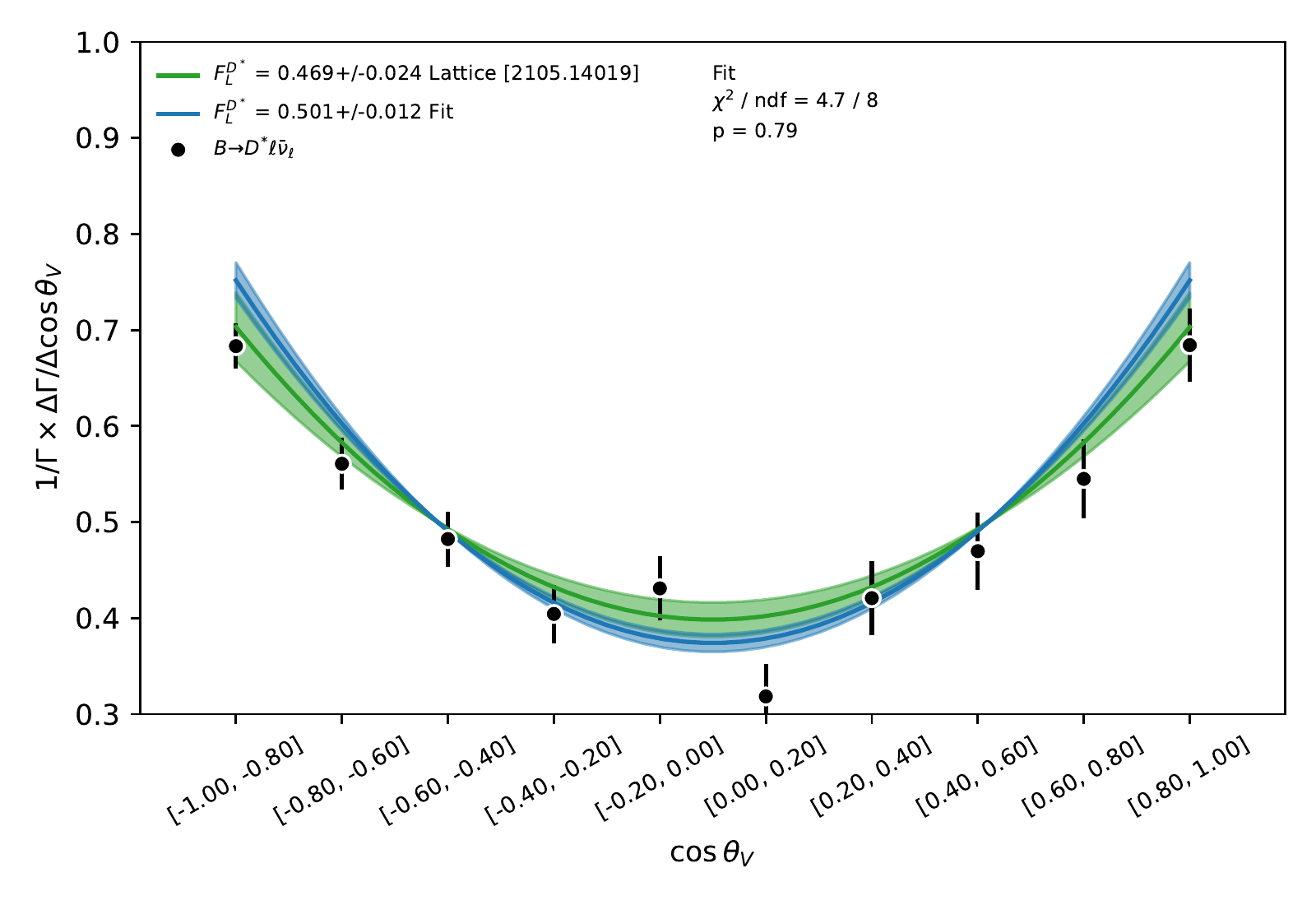}
    \caption{A representative fit of the longitudinal polarization fraction to the $\cos\theta_V$ shape of the average spectrum $\bdslnu$. The green band is the prediction using the BGL coefficients from lattice QCD calculations from \citep{FermilabLattice:2021cdg}. The blue band is our fit result.}
    \label{fig:FLFit}
\end{figure}

Using our measured $\cos\theta_V$ shapes we determine the longitudinal polarization fraction $F_L^{D^*}$ by fitting the relation~\citep{Bernlochner:2021vlv}:
\begin{equation}
    \frac{1}{\Gamma} \frac{\mathrm{d}\Gamma}{\mathrm{d}\cos \theta_V} = \frac{3}{2}\left( F_L \cos^2\theta_V + \frac{1-F_L}{2} \sin^2 \theta_V \right)\,.
\end{equation}
The fit to the fully averaged spectrum, together with the expectation from LQCD (green band) using Ref.~\cite{FermilabLattice:2021cdg}, is shown in Fig.~\ref{fig:FLFit}.
We also determine the differences
\begin{equation}
    \Delta F_L = F_L^\mu - F_L^e\,.
\end{equation}
The numerical values are tabulated in Table~\ref{tab:FL} and Table~\ref{tab:DeltaFL} for $F_L$ and $\Delta F_L$ respectively.

\begin{table}
    \centering
    \caption{The longitudinal polarization fractions for the four decay modes and various averages. The first uncertainty is statistical and the second uncertainty is systematic.}
    \label{tab:FL}    
    \begin{tabular}{lr}
\hline
\hline
{} &                $F_L^{D^{*}}$ \\
\hline
$\bar{B}^0 \to D^{*+} e \bar{\nu}_e$                &  $0.471 \pm 0.024 \pm 0.007$ \\
$\bar{B}^0 \to D^{*+} \mu \bar{\nu}_\mu$            &  $0.503 \pm 0.023 \pm 0.007$ \\
$B^- \to D^{*0} e \bar{\nu}_e$                &  $0.501 \pm 0.025 \pm 0.007$ \\
$B^- \to D^{*0} \mu \bar{\nu}_\mu$            &  $0.526 \pm 0.024 \pm 0.007$ \\
$B^{(0,-)} \to D^{*(+,0)} e \bar{\nu}_e$     &  $0.485 \pm 0.017 \pm 0.005$ \\
$B^{(0,-)} \to D^{*(+,0)} \mu \bar{\nu}_\mu$ &  $0.518 \pm 0.017 \pm 0.005$ \\
$\bar{B}^0 \to D^{*+} \ell \bar{\nu}_\ell$         &  $0.487 \pm 0.017 \pm 0.005$ \\
$B^- \to D^{*0} \ell \bar{\nu}_\ell$         &  $0.514 \pm 0.017 \pm 0.005$ \\
$B \to D^{*} \ell \bar{\nu}_\ell$            &  $0.501 \pm 0.012 \pm 0.003$ \\
\hline
\hline
\end{tabular}

\end{table}

\begin{table}
    \centering
    \caption{The difference of the longitudinal polarization fractions for the $\bar{B}^0$ and $B^-$ modes, and for the $\bar{B}^0B^-$ averages. The first uncertainty is statistical and the second uncertainty is systematics.}
    \label{tab:DeltaFL}
    \begin{tabular}{lr}
\hline
\hline
{} &         $\Delta F_L^{D^{*}}$ \\
\hline
$\bar{B}^0 \to D^{*+} \ell \bar{\nu}_\ell$ &  $0.032 \pm 0.033 \pm 0.010$ \\
$B^- \to D^{*0} \ell \bar{\nu}_\ell$ &  $0.025 \pm 0.035 \pm 0.010$ \\
$B \to D^{*} \ell \bar{\nu}_\ell$    &  $0.034 \pm 0.024 \pm 0.007$ \\
\hline
\hline
\end{tabular}

\end{table}

Finally, we determine the lepton flavor universality ratios
\begin{equation}
    R_{e\mu} = \frac{\mathcal{B}(B \to D^* e \bar{\nu}_e)}{\mathcal{B}(B \to D^* \mu \bar{\nu}_\mu)}\,,
\end{equation}
where we assume that the efficiency from the tag side reconstruction fully cancels in the ratio. The numerical values are tabulated in Table~\ref{tab:Remu}.

\begin{table}
    \centering
    \caption{The lepton flavor universality ratios for the $\bar{B}^0$ and $B^-$ modes, and for the $\bar{B}^0B^-$ average. The first uncertainty is statistical and the second uncertainty is systematic.}
    \label{tab:Remu}
    \begin{tabular}{lr}
\hline
\hline
{} &                 $R_{{e\mu}}$ \\
\hline
$\bar{B}^0 \to D^{*+} \ell \bar{\nu}_\ell$ &  $1.010 \pm 0.034 \pm 0.025$ \\
$B^- \to D^{*0} \ell \bar{\nu}_\ell$ &  $0.971 \pm 0.025 \pm 0.023$ \\
$B \to D^{*} \ell \bar{\nu}_\ell$    &  $0.990 \pm 0.021 \pm 0.023$ \\
\hline
\hline
\end{tabular}

\end{table}

\section{Summary and Conclusions}\label{sec:conclusions}

\begin{figure}
    \centering
    \includegraphics[width=\linewidth]{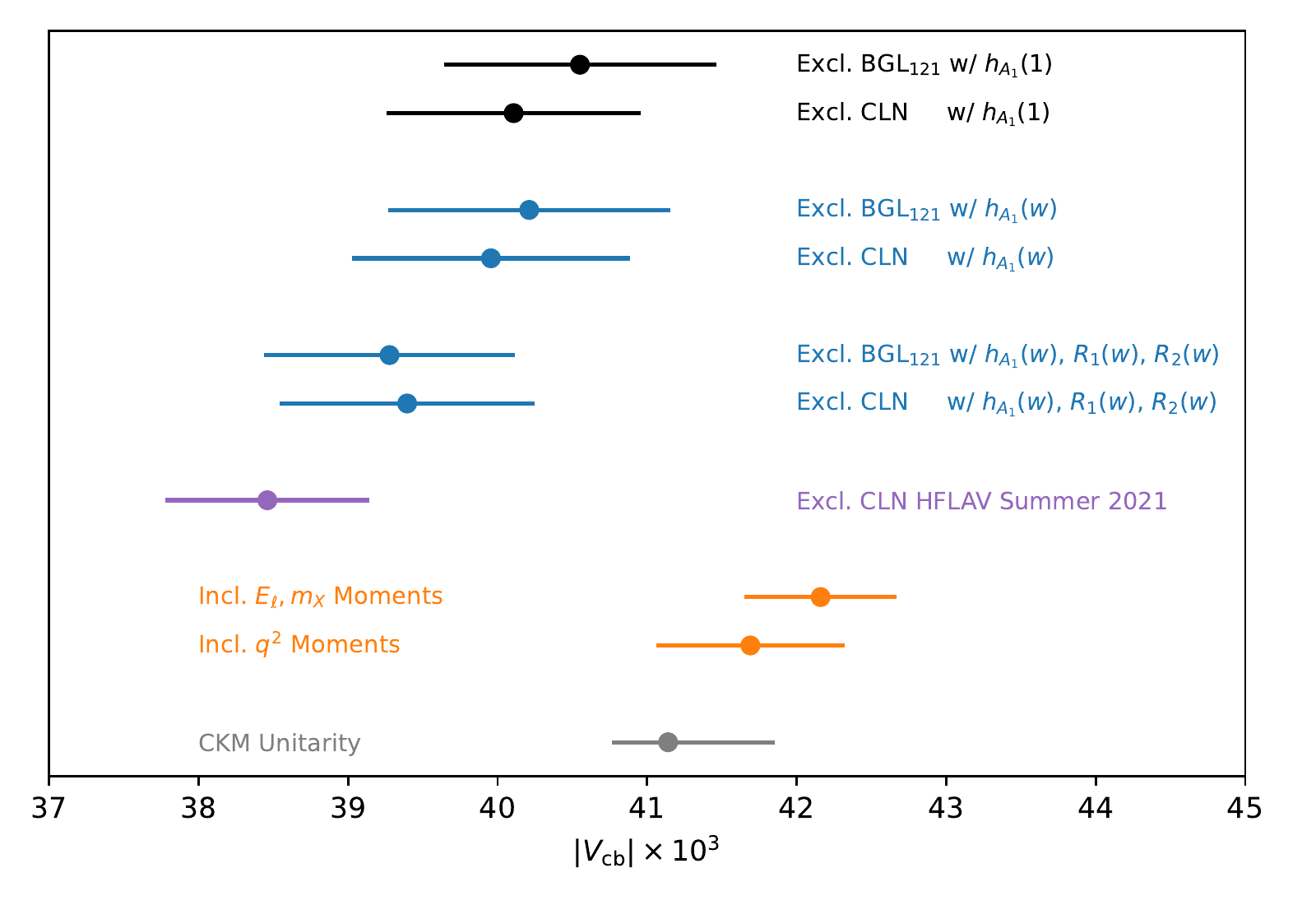}
    \caption{Our extracted \Vcba\ values using the lattice input from Ref.~\cite{FermilabLattice:2014ysv} (black) and Ref.~\cite{FermilabLattice:2021cdg} (blue), together with the latest exclusive HFLAV average \cite{HFLAV:2022pwe} (purple), determinations from inclusive approaches \cite{Bordone:2021oof,Bernlochner:2022ucr} (orange), and from CKM unitarity (grey).}
    \label{fig:VcbOverview}
\end{figure}

We presented measurements of differential distributions of \bdslnu probing both $\bar{B}^0$ and $B^-$ modes. In total, we measure the signal yield in 160 differential bins, characterizing the 1D projections of the hadronic recoil parameter $w$, and the angles $\cos \theta_\ell$, $\cos \theta_V$, and $\chi$. In addition, the full experimental correlations between the projections were determined, allowing for a simultaneous analysis of all bins. The lattice QCD calculation of Ref.~\cite{FermilabLattice:2014ysv} at zero recoil was used for the $\Vcba$\, extraction. 
The value of the CKM matrix element \Vcba\ was determined using external input for the branching fraction and we find for our fit with the BGL parameterization, with the number of floating BGL parameters determined using a nested-hypothesis test,
\begin{align}
 \Vcba & = \left(	 40.6 \pm 0.9 \right) \times 10^{-3} \, , 
\end{align}
in agreement with \Vcba\ from inclusive determinations~\cite{Bordone:2021oof,Bernlochner:2022ucr}. A study of the recent lattice QCD calculations from Ref.~\cite{FermilabLattice:2021cdg} was performed, and the impact on \Vcba\, is shown in Fig.~\ref{fig:VcbOverview}, together with other determinations of \Vcba\ .

The measured differential distribution of $\cos \theta_\ell$ is used to determine the forward-backward asymmetry $A_{\mathrm{FB}}$ for electron and muon final states, as well as their difference. We find values which are compatible with the prediction from lattice QCD from Ref.~\cite{FermilabLattice:2021cdg}, the predictions of Refs.~\cite{Bernlochner:2022ywh, Bobeth:2021lya}, and the experimental value from Ref.~\cite{Waheed:2018djm} determined in Ref.~\cite{Bobeth:2021lya}. Similarly the longitudinal $D^*$ polarization fraction can be determined from the measured distribution of $\cos \theta_V$ and we find good agreement with Refs.~\cite{FermilabLattice:2021cdg,Bernlochner:2022ywh, Bobeth:2021lya}. Lastly, we obtain the lepton-flavor universality ratio 
\begin{align}
 R_{e\mu} = \frac{\mathcal{B}(B \to D^* e \bar{\nu}_e)}{\mathcal{B}(B \to D^* \mu \bar{\nu}_\mu)} = 0.990 \pm 0.021 \pm 0.023 \, ,
\end{align}
which is in good agreement with Refs.~\cite{Bobeth:2021lya,Bernlochner:2022ywh}.

\acknowledgments

This work, based on data collected using the Belle detector, which was
operated until June 2010, was supported by 
the Ministry of Education, Culture, Sports, Science, and
Technology (MEXT) of Japan, the Japan Society for the 
Promotion of Science (JSPS), and the Tau-Lepton Physics 
Research Center of Nagoya University; 
the Australian Research Council including grants
DP180102629, 
DP170102389, 
DP170102204, 
DE220100462, 
DP150103061, 
FT130100303; 
Austrian Federal Ministry of Education, Science and Research (FWF) and
FWF Austrian Science Fund No.~P~31361-N36;
the National Natural Science Foundation of China under Contracts
No.~11675166,  
No.~11705209;  
No.~11975076;  
No.~12135005;  
No.~12175041;  
No.~12161141008; 
Key Research Program of Frontier Sciences, Chinese Academy of Sciences (CAS), Grant No.~QYZDJ-SSW-SLH011; 
Project ZR2022JQ02 supported by Shandong Provincial Natural Science Foundation;
the Ministry of Education, Youth and Sports of the Czech
Republic under Contract No.~LTT17020;
the Czech Science Foundation Grant No. 22-18469S;
Horizon 2020 ERC Advanced Grant No.~884719 and ERC Starting Grant No.~947006 ``InterLeptons'' (European Union);
the Carl Zeiss Foundation, the Deutsche Forschungsgemeinschaft, the
Excellence Cluster Universe, and the VolkswagenStiftung;
the Department of Atomic Energy (Project Identification No. RTI 4002) and the Department of Science and Technology of India; 
the Istituto Nazionale di Fisica Nucleare of Italy; 
National Research Foundation (NRF) of Korea Grant
Nos.~2016R1\-D1A1B\-02012900, 2018R1\-A2B\-3003643,
2018R1\-A6A1A\-06024970, RS\-2022\-00197659,
2019R1\-I1A3A\-01058933, 2021R1\-A6A1A\-03043957,
2021R1\-F1A\-1060423, 2021R1\-F1A\-1064008, 2022R1\-A2C\-1003993;
Radiation Science Research Institute, Foreign Large-size Research Facility Application Supporting project, the Global Science Experimental Data Hub Center of the Korea Institute of Science and Technology Information and KREONET/GLORIAD;
the Polish Ministry of Science and Higher Education and 
the National Science Center;
the Ministry of Science and Higher Education of the Russian Federation, Agreement 14.W03.31.0026, 
and the HSE University Basic Research Program, Moscow; 
University of Tabuk research grants
S-1440-0321, S-0256-1438, and S-0280-1439 (Saudi Arabia);
the Slovenian Research Agency Grant Nos. J1-9124 and P1-0135;
Ikerbasque, Basque Foundation for Science, Spain;
the Swiss National Science Foundation; 
the Ministry of Education and the Ministry of Science and Technology of Taiwan;
and the United States Department of Energy and the National Science Foundation.
These acknowledgements are not to be interpreted as an endorsement of any
statement made by any of our institutes, funding agencies, governments, or
their representatives.

We thank Danny Van Dyk, Martin Jung, Zoltan Ligeti and Dean Robinson for useful discussions that helped to improve the scientific content of this manuscript. FB thanks LBNL for its hospitality. FB is supported by DFG Emmy-Noether Grant No. BE 6075/1-1. MTP is supported by the Argelander Starter-Kit Grant of the University of Bonn. FB, MTP, and FM are supported by BMBF Grant No. 05H21PDKBA. 

We thank the KEKB group for the excellent operation of the
accelerator; the KEK cryogenics group for the efficient
operation of the solenoid; and the KEK computer group and the Pacific Northwest National
Laboratory (PNNL) Environmental Molecular Sciences Laboratory (EMSL)
computing group for strong computing support; and the National
Institute of Informatics, and Science Information NETwork 6 (SINET6) for
valuable network support.


\bibliographystyle{apsrev4-1}
\bibliography{paper}

\begin{thebibliography}{46}%
\makeatletter
\providecommand \@ifxundefined [1]{%
 \@ifx{#1\undefined}
}%
\providecommand \@ifnum [1]{%
 \ifnum #1\expandafter \@firstoftwo
 \else \expandafter \@secondoftwo
 \fi
}%
\providecommand \@ifx [1]{%
 \ifx #1\expandafter \@firstoftwo
 \else \expandafter \@secondoftwo
 \fi
}%
\providecommand \natexlab [1]{#1}%
\providecommand \enquote  [1]{``#1''}%
\providecommand \bibnamefont  [1]{#1}%
\providecommand \bibfnamefont [1]{#1}%
\providecommand \citenamefont [1]{#1}%
\providecommand \href@noop [0]{\@secondoftwo}%
\providecommand \href [0]{\begingroup \@sanitize@url \@href}%
\providecommand \@href[1]{\@@startlink{#1}\@@href}%
\providecommand \@@href[1]{\endgroup#1\@@endlink}%
\providecommand \@sanitize@url [0]{\catcode `\\12\catcode `\$12\catcode
  `\&12\catcode `\#12\catcode `\^12\catcode `\_12\catcode `\%12\relax}%
\providecommand \@@startlink[1]{}%
\providecommand \@@endlink[0]{}%
\providecommand \url  [0]{\begingroup\@sanitize@url \@url }%
\providecommand \@url [1]{\endgroup\@href {#1}{\urlprefix }}%
\providecommand \urlprefix  [0]{URL }%
\providecommand \Eprint [0]{\href }%
\providecommand \doibase [0]{http://dx.doi.org/}%
\providecommand \selectlanguage [0]{\@gobble}%
\providecommand \bibinfo  [0]{\@secondoftwo}%
\providecommand \bibfield  [0]{\@secondoftwo}%
\providecommand \translation [1]{[#1]}%
\providecommand \BibitemOpen [0]{}%
\providecommand \bibitemStop [0]{}%
\providecommand \bibitemNoStop [0]{.\EOS\space}%
\providecommand \EOS [0]{\spacefactor3000\relax}%
\providecommand \BibitemShut  [1]{\csname bibitem#1\endcsname}%
\let\auto@bib@innerbib\@empty
\bibitem [{\citenamefont {Cabibbo}(1963)}]{PhysRevLett.10.531}%
  \BibitemOpen
  \bibfield  {author} {\bibinfo {author} {\bibfnamefont {N.}~\bibnamefont
  {Cabibbo}},\ }\href {\doibase 10.1103/PhysRevLett.10.531} {\bibfield
  {journal} {\bibinfo  {journal} {Phys. Rev. Lett.}\ }\textbf {\bibinfo
  {volume} {10}},\ \bibinfo {pages} {531} (\bibinfo {year} {1963})}\BibitemShut
  {NoStop}%
\bibitem [{\citenamefont {Kobayashi}\ and\ \citenamefont
  {Maskawa}(1973)}]{km_paper}%
  \BibitemOpen
  \bibfield  {author} {\bibinfo {author} {\bibfnamefont {M.}~\bibnamefont
  {Kobayashi}}\ and\ \bibinfo {author} {\bibfnamefont {T.}~\bibnamefont
  {Maskawa}},\ }\href {\doibase 10.1143/PTP.49.652} {\bibfield  {journal}
  {\bibinfo  {journal} {Progress of Theoretical Physics}\ }\textbf {\bibinfo
  {volume} {49}},\ \bibinfo {pages} {652} (\bibinfo {year} {1973})},\ \Eprint
  {http://arxiv.org/abs/https://academic.oup.com/ptp/article-pdf/49/2/652/5257692/49-2-652.pdf}
  {https://academic.oup.com/ptp/article-pdf/49/2/652/5257692/49-2-652.pdf}
  \BibitemShut {NoStop}%
\bibitem [{\citenamefont {Workman}\ \emph {et~al.}(2022)\citenamefont {Workman}
  \emph {et~al.}}]{pdg:2022}%
  \BibitemOpen
  \bibfield  {author} {\bibinfo {author} {\bibfnamefont {R.~L.}\ \bibnamefont
  {Workman}} \emph {et~al.} (\bibinfo {collaboration} {Particle Data Group}),\
  }\href {\doibase 10.1093/ptep/ptac097} {\bibfield  {journal} {\bibinfo
  {journal} {PTEP}\ }\textbf {\bibinfo {volume} {2022}},\ \bibinfo {pages}
  {083C01} (\bibinfo {year} {2022})}\BibitemShut {NoStop}%
\bibitem [{\citenamefont {Beneke}\ \emph {et~al.}(2019)\citenamefont {Beneke},
  \citenamefont {Bobeth},\ and\ \citenamefont {Szafron}}]{Beneke:2019slt}%
  \BibitemOpen
  \bibfield  {author} {\bibinfo {author} {\bibfnamefont {M.}~\bibnamefont
  {Beneke}}, \bibinfo {author} {\bibfnamefont {C.}~\bibnamefont {Bobeth}}, \
  and\ \bibinfo {author} {\bibfnamefont {R.}~\bibnamefont {Szafron}},\ }\href
  {\doibase 10.1007/JHEP10(2019)232} {\bibfield  {journal} {\bibinfo  {journal}
  {JHEP}\ }\textbf {\bibinfo {volume} {10}},\ \bibinfo {pages} {232} (\bibinfo
  {year} {2019})},\ \Eprint {http://arxiv.org/abs/1908.07011} {arXiv:1908.07011
  [hep-ph]} \BibitemShut {NoStop}%
\bibitem [{\citenamefont {Bobeth}\ \emph {et~al.}(2014)\citenamefont {Bobeth},
  \citenamefont {Gorbahn}, \citenamefont {Hermann}, \citenamefont {Misiak},
  \citenamefont {Stamou},\ and\ \citenamefont {Steinhauser}}]{Bobeth:2013uxa}%
  \BibitemOpen
  \bibfield  {author} {\bibinfo {author} {\bibfnamefont {C.}~\bibnamefont
  {Bobeth}}, \bibinfo {author} {\bibfnamefont {M.}~\bibnamefont {Gorbahn}},
  \bibinfo {author} {\bibfnamefont {T.}~\bibnamefont {Hermann}}, \bibinfo
  {author} {\bibfnamefont {M.}~\bibnamefont {Misiak}}, \bibinfo {author}
  {\bibfnamefont {E.}~\bibnamefont {Stamou}}, \ and\ \bibinfo {author}
  {\bibfnamefont {M.}~\bibnamefont {Steinhauser}},\ }\href {\doibase
  10.1103/PhysRevLett.112.101801} {\bibfield  {journal} {\bibinfo  {journal}
  {Phys. Rev. Lett.}\ }\textbf {\bibinfo {volume} {112}},\ \bibinfo {pages}
  {101801} (\bibinfo {year} {2014})},\ \Eprint {http://arxiv.org/abs/1311.0903}
  {arXiv:1311.0903 [hep-ph]} \BibitemShut {NoStop}%
\bibitem [{\citenamefont {Zheng}\ \emph {et~al.}(2021)\citenamefont {Zheng},
  \citenamefont {Xu}, \citenamefont {Cao}, \citenamefont {Yu}, \citenamefont
  {Wang}, \citenamefont {Prell}, \citenamefont {Cheung},\ and\ \citenamefont
  {Ruan}}]{Zheng:2021xuq}%
  \BibitemOpen
  \bibfield  {author} {\bibinfo {author} {\bibfnamefont {T.}~\bibnamefont
  {Zheng}}, \bibinfo {author} {\bibfnamefont {J.}~\bibnamefont {Xu}}, \bibinfo
  {author} {\bibfnamefont {L.}~\bibnamefont {Cao}}, \bibinfo {author}
  {\bibfnamefont {D.}~\bibnamefont {Yu}}, \bibinfo {author} {\bibfnamefont
  {W.}~\bibnamefont {Wang}}, \bibinfo {author} {\bibfnamefont {S.}~\bibnamefont
  {Prell}}, \bibinfo {author} {\bibfnamefont {Y.-K.~E.}\ \bibnamefont
  {Cheung}}, \ and\ \bibinfo {author} {\bibfnamefont {M.}~\bibnamefont
  {Ruan}},\ }\href {\doibase 10.1088/1674-1137/abcf1f} {\bibfield  {journal}
  {\bibinfo  {journal} {Chin. Phys. C}\ }\textbf {\bibinfo {volume} {45}},\
  \bibinfo {pages} {023001} (\bibinfo {year} {2021})},\ \Eprint
  {http://arxiv.org/abs/2007.08234} {arXiv:2007.08234 [hep-ex]} \BibitemShut
  {NoStop}%
\bibitem [{\citenamefont {Altmannshofer}\ and\ \citenamefont
  {Lewis}(2022)}]{Altmannshofer:2021uub}%
  \BibitemOpen
  \bibfield  {author} {\bibinfo {author} {\bibfnamefont {W.}~\bibnamefont
  {Altmannshofer}}\ and\ \bibinfo {author} {\bibfnamefont {N.}~\bibnamefont
  {Lewis}},\ }\href {\doibase 10.1103/PhysRevD.105.033004} {\bibfield
  {journal} {\bibinfo  {journal} {Phys. Rev. D}\ }\textbf {\bibinfo {volume}
  {105}},\ \bibinfo {pages} {033004} (\bibinfo {year} {2022})},\ \Eprint
  {http://arxiv.org/abs/2112.03437} {arXiv:2112.03437 [hep-ph]} \BibitemShut
  {NoStop}%
\bibitem [{\citenamefont {Bordone}\ \emph {et~al.}(2021)\citenamefont
  {Bordone}, \citenamefont {Capdevila},\ and\ \citenamefont
  {Gambino}}]{Bordone:2021oof}%
  \BibitemOpen
  \bibfield  {author} {\bibinfo {author} {\bibfnamefont {M.}~\bibnamefont
  {Bordone}}, \bibinfo {author} {\bibfnamefont {B.}~\bibnamefont {Capdevila}},
  \ and\ \bibinfo {author} {\bibfnamefont {P.}~\bibnamefont {Gambino}},\ }\href
  {\doibase 10.1016/j.physletb.2021.136679} {\bibfield  {journal} {\bibinfo
  {journal} {Phys. Lett. B}\ }\textbf {\bibinfo {volume} {822}},\ \bibinfo
  {pages} {136679} (\bibinfo {year} {2021})},\ \Eprint
  {http://arxiv.org/abs/2107.00604} {arXiv:2107.00604 [hep-ph]} \BibitemShut
  {NoStop}%
\bibitem [{\citenamefont {Bernlochner}\ \emph
  {et~al.}(2022{\natexlab{a}})\citenamefont {Bernlochner}, \citenamefont
  {Fael}, \citenamefont {Olschewsky}, \citenamefont {Persson}, \citenamefont
  {van Tonder}, \citenamefont {Vos},\ and\ \citenamefont
  {Welsch}}]{Bernlochner:2022ucr}%
  \BibitemOpen
  \bibfield  {author} {\bibinfo {author} {\bibfnamefont {F.}~\bibnamefont
  {Bernlochner}}, \bibinfo {author} {\bibfnamefont {M.}~\bibnamefont {Fael}},
  \bibinfo {author} {\bibfnamefont {K.}~\bibnamefont {Olschewsky}}, \bibinfo
  {author} {\bibfnamefont {E.}~\bibnamefont {Persson}}, \bibinfo {author}
  {\bibfnamefont {R.}~\bibnamefont {van Tonder}}, \bibinfo {author}
  {\bibfnamefont {K.~K.}\ \bibnamefont {Vos}}, \ and\ \bibinfo {author}
  {\bibfnamefont {M.}~\bibnamefont {Welsch}},\ }\href@noop {} {\  (\bibinfo
  {year} {2022}{\natexlab{a}})},\ \Eprint {http://arxiv.org/abs/2205.10274}
  {arXiv:2205.10274 [hep-ph]} \BibitemShut {NoStop}%
\bibitem [{\citenamefont {Waheed}\ \emph
  {et~al.}(2019{\natexlab{a}})\citenamefont {Waheed} \emph
  {et~al.}}]{Belle:2018ezy}%
  \BibitemOpen
  \bibfield  {author} {\bibinfo {author} {\bibfnamefont {E.}~\bibnamefont
  {Waheed}} \emph {et~al.} (\bibinfo {collaboration} {Belle}),\ }\href
  {\doibase 10.1103/PhysRevD.100.052007} {\bibfield  {journal} {\bibinfo
  {journal} {Phys. Rev. D}\ }\textbf {\bibinfo {volume} {100}},\ \bibinfo
  {pages} {052007} (\bibinfo {year} {2019}{\natexlab{a}})},\ \bibinfo {note}
  {[Erratum: Phys.Rev.D 103, 079901 (2021)]},\ \Eprint
  {http://arxiv.org/abs/1809.03290} {arXiv:1809.03290 [hep-ex]} \BibitemShut
  {NoStop}%
\bibitem [{\citenamefont {Amhis}\ \emph
  {et~al.}(2022{\natexlab{a}})\citenamefont {Amhis} \emph
  {et~al.}}]{Amhis:2022mac}%
  \BibitemOpen
  \bibfield  {author} {\bibinfo {author} {\bibfnamefont {Y.}~\bibnamefont
  {Amhis}} \emph {et~al.},\ }\href@noop {} {\  (\bibinfo {year}
  {2022}{\natexlab{a}})},\ \Eprint {http://arxiv.org/abs/2206.07501}
  {arXiv:2206.07501 [hep-ex]} \BibitemShut {NoStop}%
\bibitem [{\citenamefont {Keck}\ \emph {et~al.}(2019)\citenamefont {Keck} \emph
  {et~al.}}]{Keck:2018lcd}%
  \BibitemOpen
  \bibfield  {author} {\bibinfo {author} {\bibfnamefont {T.}~\bibnamefont
  {Keck}} \emph {et~al.},\ }\href {\doibase 10.1007/s41781-019-0021-8}
  {\bibfield  {journal} {\bibinfo  {journal} {Comput. Softw. Big Sci.}\
  }\textbf {\bibinfo {volume} {3}},\ \bibinfo {pages} {6} (\bibinfo {year}
  {2019})},\ \Eprint {http://arxiv.org/abs/1807.08680} {arXiv:1807.08680
  [hep-ex]} \BibitemShut {NoStop}%
\bibitem [{\citenamefont {Manohar}\ and\ \citenamefont
  {Wise}(2000)}]{Manohar:2000dt}%
  \BibitemOpen
  \bibfield  {author} {\bibinfo {author} {\bibfnamefont {A.~V.}\ \bibnamefont
  {Manohar}}\ and\ \bibinfo {author} {\bibfnamefont {M.~B.}\ \bibnamefont
  {Wise}},\ }\href@noop {} {\bibfield  {journal} {\bibinfo  {journal} {Camb.
  Monogr. Part. Phys. Nucl. Phys. Cosmol.}\ }\textbf {\bibinfo {volume} {10}},\
  \bibinfo {pages} {1} (\bibinfo {year} {2000})}\BibitemShut {NoStop}%
\bibitem [{\citenamefont {Boyd}\ \emph {et~al.}(1996)\citenamefont {Boyd},
  \citenamefont {Grinstein},\ and\ \citenamefont {Lebed}}]{Boyd:1995sq}%
  \BibitemOpen
  \bibfield  {author} {\bibinfo {author} {\bibfnamefont {C.~G.}\ \bibnamefont
  {Boyd}}, \bibinfo {author} {\bibfnamefont {B.}~\bibnamefont {Grinstein}}, \
  and\ \bibinfo {author} {\bibfnamefont {R.~F.}\ \bibnamefont {Lebed}},\ }\href
  {\doibase 10.1016/0550-3213(95)00653-2} {\bibfield  {journal} {\bibinfo
  {journal} {Nucl. Phys. B}\ }\textbf {\bibinfo {volume} {461}},\ \bibinfo
  {pages} {493} (\bibinfo {year} {1996})},\ \Eprint
  {http://arxiv.org/abs/hep-ph/9508211} {arXiv:hep-ph/9508211} \BibitemShut
  {NoStop}%
\bibitem [{\citenamefont {Boyd}\ \emph {et~al.}(1997)\citenamefont {Boyd},
  \citenamefont {Grinstein},\ and\ \citenamefont {Lebed}}]{Boyd:1997kz}%
  \BibitemOpen
  \bibfield  {author} {\bibinfo {author} {\bibfnamefont {C.~G.}\ \bibnamefont
  {Boyd}}, \bibinfo {author} {\bibfnamefont {B.}~\bibnamefont {Grinstein}}, \
  and\ \bibinfo {author} {\bibfnamefont {R.~F.}\ \bibnamefont {Lebed}},\ }\href
  {\doibase 10.1103/PhysRevD.56.6895} {\bibfield  {journal} {\bibinfo
  {journal} {Phys. Rev. D}\ }\textbf {\bibinfo {volume} {56}},\ \bibinfo
  {pages} {6895} (\bibinfo {year} {1997})},\ \Eprint
  {http://arxiv.org/abs/hep-ph/9705252} {arXiv:hep-ph/9705252} \BibitemShut
  {NoStop}%
\bibitem [{\citenamefont {Bazavov}\ \emph {et~al.}(2022)\citenamefont {Bazavov}
  \emph {et~al.}}]{FermilabLattice:2021cdg}%
  \BibitemOpen
  \bibfield  {author} {\bibinfo {author} {\bibfnamefont {A.}~\bibnamefont
  {Bazavov}} \emph {et~al.} (\bibinfo {collaboration} {Fermilab Lattice,
  MILC}),\ }\href {\doibase 10.1140/epjc/s10052-022-10984-9} {\bibfield
  {journal} {\bibinfo  {journal} {Eur. Phys. J. C}\ }\textbf {\bibinfo {volume}
  {82}},\ \bibinfo {pages} {1141} (\bibinfo {year} {2022})},\ \Eprint
  {http://arxiv.org/abs/2105.14019} {arXiv:2105.14019 [hep-lat]} \BibitemShut
  {NoStop}%
\bibitem [{\citenamefont {Bailey}\ \emph {et~al.}(2014)\citenamefont {Bailey}
  \emph {et~al.}}]{FermilabLattice:2014ysv}%
  \BibitemOpen
  \bibfield  {author} {\bibinfo {author} {\bibfnamefont {J.~A.}\ \bibnamefont
  {Bailey}} \emph {et~al.} (\bibinfo {collaboration} {Fermilab Lattice,
  MILC}),\ }\href {\doibase 10.1103/PhysRevD.89.114504} {\bibfield  {journal}
  {\bibinfo  {journal} {Phys. Rev. D}\ }\textbf {\bibinfo {volume} {89}},\
  \bibinfo {pages} {114504} (\bibinfo {year} {2014})},\ \Eprint
  {http://arxiv.org/abs/1403.0635} {arXiv:1403.0635 [hep-lat]} \BibitemShut
  {NoStop}%
\bibitem [{\citenamefont {Grinstein}\ and\ \citenamefont
  {Kobach}(2017)}]{Grinstein:2017nlq}%
  \BibitemOpen
  \bibfield  {author} {\bibinfo {author} {\bibfnamefont {B.}~\bibnamefont
  {Grinstein}}\ and\ \bibinfo {author} {\bibfnamefont {A.}~\bibnamefont
  {Kobach}},\ }\href {\doibase 10.1016/j.physletb.2017.05.078} {\bibfield
  {journal} {\bibinfo  {journal} {Phys. Lett. B}\ }\textbf {\bibinfo {volume}
  {771}},\ \bibinfo {pages} {359} (\bibinfo {year} {2017})},\ \Eprint
  {http://arxiv.org/abs/1703.08170} {arXiv:1703.08170 [hep-ph]} \BibitemShut
  {NoStop}%
\bibitem [{\citenamefont {Caprini}\ \emph {et~al.}(1998)\citenamefont
  {Caprini}, \citenamefont {Lellouch},\ and\ \citenamefont
  {Neubert}}]{Caprini:1997mu}%
  \BibitemOpen
  \bibfield  {author} {\bibinfo {author} {\bibfnamefont {I.}~\bibnamefont
  {Caprini}}, \bibinfo {author} {\bibfnamefont {L.}~\bibnamefont {Lellouch}}, \
  and\ \bibinfo {author} {\bibfnamefont {M.}~\bibnamefont {Neubert}},\ }\href
  {\doibase 10.1016/S0550-3213(98)00350-2} {\bibfield  {journal} {\bibinfo
  {journal} {Nucl. Phys. B}\ }\textbf {\bibinfo {volume} {530}},\ \bibinfo
  {pages} {153} (\bibinfo {year} {1998})},\ \Eprint
  {http://arxiv.org/abs/hep-ph/9712417} {arXiv:hep-ph/9712417} \BibitemShut
  {NoStop}%
\bibitem [{\citenamefont {Kurokawa}\ and\ \citenamefont
  {Kikutani}(2003)}]{KEKB}%
  \BibitemOpen
  \bibfield  {author} {\bibinfo {author} {\bibfnamefont {S.}~\bibnamefont
  {Kurokawa}}\ and\ \bibinfo {author} {\bibfnamefont {E.}~\bibnamefont
  {Kikutani}},\ }\href {\doibase https://doi.org/10.1016/S0168-9002(02)01771-0}
  {\bibfield  {journal} {\bibinfo  {journal} {Nucl. Instr. and. Meth.}\
  }\textbf {\bibinfo {volume} {A499}},\ \bibinfo {pages} {1 } (\bibinfo {year}
  {2003})},\ \bibinfo {note} {{and other papers included in this Volume; T.~Abe
  {\it et al.}, Prog. Theor. Exp. Phys. {\bf 2013}, 03A001 (2013) and
  references therein.}}\BibitemShut {Stop}%
\bibitem [{\citenamefont {Abashian}\ \emph
  {et~al.}(2002{\natexlab{a}})\citenamefont {Abashian} \emph
  {et~al.}}]{Abashian:2000cg}%
  \BibitemOpen
  \bibfield  {author} {\bibinfo {author} {\bibfnamefont {A.}~\bibnamefont
  {Abashian}} \emph {et~al.},\ }\href {\doibase 10.1016/S0168-9002(01)02013-7}
  {\bibfield  {journal} {\bibinfo  {journal} {Nucl. Instrum. Meth.}\ }\textbf
  {\bibinfo {volume} {A479}},\ \bibinfo {pages} {117} (\bibinfo {year}
  {2002}{\natexlab{a}})},\ \bibinfo {note} {also see detector section in
  J.~Brodzicka {\it et al.}, Prog. Theor. Exp. Phys. {\bf 2012}, 04D001
  (2012).}\BibitemShut {Stop}%
\bibitem [{\citenamefont {Hanagaki}\ \emph {et~al.}(2002)\citenamefont
  {Hanagaki}, \citenamefont {Kakuno}, \citenamefont {Ikeda}, \citenamefont
  {Iijima},\ and\ \citenamefont {Tsukamoto}}]{HANAGAKI2002490}%
  \BibitemOpen
  \bibfield  {author} {\bibinfo {author} {\bibfnamefont {K.}~\bibnamefont
  {Hanagaki}}, \bibinfo {author} {\bibfnamefont {H.}~\bibnamefont {Kakuno}},
  \bibinfo {author} {\bibfnamefont {H.}~\bibnamefont {Ikeda}}, \bibinfo
  {author} {\bibfnamefont {T.}~\bibnamefont {Iijima}}, \ and\ \bibinfo {author}
  {\bibfnamefont {T.}~\bibnamefont {Tsukamoto}},\ }\href {\doibase
  https://doi.org/10.1016/S0168-9002(01)02113-1} {\bibfield  {journal}
  {\bibinfo  {journal} {Nucl. Instr. and. Meth.}\ }\textbf {\bibinfo {volume}
  {A485}},\ \bibinfo {pages} {490 } (\bibinfo {year} {2002})}\BibitemShut
  {NoStop}%
\bibitem [{\citenamefont {Abashian}\ \emph
  {et~al.}(2002{\natexlab{b}})\citenamefont {Abashian} \emph
  {et~al.}}]{ABASHIAN200269}%
  \BibitemOpen
  \bibfield  {author} {\bibinfo {author} {\bibfnamefont {A.}~\bibnamefont
  {Abashian}} \emph {et~al.},\ }\href {\doibase
  https://doi.org/10.1016/S0168-9002(02)01164-6} {\bibfield  {journal}
  {\bibinfo  {journal} {Nucl. Instr. and. Meth.}\ }\textbf {\bibinfo {volume}
  {A491}},\ \bibinfo {pages} {69 } (\bibinfo {year}
  {2002}{\natexlab{b}})}\BibitemShut {NoStop}%
\bibitem [{\citenamefont {Kuhr}\ \emph {et~al.}(2019)\citenamefont {Kuhr},
  \citenamefont {Pulvermacher}, \citenamefont {Ritter}, \citenamefont {Hauth},\
  and\ \citenamefont {Braun}}]{basf2}%
  \BibitemOpen
  \bibfield  {author} {\bibinfo {author} {\bibfnamefont {T.}~\bibnamefont
  {Kuhr}}, \bibinfo {author} {\bibfnamefont {C.}~\bibnamefont {Pulvermacher}},
  \bibinfo {author} {\bibfnamefont {M.}~\bibnamefont {Ritter}}, \bibinfo
  {author} {\bibfnamefont {T.}~\bibnamefont {Hauth}}, \ and\ \bibinfo {author}
  {\bibfnamefont {N.}~\bibnamefont {Braun}} (\bibinfo {collaboration} {Belle II
  Framework Software Group}),\ }\href {\doibase 10.1007/s41781-018-0017-9}
  {\bibfield  {journal} {\bibinfo  {journal} {Comput. Softw. Big Sci.}\
  }\textbf {\bibinfo {volume} {3}},\ \bibinfo {pages} {1} (\bibinfo {year}
  {2019})},\ \Eprint {http://arxiv.org/abs/1809.04299} {arXiv:1809.04299
  [physics.comp-ph]} \BibitemShut {NoStop}%
\bibitem [{\citenamefont {Gelb}\ \emph {et~al.}(2018)\citenamefont {Gelb} \emph
  {et~al.}}]{b2b2}%
  \BibitemOpen
  \bibfield  {author} {\bibinfo {author} {\bibfnamefont {M.}~\bibnamefont
  {Gelb}} \emph {et~al.},\ }\href {\doibase 10.1007/s41781-018-0016-x}
  {\bibfield  {journal} {\bibinfo  {journal} {Comput. Softw. Big Sci.}\
  }\textbf {\bibinfo {volume} {2}},\ \bibinfo {pages} {9} (\bibinfo {year}
  {2018})},\ \Eprint {http://arxiv.org/abs/1810.00019} {arXiv:1810.00019
  [hep-ex]} \BibitemShut {NoStop}%
\bibitem [{\citenamefont {Lange}(2001)}]{EvtGen}%
  \BibitemOpen
  \bibfield  {author} {\bibinfo {author} {\bibfnamefont {D.~J.}\ \bibnamefont
  {Lange}},\ }\href
  {http://www.sciencedirect.com/science/article/pii/S0168900201000894}
  {\bibfield  {journal} {\bibinfo  {journal} {Nucl. Instr. and. Meth.}\
  }\textbf {\bibinfo {volume} {A462}},\ \bibinfo {pages} {152 } (\bibinfo
  {year} {2001})}\BibitemShut {NoStop}%
\bibitem [{\citenamefont {Brun}\ \emph {et~al.}(1987)\citenamefont {Brun},
  \citenamefont {Bruyant}, \citenamefont {Maire}, \citenamefont {McPherson},\
  and\ \citenamefont {Zanarini}}]{Geant3}%
  \BibitemOpen
  \bibfield  {author} {\bibinfo {author} {\bibfnamefont {R.}~\bibnamefont
  {Brun}}, \bibinfo {author} {\bibfnamefont {F.}~\bibnamefont {Bruyant}},
  \bibinfo {author} {\bibfnamefont {M.}~\bibnamefont {Maire}}, \bibinfo
  {author} {\bibfnamefont {A.~C.}\ \bibnamefont {McPherson}}, \ and\ \bibinfo
  {author} {\bibfnamefont {P.}~\bibnamefont {Zanarini}},\ }\href
  {http://inspirehep.net/record/252007?ln=en} {\bibfield  {journal} {\bibinfo
  {journal} {{CERN-DD-EE-84-1}}\ } (\bibinfo {year} {1987})}\BibitemShut
  {NoStop}%
\bibitem [{\citenamefont {Barberio}\ \emph {et~al.}(1991)\citenamefont
  {Barberio}, \citenamefont {van Eijk},\ and\ \citenamefont {Was}}]{Photos}%
  \BibitemOpen
  \bibfield  {author} {\bibinfo {author} {\bibfnamefont {E.}~\bibnamefont
  {Barberio}}, \bibinfo {author} {\bibfnamefont {B.}~\bibnamefont {van Eijk}},
  \ and\ \bibinfo {author} {\bibfnamefont {Z.}~\bibnamefont {Was}},\ }\href
  {\doibase 10.1016/0010-4655(91)90012-A} {\bibfield  {journal} {\bibinfo
  {journal} {Comput. Phys. Commun.}\ }\textbf {\bibinfo {volume} {66}},\
  \bibinfo {pages} {115} (\bibinfo {year} {1991})}\BibitemShut {NoStop}%
\bibitem [{\citenamefont {Glattauer}\ \emph {et~al.}(2016)\citenamefont
  {Glattauer}, \citenamefont {Schwanda}, \citenamefont {Abdesselam},
  \citenamefont {Adachi} \emph {et~al.}}]{glattauer_BGL_params}%
  \BibitemOpen
  \bibfield  {author} {\bibinfo {author} {\bibfnamefont {R.}~\bibnamefont
  {Glattauer}}, \bibinfo {author} {\bibfnamefont {C.}~\bibnamefont {Schwanda}},
  \bibinfo {author} {\bibfnamefont {A.}~\bibnamefont {Abdesselam}}, \bibinfo
  {author} {\bibfnamefont {I.}~\bibnamefont {Adachi}},  \emph {et~al.}
  (\bibinfo {collaboration} {Belle Collaboration}),\ }\href {\doibase
  10.1103/PhysRevD.93.032006} {\bibfield  {journal} {\bibinfo  {journal} {Phys.
  Rev. D}\ }\textbf {\bibinfo {volume} {93}},\ \bibinfo {pages} {032006}
  (\bibinfo {year} {2016})}\BibitemShut {NoStop}%
\bibitem [{\citenamefont {Ferlewicz}\ \emph {et~al.}(2021)\citenamefont
  {Ferlewicz}, \citenamefont {Urquijo},\ and\ \citenamefont
  {Waheed}}]{Ferlewicz:2020lxm}%
  \BibitemOpen
  \bibfield  {author} {\bibinfo {author} {\bibfnamefont {D.}~\bibnamefont
  {Ferlewicz}}, \bibinfo {author} {\bibfnamefont {P.}~\bibnamefont {Urquijo}},
  \ and\ \bibinfo {author} {\bibfnamefont {E.}~\bibnamefont {Waheed}},\ }\href
  {\doibase 10.1103/PhysRevD.103.073005} {\bibfield  {journal} {\bibinfo
  {journal} {Phys. Rev. D}\ }\textbf {\bibinfo {volume} {103}},\ \bibinfo
  {pages} {073005} (\bibinfo {year} {2021})},\ \Eprint
  {http://arxiv.org/abs/2008.09341} {arXiv:2008.09341 [hep-ph]} \BibitemShut
  {NoStop}%
\bibitem [{\citenamefont {Bernlochner}\ and\ \citenamefont
  {Ligeti}(2017)}]{Bernlochner:2016bci}%
  \BibitemOpen
  \bibfield  {author} {\bibinfo {author} {\bibfnamefont {F.~U.}\ \bibnamefont
  {Bernlochner}}\ and\ \bibinfo {author} {\bibfnamefont {Z.}~\bibnamefont
  {Ligeti}},\ }\href {\doibase 10.1103/PhysRevD.95.014022} {\bibfield
  {journal} {\bibinfo  {journal} {Phys. Rev. D}\ }\textbf {\bibinfo {volume}
  {95}},\ \bibinfo {pages} {014022} (\bibinfo {year} {2017})},\ \Eprint
  {http://arxiv.org/abs/1606.09300} {arXiv:1606.09300 [hep-ph]} \BibitemShut
  {NoStop}%
\bibitem [{\citenamefont {Nakano}\ \emph {et~al.}(2018)\citenamefont {Nakano}
  \emph {et~al.}}]{Belle:2018xst}%
  \BibitemOpen
  \bibfield  {author} {\bibinfo {author} {\bibfnamefont {H.}~\bibnamefont
  {Nakano}} \emph {et~al.} (\bibinfo {collaboration} {Belle}),\ }\href
  {\doibase 10.1103/PhysRevD.97.092003} {\bibfield  {journal} {\bibinfo
  {journal} {Phys. Rev. D}\ }\textbf {\bibinfo {volume} {97}},\ \bibinfo
  {pages} {092003} (\bibinfo {year} {2018})},\ \Eprint
  {http://arxiv.org/abs/1803.07774} {arXiv:1803.07774 [hep-ex]} \BibitemShut
  {NoStop}%
\bibitem [{\citenamefont {Krohn}\ \emph {et~al.}(2020)\citenamefont {Krohn}
  \emph {et~al.}}]{Belle-IIanalysissoftwareGroup:2019dlq}%
  \BibitemOpen
  \bibfield  {author} {\bibinfo {author} {\bibfnamefont {J.~F.}\ \bibnamefont
  {Krohn}} \emph {et~al.} (\bibinfo {collaboration} {Belle-II analysis software
  Group}),\ }\href {\doibase 10.1016/j.nima.2020.164269} {\bibfield  {journal}
  {\bibinfo  {journal} {Nucl. Instrum. Meth. A}\ }\textbf {\bibinfo {volume}
  {976}},\ \bibinfo {pages} {164269} (\bibinfo {year} {2020})},\ \Eprint
  {http://arxiv.org/abs/1901.11198} {arXiv:1901.11198 [hep-ex]} \BibitemShut
  {NoStop}%
\bibitem [{SFW()}]{SFW}%
  \BibitemOpen
  \href@noop {} {}\bibinfo {note} {{The Fox-Wolfram moments were introduced in
  G.~C.~Fox and S.~Wolfram, Phys. Rev. Lett. {\bf 41}, 1581 (1978). The
  modified Fox-Wolfram moments (SFW) used by Belle are described in K.~Abe {\it
  et al.} (Belle Collaboration), Phys. Rev. Lett. {\bf 87}, 101801 (2001) and
  K.~Abe {\it et al.} (Belle Collaboration), Phys. Lett. {\bf B 511}, 151
  (2001).}}\BibitemShut {Stop}%
\bibitem [{\citenamefont {Asner}\ \emph {et~al.}(1996)\citenamefont {Asner}
  \emph {et~al.}}]{cleocones}%
  \BibitemOpen
  \bibfield  {author} {\bibinfo {author} {\bibfnamefont {D.~M.}\ \bibnamefont
  {Asner}} \emph {et~al.} (\bibinfo {collaboration} {CLEO Collaboration}),\
  }\href {https://link.aps.org/doi/10.1103/PhysRevD.53.1039} {\bibfield
  {journal} {\bibinfo  {journal} {Phys. Rev.}\ }\textbf {\bibinfo {volume} {D
  53}},\ \bibinfo {pages} {1039} (\bibinfo {year} {1996})}\BibitemShut
  {NoStop}%
\bibitem [{\citenamefont {Keck}(2017)}]{Keck:2017gsv}%
  \BibitemOpen
  \bibfield  {author} {\bibinfo {author} {\bibfnamefont {T.}~\bibnamefont
  {Keck}},\ }\href {\doibase 10.1007/s41781-017-0002-8} {\bibfield  {journal}
  {\bibinfo  {journal} {Comput. Softw. Big Sci.}\ }\textbf {\bibinfo {volume}
  {1}},\ \bibinfo {pages} {2} (\bibinfo {year} {2017})}\BibitemShut {NoStop}%
\bibitem [{\citenamefont {James}\ and\ \citenamefont
  {Roos}(1975)}]{James:1975dr}%
  \BibitemOpen
  \bibfield  {author} {\bibinfo {author} {\bibfnamefont {F.}~\bibnamefont
  {James}}\ and\ \bibinfo {author} {\bibfnamefont {M.}~\bibnamefont {Roos}},\
  }\href {\doibase 10.1016/0010-4655(75)90039-9} {\bibfield  {journal}
  {\bibinfo  {journal} {Comput. Phys. Commun.}\ }\textbf {\bibinfo {volume}
  {10}},\ \bibinfo {pages} {343} (\bibinfo {year} {1975})}\BibitemShut
  {NoStop}%
\bibitem [{\citenamefont {Ongmongkolkul}\ \emph {et~al.}(12  )\citenamefont
  {Ongmongkolkul}, \citenamefont {Deil}, \citenamefont {Dembinski},
  \citenamefont {Dapid}, \citenamefont {Burr}, \citenamefont {Andrew},
  \citenamefont {Rost}, \citenamefont {Pearce}, \citenamefont {Geiger},\ and\
  \citenamefont {Zapata}}]{iminuit}%
  \BibitemOpen
  \bibfield  {author} {\bibinfo {author} {\bibfnamefont {P.}~\bibnamefont
  {Ongmongkolkul}}, \bibinfo {author} {\bibfnamefont {C.}~\bibnamefont {Deil}},
  \bibinfo {author} {\bibfnamefont {H.}~\bibnamefont {Dembinski}}, \bibinfo
  {author} {\bibnamefont {Dapid}}, \bibinfo {author} {\bibfnamefont
  {C.}~\bibnamefont {Burr}}, \bibinfo {author} {\bibnamefont {Andrew}},
  \bibinfo {author} {\bibfnamefont {F.}~\bibnamefont {Rost}}, \bibinfo {author}
  {\bibfnamefont {A.}~\bibnamefont {Pearce}}, \bibinfo {author} {\bibfnamefont
  {L.}~\bibnamefont {Geiger}}, \ and\ \bibinfo {author} {\bibfnamefont
  {O.}~\bibnamefont {Zapata}},\ }\href {https://github.com/iminuit/iminuit}
  {\enquote {\bibinfo {title} {iminuit - minuit from python},}\ } (\bibinfo
  {year} {2012--}),\ \bibinfo {note} {[Online; accessed
  2018.03.05]}\BibitemShut {NoStop}%
\bibitem [{\citenamefont {Baker}\ and\ \citenamefont
  {Cousins}(1984)}]{BAKER1984437}%
  \BibitemOpen
  \bibfield  {author} {\bibinfo {author} {\bibfnamefont {S.}~\bibnamefont
  {Baker}}\ and\ \bibinfo {author} {\bibfnamefont {R.~D.}\ \bibnamefont
  {Cousins}},\ }\href {\doibase https://doi.org/10.1016/0167-5087(84)90016-4}
  {\bibfield  {journal} {\bibinfo  {journal} {Nuclear Instruments and Methods
  in Physics Research}\ }\textbf {\bibinfo {volume} {221}},\ \bibinfo {pages}
  {437} (\bibinfo {year} {1984})}\BibitemShut {NoStop}%
\bibitem [{\citenamefont {Bernlochner}\ \emph {et~al.}(2019)\citenamefont
  {Bernlochner}, \citenamefont {Ligeti},\ and\ \citenamefont
  {Robinson}}]{Bernlochner:2019ldg}%
  \BibitemOpen
  \bibfield  {author} {\bibinfo {author} {\bibfnamefont {F.~U.}\ \bibnamefont
  {Bernlochner}}, \bibinfo {author} {\bibfnamefont {Z.}~\bibnamefont {Ligeti}},
  \ and\ \bibinfo {author} {\bibfnamefont {D.~J.}\ \bibnamefont {Robinson}},\
  }\href {\doibase 10.1103/PhysRevD.100.013005} {\bibfield  {journal} {\bibinfo
   {journal} {Phys. Rev. D}\ }\textbf {\bibinfo {volume} {100}},\ \bibinfo
  {pages} {013005} (\bibinfo {year} {2019})},\ \Eprint
  {http://arxiv.org/abs/1902.09553} {arXiv:1902.09553 [hep-ph]} \BibitemShut
  {NoStop}%
\bibitem [{\citenamefont {D'Agostini}(1994)}]{DAgostini:1993arp}%
  \BibitemOpen
  \bibfield  {author} {\bibinfo {author} {\bibfnamefont {G.}~\bibnamefont
  {D'Agostini}},\ }\href {\doibase 10.1016/0168-9002(94)90719-6} {\bibfield
  {journal} {\bibinfo  {journal} {Nucl. Instrum. Meth. A}\ }\textbf {\bibinfo
  {volume} {346}},\ \bibinfo {pages} {306} (\bibinfo {year}
  {1994})}\BibitemShut {NoStop}%
\bibitem [{\citenamefont {Bernlochner}\ \emph
  {et~al.}(2022{\natexlab{b}})\citenamefont {Bernlochner}, \citenamefont
  {Sevilla}, \citenamefont {Robinson},\ and\ \citenamefont
  {Wormser}}]{Bernlochner:2021vlv}%
  \BibitemOpen
  \bibfield  {author} {\bibinfo {author} {\bibfnamefont {F.~U.}\ \bibnamefont
  {Bernlochner}}, \bibinfo {author} {\bibfnamefont {M.~F.}\ \bibnamefont
  {Sevilla}}, \bibinfo {author} {\bibfnamefont {D.~J.}\ \bibnamefont
  {Robinson}}, \ and\ \bibinfo {author} {\bibfnamefont {G.}~\bibnamefont
  {Wormser}},\ }\href {\doibase 10.1103/RevModPhys.94.015003} {\bibfield
  {journal} {\bibinfo  {journal} {Rev. Mod. Phys.}\ }\textbf {\bibinfo {volume}
  {94}},\ \bibinfo {pages} {015003} (\bibinfo {year} {2022}{\natexlab{b}})},\
  \Eprint {http://arxiv.org/abs/2101.08326} {arXiv:2101.08326 [hep-ex]}
  \BibitemShut {NoStop}%
\bibitem [{\citenamefont {Amhis}\ \emph
  {et~al.}(2022{\natexlab{b}})\citenamefont {Amhis} \emph
  {et~al.}}]{HFLAV:2022pwe}%
  \BibitemOpen
  \bibfield  {author} {\bibinfo {author} {\bibfnamefont {Y.}~\bibnamefont
  {Amhis}} \emph {et~al.} (\bibinfo {collaboration} {HFLAV}),\ }\href@noop {}
  {\  (\bibinfo {year} {2022}{\natexlab{b}})},\ \Eprint
  {http://arxiv.org/abs/2206.07501} {arXiv:2206.07501 [hep-ex]} \BibitemShut
  {NoStop}%
\bibitem [{\citenamefont {Bernlochner}\ \emph
  {et~al.}(2022{\natexlab{c}})\citenamefont {Bernlochner}, \citenamefont
  {Ligeti}, \citenamefont {Papucci}, \citenamefont {Prim}, \citenamefont
  {Robinson},\ and\ \citenamefont {Xiong}}]{Bernlochner:2022ywh}%
  \BibitemOpen
  \bibfield  {author} {\bibinfo {author} {\bibfnamefont {F.~U.}\ \bibnamefont
  {Bernlochner}}, \bibinfo {author} {\bibfnamefont {Z.}~\bibnamefont {Ligeti}},
  \bibinfo {author} {\bibfnamefont {M.}~\bibnamefont {Papucci}}, \bibinfo
  {author} {\bibfnamefont {M.~T.}\ \bibnamefont {Prim}}, \bibinfo {author}
  {\bibfnamefont {D.~J.}\ \bibnamefont {Robinson}}, \ and\ \bibinfo {author}
  {\bibfnamefont {C.}~\bibnamefont {Xiong}},\ }\href {\doibase
  10.1103/PhysRevD.106.096015} {\bibfield  {journal} {\bibinfo  {journal}
  {Phys. Rev. D}\ }\textbf {\bibinfo {volume} {106}},\ \bibinfo {pages}
  {096015} (\bibinfo {year} {2022}{\natexlab{c}})},\ \Eprint
  {http://arxiv.org/abs/2206.11281} {arXiv:2206.11281 [hep-ph]} \BibitemShut
  {NoStop}%
\bibitem [{\citenamefont {Bobeth}\ \emph {et~al.}(2021)\citenamefont {Bobeth},
  \citenamefont {Bordone}, \citenamefont {Gubernari}, \citenamefont {Jung},\
  and\ \citenamefont {van Dyk}}]{Bobeth:2021lya}%
  \BibitemOpen
  \bibfield  {author} {\bibinfo {author} {\bibfnamefont {C.}~\bibnamefont
  {Bobeth}}, \bibinfo {author} {\bibfnamefont {M.}~\bibnamefont {Bordone}},
  \bibinfo {author} {\bibfnamefont {N.}~\bibnamefont {Gubernari}}, \bibinfo
  {author} {\bibfnamefont {M.}~\bibnamefont {Jung}}, \ and\ \bibinfo {author}
  {\bibfnamefont {D.}~\bibnamefont {van Dyk}},\ }\href {\doibase
  10.1140/epjc/s10052-021-09724-2} {\bibfield  {journal} {\bibinfo  {journal}
  {Eur. Phys. J. C}\ }\textbf {\bibinfo {volume} {81}},\ \bibinfo {pages} {984}
  (\bibinfo {year} {2021})},\ \Eprint {http://arxiv.org/abs/2104.02094}
  {arXiv:2104.02094 [hep-ph]} \BibitemShut {NoStop}%
\bibitem [{\citenamefont {Waheed}\ \emph
  {et~al.}(2019{\natexlab{b}})\citenamefont {Waheed} \emph
  {et~al.}}]{Waheed:2018djm}%
  \BibitemOpen
  \bibfield  {author} {\bibinfo {author} {\bibfnamefont {E.}~\bibnamefont
  {Waheed}} \emph {et~al.} (\bibinfo {collaboration} {Belle Collaboration}),\
  }\href {\doibase 10.1103/PhysRevD.100.052007} {\bibfield  {journal} {\bibinfo
   {journal} {Phys. Rev. D}\ }\textbf {\bibinfo {volume} {100}},\ \bibinfo
  {pages} {052007} (\bibinfo {year} {2019}{\natexlab{b}})},\ \Eprint
  {http://arxiv.org/abs/1809.03290} {arXiv:1809.03290 [hep-ex]} \BibitemShut
  {NoStop}%
\end{thebibliography}%

\clearpage

\appendix

\onecolumngrid

\section{Systematic Tables}
\label{app:systematics}
Tables~\ref{tab:systematics15}, \ref{tab:systematics16}, \ref{tab:systematics17}, and \ref{tab:systematics18}, provide the individual contributions of each uncertainty to the normalized shape of $\bar{B}^0 \to D^* e \bar{\nu}_e$, $\bar{B}^0 \to D^* \mu \bar {\nu}_\mu$, $B^- \to D^* e \bar{\nu}_e$, and $B^- \to D^* \mu \bar{\nu}_\mu$. The columns in the tables are the total uncertainty, the uncertainty from the $M_\mathrm{miss}^2$ fits, from the \bdslnu form factors, from the $D$ branching fractions, from the limited MC statistics, from the slow pion efficiency, from the lepton identification, on the $\pi^0$ efficiency, from the tracking efficiency, and from the $K_\mathrm{S}^0$ efficiency. 

\begin{table*}
    \centering
    \caption{Uncertainties in \% for the $\bar{B}^0 \to D^* e \bar{\nu}_e$ channel.}
    \label{tab:systematics15}
    \resizebox{0.9\textwidth}{!}{
    \begin{tabular}{llrrrrrrrrrrr}
\hline
\hline
       &              &  total & $M_\mathrm{miss}^2$ fit & FF$(B\to D^*\ell\bar{\nu}_\ell)$ & $\mathcal{B}(D\to X)$ & MC stat. & $\epsilon(\pi_\mathrm{slow})$ & $\epsilon(\mathrm{LID})$ & $\epsilon(\pi^0)$ & $\epsilon$(Tracking) & $\epsilon(K_S^0)$ & FEI Shape \\
Projection & Bin &        &                         &                       &                       &          &                               &                          &                   &                      &                   &           \\
\hline
$w$ & [1.00, 1.05) &  16.99 &                   16.12 &                  1.48 &                  1.02 &     4.90 &                          0.83 &                     0.32 &              0.19 &                 0.08 &              0.02 &      0.83 \\
       & [1.05, 1.10) &  15.84 &                   15.26 &                  0.63 &                  1.00 &     4.01 &                          0.65 &                     0.20 &              0.13 &                 0.07 &              0.01 &      0.48 \\
       & [1.10, 1.15) &  13.07 &                   12.61 &                  0.47 &                  0.39 &     3.33 &                          0.20 &                     0.15 &              0.10 &                 0.04 &              0.01 &      0.23 \\
       & [1.15, 1.20) &  10.36 &                   10.02 &                  0.52 &                  0.16 &     2.57 &                          0.12 &                     0.09 &              0.02 &                 0.01 &              0.02 &      0.24 \\
       & [1.20, 1.25) &   9.95 &                    9.59 &                  0.52 &                  0.17 &     2.56 &                          0.17 &                     0.04 &              0.01 &                 0.01 &              0.00 &      0.28 \\
       & [1.25, 1.30) &   9.32 &                    9.00 &                  0.59 &                  0.22 &     2.33 &                          0.17 &                     0.05 &              0.04 &                 0.03 &              0.01 &      0.22 \\
       & [1.30, 1.35) &   9.79 &                    9.43 &                  0.41 &                  0.39 &     2.49 &                          0.24 &                     0.10 &              0.07 &                 0.02 &              0.01 &      0.43 \\
       & [1.35, 1.40) &  10.31 &                   10.01 &                  0.23 &                  0.44 &     2.37 &                          0.26 &                     0.18 &              0.08 &                 0.04 &              0.01 &      0.43 \\
       & [1.40, 1.45) &   9.55 &                    9.27 &                  0.61 &                  0.39 &     2.16 &                          0.29 &                     0.21 &              0.10 &                 0.03 &              0.01 &      0.09 \\
       & [1.45, 1.50) &  10.87 &                   10.56 &                  1.43 &                  0.60 &     1.99 &                          0.34 &                     0.25 &              0.08 &                 0.04 &              0.02 &      0.04 \\
\hline
$\cos \theta_\ell$ & [-1.00, -0.80) &  23.89 &                   23.34 &                  2.19 &                  0.24 &     4.43 &                          0.16 &                     0.89 &              0.04 &                 0.01 &              0.01 &      0.68 \\
       & [-0.80, -0.60) &  15.03 &                   14.57 &                  0.58 &                  0.15 &     3.55 &                          0.10 &                     0.81 &              0.05 &                 0.01 &              0.00 &      0.26 \\
       & [-0.60, -0.40) &  16.55 &                   16.11 &                  0.40 &                  0.11 &     3.65 &                          0.08 &                     0.80 &              0.02 &                 0.00 &              0.01 &      0.49 \\
       & [-0.40, -0.20) &  13.00 &                   12.56 &                  0.30 &                  0.09 &     3.30 &                          0.05 &                     0.47 &              0.04 &                 0.00 &              0.00 &      0.12 \\
       & [-0.20, 0.00) &  12.95 &                   12.52 &                  0.35 &                  0.13 &     3.13 &                          0.10 &                     0.16 &              0.00 &                 0.01 &              0.01 &      0.95 \\
       & [0.00, 0.20) &  17.23 &                   16.63 &                  0.45 &                  0.13 &     4.32 &                          0.08 &                     0.34 &              0.01 &                 0.02 &              0.01 &      1.13 \\
       & [0.20, 0.40) &  10.94 &                   10.53 &                  0.41 &                  0.13 &     2.92 &                          0.03 &                     0.32 &              0.05 &                 0.01 &              0.00 &      0.36 \\
       & [0.40, 0.60) &  11.67 &                   11.19 &                  0.32 &                  0.06 &     3.26 &                          0.07 &                     0.37 &              0.01 &                 0.01 &              0.01 &      0.32 \\
       & [0.60, 0.80) &  10.30 &                   10.01 &                  0.38 &                  0.10 &     2.35 &                          0.05 &                     0.33 &              0.06 &                 0.00 &              0.01 &      0.30 \\
       & [0.80, 1.00) &   7.85 &                    7.56 &                  1.01 &                  0.06 &     1.82 &                          0.05 &                     0.34 &              0.01 &                 0.00 &              0.00 &      0.03 \\
\hline
$\cos \theta_V$ & [-1.00, -0.80) &   6.63 &                    6.40 &                  0.41 &                  0.50 &     1.54 &                          0.34 &                     0.12 &              0.09 &                 0.04 &              0.00 &      0.03 \\
       & [-0.80, -0.60) &   8.17 &                    7.84 &                  0.74 &                  0.39 &     2.09 &                          0.28 &                     0.06 &              0.05 &                 0.04 &              0.00 &      0.29 \\
       & [-0.60, -0.40) &  11.32 &                   10.91 &                  0.69 &                  0.48 &     2.86 &                          0.28 &                     0.04 &              0.08 &                 0.03 &              0.00 &      0.08 \\
       & [-0.40, -0.20) &  12.96 &                   12.50 &                  0.47 &                  0.31 &     3.40 &                          0.25 &                     0.02 &              0.03 &                 0.03 &              0.02 &      0.04 \\
       & [-0.20, 0.00) &  14.86 &                   14.38 &                  1.14 &                  0.25 &     3.54 &                          0.16 &                     0.17 &              0.09 &                 0.02 &              0.00 &      0.23 \\
       & [0.00, 0.20) &  21.80 &                   21.10 &                  1.15 &                  0.18 &     5.34 &                          0.21 &                     0.08 &              0.04 &                 0.02 &              0.01 &      0.19 \\
       & [0.20, 0.40) &  17.12 &                   16.69 &                  0.51 &                  0.28 &     3.74 &                          0.17 &                     0.15 &              0.05 &                 0.00 &              0.02 &      0.35 \\
       & [0.40, 0.60) &  16.98 &                   16.51 &                  0.80 &                  0.17 &     3.84 &                          0.24 &                     0.01 &              0.02 &                 0.03 &              0.01 &      0.20 \\
       & [0.60, 0.80) &  26.00 &                   25.54 &                  0.38 &                  0.57 &     4.75 &                          0.45 &                     0.06 &              0.08 &                 0.05 &              0.01 &      0.26 \\
       & [0.80, 1.00) &  13.61 &                   13.22 &                  0.32 &                  0.93 &     3.04 &                          0.60 &                     0.13 &              0.20 &                 0.06 &              0.01 &      0.07 \\
\hline
$\chi$ & [0.00, 0.63) &  15.32 &                   14.90 &                  0.34 &                  0.22 &     3.50 &                          0.11 &                     0.08 &              0.02 &                 0.00 &              0.01 &      0.19 \\
       & [0.63, 1.26) &  15.17 &                   14.76 &                  0.27 &                  0.22 &     3.43 &                          0.09 &                     0.01 &              0.00 &                 0.01 &              0.01 &      0.43 \\
       & [1.26, 1.88) &  12.56 &                   12.24 &                  0.40 &                  0.15 &     2.79 &                          0.05 &                     0.04 &              0.01 &                 0.01 &              0.01 &      0.22 \\
       & [1.88, 2.51) &  10.47 &                   10.17 &                  0.18 &                  0.10 &     2.41 &                          0.06 &                     0.01 &              0.02 &                 0.00 &              0.01 &      0.61 \\
       & [2.51, 3.14) &  16.26 &                   15.74 &                  0.55 &                  0.21 &     3.99 &                          0.06 &                     0.05 &              0.06 &                 0.01 &              0.01 &      0.59 \\
       & [3.14, 3.77) &  11.39 &                   11.01 &                  0.58 &                  0.15 &     2.83 &                          0.06 &                     0.09 &              0.01 &                 0.03 &              0.01 &      0.19 \\
       & [3.77, 4.40) &  11.70 &                   11.26 &                  0.17 &                  0.04 &     3.18 &                          0.09 &                     0.01 &              0.01 &                 0.01 &              0.00 &      0.03 \\
       & [4.40, 5.03) &  11.66 &                   11.26 &                  0.34 &                  0.09 &     3.00 &                          0.07 &                     0.01 &              0.03 &                 0.00 &              0.00 &      0.33 \\
       & [5.03, 5.65) &  12.24 &                   11.83 &                  0.29 &                  0.10 &     3.11 &                          0.06 &                     0.04 &              0.01 &                 0.00 &              0.00 &      0.03 \\
       & [5.65, 6.28) &  13.85 &                   13.45 &                  0.31 &                  0.09 &     3.30 &                          0.09 &                     0.05 &              0.01 &                 0.02 &              0.00 &      0.21 \\
\hline
\hline
\end{tabular}

    }
\end{table*}

\begin{table*}
    \centering
    \caption{Uncertainties in \% for the $\bar{B}^0 \to D^* \mu \bar{\nu}_\mu$ channel.}
    \label{tab:systematics16}
    \resizebox{0.9\textwidth}{!}{
    \begin{tabular}{llrrrrrrrrrrr}
\hline
\hline
       &              &  total & $M_\mathrm{miss}^2$ fit & FF$(B\to D^*\ell\bar{\nu}_\ell)$ & $\mathcal{B}(D\to X)$ & MC stat. & $\epsilon(\pi_\mathrm{slow})$ & $\epsilon(\mathrm{LID})$ & $\epsilon(\pi^0)$ & $\epsilon$(Tracking) & $\epsilon(K_S^0)$ & FEI Shape \\
Projection & Bin &        &                         &                       &                       &          &                               &                          &                   &                      &                   &           \\
\hline
$w$ & [1.00, 1.05) &  17.71 &                   16.97 &                  1.52 &                  0.95 &     4.67 &                          0.64 &                     0.19 &              0.22 &                 0.08 &              0.02 &      0.38 \\
       & [1.05, 1.10) &  13.31 &                   12.68 &                  0.61 &                  0.99 &     3.82 &                          0.64 &                     0.09 &              0.21 &                 0.08 &              0.01 &      0.33 \\
       & [1.10, 1.15) &  15.02 &                   14.64 &                  0.52 &                  0.49 &     3.18 &                          0.32 &                     0.18 &              0.09 &                 0.05 &              0.01 &      0.70 \\
       & [1.15, 1.20) &  10.25 &                    9.88 &                  0.56 &                  0.10 &     2.68 &                          0.14 &                     0.05 &              0.00 &                 0.01 &              0.00 &      0.19 \\
       & [1.20, 1.25) &   9.34 &                    8.88 &                  0.61 &                  0.17 &     2.83 &                          0.13 &                     0.00 &              0.04 &                 0.02 &              0.01 &      0.14 \\
       & [1.25, 1.30) &   9.78 &                    9.40 &                  0.60 &                  0.26 &     2.62 &                          0.16 &                     0.01 &              0.05 &                 0.02 &              0.01 &      0.10 \\
       & [1.30, 1.35) &   8.93 &                    8.52 &                  0.38 &                  0.33 &     2.60 &                          0.24 &                     0.06 &              0.05 &                 0.03 &              0.01 &      0.15 \\
       & [1.35, 1.40) &  10.55 &                   10.19 &                  0.28 &                  0.38 &     2.66 &                          0.26 &                     0.13 &              0.07 &                 0.02 &              0.01 &      0.32 \\
       & [1.40, 1.45) &  10.55 &                   10.20 &                  0.61 &                  0.51 &     2.48 &                          0.29 &                     0.11 &              0.10 &                 0.04 &              0.00 &      0.60 \\
       & [1.45, 1.50) &  10.18 &                    9.81 &                  1.30 &                  0.49 &     2.32 &                          0.34 &                     0.14 &              0.11 &                 0.03 &              0.01 &      0.09 \\
\hline
$\cos \theta_\ell$ & [-1.00, -0.80) &  24.01 &                   23.32 &                  2.17 &                  0.45 &     5.21 &                          0.37 &                     0.05 &              0.12 &                 0.02 &              0.01 &      0.72 \\
       & [-0.80, -0.60) &  27.29 &                   26.57 &                  0.64 &                  0.37 &     6.18 &                          0.16 &                     0.08 &              0.08 &                 0.03 &              0.01 &      0.01 \\
       & [-0.60, -0.40) &  18.03 &                   17.53 &                  0.38 &                  0.22 &     4.21 &                          0.08 &                     0.05 &              0.05 &                 0.01 &              0.01 &      0.34 \\
       & [-0.40, -0.20) &  14.09 &                   13.74 &                  0.34 &                  0.15 &     3.07 &                          0.11 &                     0.05 &              0.04 &                 0.02 &              0.01 &      0.22 \\
       & [-0.20, 0.00) &  12.75 &                   12.41 &                  0.37 &                  0.14 &     2.89 &                          0.07 &                     0.08 &              0.01 &                 0.02 &              0.00 &      0.02 \\
       & [0.00, 0.20) &  13.24 &                   12.87 &                  0.48 &                  0.12 &     3.03 &                          0.04 &                     0.02 &              0.01 &                 0.01 &              0.01 &      0.04 \\
       & [0.20, 0.40) &  11.93 &                   11.43 &                  0.43 &                  0.08 &     3.37 &                          0.06 &                     0.04 &              0.04 &                 0.00 &              0.01 &      0.15 \\
       & [0.40, 0.60) &  10.07 &                    9.63 &                  0.29 &                  0.20 &     2.92 &                          0.04 &                     0.03 &              0.08 &                 0.01 &              0.01 &      0.04 \\
       & [0.60, 0.80) &   9.87 &                    9.46 &                  0.35 &                  0.21 &     2.76 &                          0.05 &                     0.03 &              0.00 &                 0.01 &              0.00 &      0.29 \\
       & [0.80, 1.00) &   7.47 &                    7.10 &                  1.01 &                  0.14 &     2.07 &                          0.06 &                     0.13 &              0.02 &                 0.00 &              0.00 &      0.18 \\
\hline
$\cos \theta_V$ & [-1.00, -0.80) &   6.69 &                    6.39 &                  0.44 &                  0.49 &     1.83 &                          0.34 &                     0.04 &              0.10 &                 0.04 &              0.01 &      0.19 \\
       & [-0.80, -0.60) &   9.31 &                    8.98 &                  0.72 &                  0.48 &     2.25 &                          0.30 &                     0.02 &              0.08 &                 0.03 &              0.00 &      0.15 \\
       & [-0.60, -0.40) &  10.17 &                    9.74 &                  0.62 &                  0.51 &     2.79 &                          0.29 &                     0.06 &              0.11 &                 0.02 &              0.00 &      0.20 \\
       & [-0.40, -0.20) &  16.39 &                   15.84 &                  0.45 &                  0.25 &     4.18 &                          0.24 &                     0.01 &              0.01 &                 0.03 &              0.02 &      0.17 \\
       & [-0.20, 0.00) &  13.92 &                   13.30 &                  1.12 &                  0.30 &     3.94 &                          0.26 &                     0.04 &              0.06 &                 0.03 &              0.01 &      0.24 \\
       & [0.00, 0.20) &  20.69 &                   19.89 &                  1.07 &                  0.21 &     5.59 &                          0.19 &                     0.07 &              0.03 &                 0.02 &              0.01 &      0.42 \\
       & [0.20, 0.40) &  20.55 &                   19.91 &                  0.47 &                  0.17 &     5.02 &                          0.20 &                     0.05 &              0.07 &                 0.02 &              0.02 &      0.47 \\
       & [0.40, 0.60) &  18.12 &                   17.58 &                  0.74 &                  0.22 &     4.19 &                          0.19 &                     0.12 &              0.06 &                 0.01 &              0.03 &      0.98 \\
       & [0.60, 0.80) &  16.07 &                   15.52 &                  0.53 &                  0.38 &     4.08 &                          0.38 &                     0.01 &              0.13 &                 0.03 &              0.01 &      0.49 \\
       & [0.80, 1.00) &  13.62 &                   13.17 &                  0.24 &                  0.93 &     3.30 &                          0.56 &                     0.02 &              0.14 &                 0.06 &              0.01 &      0.16 \\
\hline
$\chi$ & [0.00, 0.63) &  15.80 &                   15.42 &                  0.26 &                  0.15 &     3.41 &                          0.19 &                     0.12 &              0.02 &                 0.02 &              0.01 &      0.11 \\
       & [0.63, 1.26) &  15.16 &                   14.66 &                  0.27 &                  0.19 &     3.86 &                          0.10 &                     0.01 &              0.08 &                 0.01 &              0.01 &      0.18 \\
       & [1.26, 1.88) &  10.69 &                   10.44 &                  0.35 &                  0.14 &     2.29 &                          0.06 &                     0.01 &              0.02 &                 0.02 &              0.01 &      0.23 \\
       & [1.88, 2.51) &  13.18 &                   12.80 &                  0.19 &                  0.17 &     3.14 &                          0.06 &                     0.05 &              0.01 &                 0.01 &              0.01 &      0.07 \\
       & [2.51, 3.14) &  14.20 &                   13.68 &                  0.62 &                  0.24 &     3.70 &                          0.08 &                     0.05 &              0.03 &                 0.00 &              0.00 &      0.50 \\
       & [3.14, 3.77) &  11.60 &                   11.17 &                  0.58 &                  0.11 &     3.05 &                          0.10 &                     0.03 &              0.07 &                 0.00 &              0.01 &      0.01 \\
       & [3.77, 4.40) &  11.97 &                   11.61 &                  0.16 &                  0.10 &     2.88 &                          0.08 &                     0.06 &              0.08 &                 0.01 &              0.01 &      0.24 \\
       & [4.40, 5.03) &  10.45 &                   10.04 &                  0.37 &                  0.08 &     2.87 &                          0.04 &                     0.03 &              0.02 &                 0.00 &              0.01 &      0.02 \\
       & [5.03, 5.65) &  12.64 &                   12.12 &                  0.30 &                  0.11 &     3.58 &                          0.07 &                     0.02 &              0.03 &                 0.02 &              0.01 &      0.10 \\
       & [5.65, 6.28) &  14.51 &                   13.95 &                  0.33 &                  0.15 &     3.97 &                          0.09 &                     0.11 &              0.03 &                 0.03 &              0.01 &      0.25 \\
\hline
\hline
\end{tabular}

    }
\end{table*}

\begin{table*}
    \centering
    \caption{Uncertainties in \% for the $B^- \to D^* e \bar{\nu}_e$ channel.}
    \label{tab:systematics17}
    \resizebox{0.9\textwidth}{!}{
    \begin{tabular}{llrrrrrrrrrrr}
\hline
\hline
       &              &  total & $M_\mathrm{miss}^2$ fit & FF$(B\to D^*\ell\bar{\nu}_\ell)$ & $\mathcal{B}(D\to X)$ & MC stat. & $\epsilon(\pi_\mathrm{slow})$ & $\epsilon(\mathrm{LID})$ & $\epsilon(\pi^0)$ & $\epsilon$(Tracking) & $\epsilon(K_S^0)$ & FEI Shape \\
Projection & Bin &        &                         &                       &                       &          &                               &                          &                   &                      &                   &           \\
\hline
$w$ & [1.00, 1.05) &   7.69 &                    7.29 &                  1.51 &                  0.13 &     1.81 &                          0.44 &                     0.30 &              0.01 &                 0.01 &              0.01 &      0.39 \\
       & [1.05, 1.10) &   7.22 &                    6.91 &                  0.66 &                  0.06 &     1.92 &                          0.35 &                     0.25 &              0.03 &                 0.01 &              0.01 &      0.12 \\
       & [1.10, 1.15) &   7.34 &                    7.12 &                  0.33 &                  0.10 &     1.73 &                          0.24 &                     0.15 &              0.00 &                 0.00 &              0.00 &      0.02 \\
       & [1.15, 1.20) &   7.07 &                    6.78 &                  0.56 &                  0.08 &     1.89 &                          0.11 &                     0.12 &              0.02 &                 0.00 &              0.01 &      0.05 \\
       & [1.20, 1.25) &   9.17 &                    8.90 &                  0.55 &                  0.12 &     2.13 &                          0.09 &                     0.02 &              0.02 &                 0.00 &              0.00 &      0.12 \\
       & [1.25, 1.30) &   7.88 &                    7.56 &                  0.63 &                  0.08 &     2.13 &                          0.15 &                     0.03 &              0.01 &                 0.01 &              0.01 &      0.19 \\
       & [1.30, 1.35) &   9.37 &                    9.06 &                  0.36 &                  0.03 &     2.36 &                          0.18 &                     0.06 &              0.00 &                 0.01 &              0.00 &      0.06 \\
       & [1.35, 1.40) &  10.94 &                   10.67 &                  0.23 &                  0.07 &     2.39 &                          0.23 &                     0.11 &              0.01 &                 0.00 &              0.01 &      0.11 \\
       & [1.40, 1.45) &   9.88 &                    9.66 &                  0.65 &                  0.13 &     1.94 &                          0.23 &                     0.26 &              0.00 &                 0.01 &              0.00 &      0.25 \\
       & [1.45, 1.51) &  10.63 &                   10.26 &                  1.37 &                  0.08 &     2.36 &                          0.28 &                     0.34 &              0.05 &                 0.01 &              0.01 &      0.29 \\
\hline
$\cos \theta_\ell$ & [-1.00, -0.80) &  13.70 &                   13.13 &                  2.28 &                  0.15 &     3.08 &                          0.10 &                     0.62 &              0.01 &                 0.02 &              0.00 &      0.37 \\
       & [-0.80, -0.60) &  11.33 &                   11.05 &                  0.57 &                  0.06 &     2.31 &                          0.08 &                     0.76 &              0.02 &                 0.00 &              0.01 &      0.05 \\
       & [-0.60, -0.40) &   9.95 &                    9.69 &                  0.25 &                  0.12 &     2.16 &                          0.09 &                     0.61 &              0.04 &                 0.00 &              0.00 &      0.07 \\
       & [-0.40, -0.20) &  12.57 &                   12.20 &                  0.21 &                  0.18 &     3.00 &                          0.09 &                     0.40 &              0.01 &                 0.00 &              0.01 &      0.15 \\
       & [-0.20, 0.00) &   9.66 &                    9.42 &                  0.32 &                  0.14 &     2.11 &                          0.04 &                     0.03 &              0.00 &                 0.01 &              0.01 &      0.19 \\
       & [0.00, 0.20) &  10.59 &                   10.28 &                  0.38 &                  0.14 &     2.51 &                          0.06 &                     0.25 &              0.03 &                 0.01 &              0.00 &      0.31 \\
       & [0.20, 0.40) &   9.09 &                    8.78 &                  0.33 &                  0.21 &     2.32 &                          0.05 &                     0.27 &              0.06 &                 0.01 &              0.01 &      0.05 \\
       & [0.40, 0.60) &   8.54 &                    8.22 &                  0.36 &                  0.13 &     2.26 &                          0.04 &                     0.25 &              0.04 &                 0.01 &              0.01 &      0.00 \\
       & [0.60, 0.80) &   7.86 &                    7.58 &                  0.27 &                  0.05 &     2.02 &                          0.07 &                     0.27 &              0.00 &                 0.01 &              0.01 &      0.23 \\
       & [0.80, 1.00) &   6.34 &                    6.10 &                  0.64 &                  0.07 &     1.62 &                          0.06 &                     0.22 &              0.01 &                 0.01 &              0.01 &      0.10 \\
\hline
$\cos \theta_V$ & [-1.00, -0.80) &   7.66 &                    7.43 &                  0.36 &                  0.07 &     1.77 &                          0.44 &                     0.09 &              0.03 &                 0.00 &              0.00 &      0.05 \\
       & [-0.80, -0.60) &  10.63 &                   10.28 &                  0.46 &                  0.16 &     2.60 &                          0.42 &                     0.05 &              0.02 &                 0.01 &              0.00 &      0.11 \\
       & [-0.60, -0.40) &  14.04 &                   13.57 &                  0.72 &                  0.14 &     3.51 &                          0.36 &                     0.05 &              0.01 &                 0.01 &              0.01 &      0.16 \\
       & [-0.40, -0.20) &  18.57 &                   17.91 &                  0.23 &                  0.19 &     4.89 &                          0.46 &                     0.03 &              0.05 &                 0.01 &              0.01 &      0.15 \\
       & [-0.20, 0.00) &  17.95 &                   17.36 &                  0.70 &                  0.29 &     4.52 &                          0.22 &                     0.14 &              0.01 &                 0.01 &              0.01 &      0.13 \\
       & [0.00, 0.20) &  20.03 &                   19.42 &                  0.65 &                  0.43 &     4.83 &                          0.29 &                     0.22 &              0.07 &                 0.02 &              0.01 &      0.13 \\
       & [0.20, 0.40) &  19.40 &                   18.76 &                  0.19 &                  0.50 &     4.87 &                          0.24 &                     0.02 &              0.09 &                 0.02 &              0.01 &      0.33 \\
       & [0.40, 0.60) &  17.74 &                   17.08 &                  0.65 &                  0.07 &     4.74 &                          0.34 &                     0.04 &              0.01 &                 0.00 &              0.00 &      0.33 \\
       & [0.60, 0.80) &  12.43 &                   12.12 &                  0.39 &                  0.13 &     2.69 &                          0.48 &                     0.03 &              0.06 &                 0.01 &              0.01 &      0.21 \\
       & [0.80, 1.00) &   9.88 &                    9.53 &                  0.34 &                  0.09 &     2.41 &                          0.91 &                     0.14 &              0.04 &                 0.02 &              0.00 &      0.07 \\
\hline
$\chi$ & [0.00, 0.63) &  13.29 &                   12.87 &                  0.17 &                  0.18 &     3.32 &                          0.10 &                     0.03 &              0.02 &                 0.03 &              0.00 &      0.02 \\
       & [0.63, 1.26) &  15.44 &                   14.99 &                  0.23 &                  0.18 &     3.68 &                          0.10 &                     0.07 &              0.00 &                 0.03 &              0.00 &      0.45 \\
       & [1.26, 1.88) &  10.95 &                   10.63 &                  0.12 &                  0.08 &     2.63 &                          0.07 &                     0.06 &              0.02 &                 0.00 &              0.00 &      0.35 \\
       & [1.88, 2.51) &  15.45 &                   15.01 &                  0.30 &                  0.11 &     3.63 &                          0.10 &                     0.03 &              0.04 &                 0.00 &              0.01 &      0.32 \\
       & [2.51, 3.14) &  14.06 &                   13.65 &                  0.20 &                  0.15 &     3.38 &                          0.12 &                     0.03 &              0.04 &                 0.01 &              0.01 &      0.28 \\
       & [3.14, 3.77) &  16.16 &                   15.63 &                  0.16 &                  0.12 &     4.08 &                          0.08 &                     0.01 &              0.06 &                 0.01 &              0.00 &      0.41 \\
       & [3.77, 4.40) &  16.04 &                   15.55 &                  0.26 &                  0.16 &     3.93 &                          0.09 &                     0.03 &              0.00 &                 0.01 &              0.01 &      0.15 \\
       & [4.40, 5.03) &   9.57 &                    9.24 &                  0.13 &                  0.08 &     2.48 &                          0.07 &                     0.01 &              0.01 &                 0.00 &              0.00 &      0.33 \\
       & [5.03, 5.65) &  14.58 &                   14.17 &                  0.23 &                  0.08 &     3.42 &                          0.09 &                     0.01 &              0.03 &                 0.00 &              0.01 &      0.11 \\
       & [5.65, 6.28) &  15.59 &                   15.05 &                  0.17 &                  0.22 &     4.03 &                          0.10 &                     0.01 &              0.07 &                 0.02 &              0.00 &      0.07 \\
\hline
\hline
\end{tabular}

    }
\end{table*}

\begin{table*}
    \centering
    \caption{Uncertainties in \% for the $B^- \to D^* \mu \bar{\nu}_\mu$ channel.}
    \label{tab:systematics18}    
    \resizebox{0.9\textwidth}{!}{
    \begin{tabular}{llrrrrrrrrrrr}
\hline
\hline
       &              &  total & $M_\mathrm{miss}^2$ fit & FF$(B\to D^*\ell\bar{\nu}_\ell)$ & $\mathcal{B}(D\to X)$ & MC stat. & $\epsilon(\pi_\mathrm{slow})$ & $\epsilon(\mathrm{LID})$ & $\epsilon(\pi^0)$ & $\epsilon$(Tracking) & $\epsilon(K_S^0)$ & FEI Shape \\
Projection & Bin &        &                         &                       &                       &          &                               &                          &                   &                      &                   &           \\
\hline
$w$ & [1.00, 1.05) &   7.54 &                    7.04 &                  1.52 &                  0.05 &     2.13 &                          0.40 &                     0.07 &              0.01 &                 0.00 &              0.01 &      0.43 \\
       & [1.05, 1.10) &   7.03 &                    6.70 &                  0.69 &                  0.07 &     1.96 &                          0.32 &                     0.06 &              0.03 &                 0.01 &              0.01 &      0.17 \\
       & [1.10, 1.15) &   5.82 &                    5.55 &                  0.32 &                  0.11 &     1.69 &                          0.21 &                     0.05 &              0.01 &                 0.00 &              0.01 &      0.00 \\
       & [1.15, 1.20) &   6.56 &                    6.24 &                  0.59 &                  0.10 &     1.92 &                          0.09 &                     0.04 &              0.01 &                 0.01 &              0.01 &      0.13 \\
       & [1.20, 1.25) &   7.47 &                    7.16 &                  0.53 &                  0.05 &     2.04 &                          0.10 &                     0.03 &              0.02 &                 0.00 &              0.01 &      0.13 \\
       & [1.25, 1.30) &   8.09 &                    7.79 &                  0.60 &                  0.08 &     2.07 &                          0.12 &                     0.02 &              0.00 &                 0.00 &              0.01 &      0.15 \\
       & [1.30, 1.35) &   9.69 &                    9.29 &                  0.35 &                  0.08 &     2.74 &                          0.14 &                     0.01 &              0.01 &                 0.01 &              0.00 &      0.10 \\
       & [1.35, 1.40) &  10.62 &                   10.25 &                  0.23 &                  0.07 &     2.75 &                          0.21 &                     0.05 &              0.01 &                 0.01 &              0.01 &      0.18 \\
       & [1.40, 1.45) &  11.19 &                   10.84 &                  0.61 &                  0.07 &     2.68 &                          0.25 &                     0.13 &              0.03 &                 0.01 &              0.01 &      0.16 \\
       & [1.45, 1.51) &  14.00 &                   13.69 &                  1.28 &                  0.15 &     2.63 &                          0.28 &                     0.10 &              0.03 &                 0.01 &              0.02 &      0.21 \\
\hline
$\cos \theta_\ell$ & [-1.00, -0.80) &  15.29 &                   14.74 &                  2.48 &                  0.32 &     3.19 &                          0.15 &                     0.11 &              0.06 &                 0.02 &              0.01 &      0.13 \\
       & [-0.80, -0.60) &  12.43 &                   12.03 &                  1.08 &                  0.09 &     2.96 &                          0.09 &                     0.04 &              0.04 &                 0.00 &              0.00 &      0.13 \\
       & [-0.60, -0.40) &  10.89 &                   10.55 &                  0.34 &                  0.05 &     2.66 &                          0.13 &                     0.09 &              0.01 &                 0.02 &              0.01 &      0.26 \\
       & [-0.40, -0.20) &   9.55 &                    9.25 &                  0.27 &                  0.07 &     2.37 &                          0.03 &                     0.02 &              0.04 &                 0.01 &              0.01 &      0.09 \\
       & [-0.20, 0.00) &  12.17 &                   11.84 &                  0.34 &                  0.17 &     2.76 &                          0.05 &                     0.03 &              0.05 &                 0.03 &              0.00 &      0.24 \\
       & [0.00, 0.20) &   8.63 &                    8.35 &                  0.42 &                  0.06 &     2.12 &                          0.07 &                     0.02 &              0.01 &                 0.02 &              0.00 &      0.28 \\
       & [0.20, 0.40) &   8.03 &                    7.75 &                  0.40 &                  0.05 &     2.06 &                          0.05 &                     0.03 &              0.03 &                 0.01 &              0.01 &      0.08 \\
       & [0.40, 0.60) &   8.86 &                    8.50 &                  0.42 &                  0.08 &     2.47 &                          0.05 &                     0.01 &              0.00 &                 0.00 &              0.01 &      0.12 \\
       & [0.60, 0.80) &   7.32 &                    6.97 &                  0.30 &                  0.05 &     2.23 &                          0.08 &                     0.02 &              0.00 &                 0.00 &              0.01 &      0.13 \\
       & [0.80, 1.00) &   5.76 &                    5.54 &                  0.59 &                  0.12 &     1.44 &                          0.06 &                     0.08 &              0.01 &                 0.02 &              0.01 &      0.11 \\
\hline
$\cos \theta_V$ & [-1.00, -0.80) &   7.33 &                    6.98 &                  0.40 &                  0.07 &     2.18 &                          0.43 &                     0.02 &              0.03 &                 0.00 &              0.00 &      0.05 \\
       & [-0.80, -0.60) &  11.91 &                   11.50 &                  0.39 &                  0.33 &     3.04 &                          0.40 &                     0.10 &              0.03 &                 0.02 &              0.01 &      0.01 \\
       & [-0.60, -0.40) &  13.85 &                   13.33 &                  0.60 &                  0.20 &     3.68 &                          0.33 &                     0.05 &              0.06 &                 0.00 &              0.01 &      0.05 \\
       & [-0.40, -0.20) &  14.21 &                   13.61 &                  0.28 &                  0.21 &     4.04 &                          0.33 &                     0.02 &              0.03 &                 0.02 &              0.01 &      0.18 \\
       & [-0.20, 0.00) &  17.32 &                   16.50 &                  0.65 &                  0.14 &     5.23 &                          0.29 &                     0.02 &              0.00 &                 0.03 &              0.01 &      0.33 \\
       & [0.00, 0.20) &  22.95 &                   22.24 &                  0.57 &                  0.26 &     5.60 &                          0.29 &                     0.03 &              0.05 &                 0.00 &              0.01 &      0.34 \\
       & [0.20, 0.40) &  16.51 &                   16.03 &                  0.18 &                  0.13 &     3.90 &                          0.32 &                     0.06 &              0.01 &                 0.00 &              0.00 &      0.51 \\
       & [0.40, 0.60) &  17.25 &                   16.71 &                  0.70 &                  0.15 &     4.18 &                          0.36 &                     0.02 &              0.02 &                 0.01 &              0.01 &      0.17 \\
       & [0.60, 0.80) &  12.47 &                   12.17 &                  0.45 &                  0.30 &     2.61 &                          0.48 &                     0.05 &              0.04 &                 0.01 &              0.01 &      0.23 \\
       & [0.80, 1.00) &   9.26 &                    8.89 &                  0.35 &                  0.11 &     2.40 &                          0.83 &                     0.03 &              0.01 &                 0.01 &              0.01 &      0.00 \\
\hline
$\chi$ & [0.00, 0.63) &  19.29 &                   18.65 &                  0.24 &                  0.20 &     4.80 &                          0.29 &                     0.01 &              0.14 &                 0.03 &              0.01 &      0.91 \\
       & [0.63, 1.26) &  14.22 &                   13.78 &                  0.24 &                  0.07 &     3.52 &                          0.13 &                     0.03 &              0.05 &                 0.00 &              0.00 &      0.08 \\
       & [1.26, 1.88) &  10.98 &                   10.60 &                  0.14 &                  0.10 &     2.84 &                          0.05 &                     0.04 &              0.00 &                 0.01 &              0.01 &      0.14 \\
       & [1.88, 2.51) &  12.27 &                   11.87 &                  0.22 &                  0.13 &     3.07 &                          0.08 &                     0.04 &              0.03 &                 0.02 &              0.01 &      0.10 \\
       & [2.51, 3.14) &  13.73 &                   13.28 &                  0.27 &                  0.20 &     3.44 &                          0.21 &                     0.01 &              0.01 &                 0.00 &              0.01 &      0.01 \\
       & [3.14, 3.77) &  12.78 &                   12.25 &                  0.18 &                  0.19 &     3.62 &                          0.16 &                     0.01 &              0.01 &                 0.00 &              0.00 &      0.00 \\
       & [3.77, 4.40) &  13.21 &                   12.72 &                  0.17 &                  0.08 &     3.53 &                          0.06 &                     0.06 &              0.01 &                 0.01 &              0.01 &      0.49 \\
       & [4.40, 5.03) &  10.68 &                   10.32 &                  0.11 &                  0.15 &     2.72 &                          0.06 &                     0.04 &              0.01 &                 0.00 &              0.01 &      0.38 \\
       & [5.03, 5.65) &  12.72 &                   12.29 &                  0.19 &                  0.09 &     3.28 &                          0.06 &                     0.02 &              0.00 &                 0.00 &              0.01 &      0.11 \\
       & [5.65, 6.28) &  13.42 &                   13.02 &                  0.21 &                  0.16 &     3.24 &                          0.10 &                     0.03 &              0.03 &                 0.02 &              0.02 &      0.12 \\
\hline
\hline
\end{tabular}

    }
\end{table*}

\clearpage

\section{Nested Hypothesis Test (NHT)}
\label{app:nht}
We perform a NHT to determine the optimal number of coefficients in the BGL form factor expansion. Starting point for the NHT is $N_a=1$, $N_b=1$, and $N_c=1$ to allow at least one degree of freedom to each contributing form factor. To truncate the series, we reject hypotheses with $\Delta \chi^2<1$ when moving from $N$ to $N+1$ free parameters in the fit. Additionally, we reject hypotheses that introduce correlations over 95\% in the free parameters to avoid blind directions in the fit. The NHT converges to the choice of $N_a=1$, $N_b=1$, and $N_c=1$. For our nominal fit scenario we choose the fit one order higher to estimate truncation related uncertainties. The fit with $N_a=1$, $N_b=2$, and $N_c=1$ is the only $N+1$ hypothesis that does not introduce larger than 95\% correlations. The full set of NHT results are tabulated in Table~\ref{tab:nht}.

\vspace{2ex}

We also perform the NHT as a cross-check by enforcing unitarity bounds of the form
\begin{equation}
\begin{aligned}
    \sum_{n=0}^N |a_n|^2 &\leq 1, \\
    \sum_{n=0}^N (|b_n|^2 + |c_n|^2) &\leq 1,
\end{aligned}
\label{eq:unitarity_bound}
\end{equation}
on the coefficients $a_n$, $b_n$, and $c_n$. The NHT results with the unitarity bound are tabulated in Table~\ref{tab:nhtu}.

\begin{table}
    \centering
    \caption{The result of the NHT without the unitarity bound on the coefficients $a_n$, $b_n$, and $c_n$. The $\rho_\mathrm{max}$ columns is the largest off-diagonal correlation coefficients and used to reject hypothesis if $\rho_\mathrm{max} \geq 0.95$. Highlighted in bold is the expansion order used in the main text.}
    \label{tab:nht}
    \begin{tabular}{llllll}
\hline
\hline
{} & $|V_{\mathrm{cb}}|$ & $\chi^2$ & dof &  N & $\rho_\mathrm{max}$ \\
\hline
BGL$_{111}$ &      $40.3 \pm 0.8$ &     45.7 &  32 &  3 &                0.71 \\
BGL$_{112}$ &      $40.8 \pm 0.9$ &     43.8 &  31 &  4 &                0.98 \\
\textbf{BGL$_{121}$} &      \textbf{$40.6 \pm 0.9$} &     \textbf{45.3} &  \textbf{31} &  \textbf{4} &                \textbf{0.62} \\
BGL$_{122}$ &      $41.3 \pm 1.0$ &     42.8 &  30 &  5 &                0.98 \\
BGL$_{131}$ &      $38.6 \pm 1.5$ &     42.7 &  30 &  5 &                0.98 \\
BGL$_{132}$ &      $39.1 \pm 1.5$ &     38.4 &  29 &  6 &                0.98 \\
BGL$_{211}$ &      $39.8 \pm 0.9$ &     43.5 &  31 &  4 &                0.99 \\
BGL$_{212}$ &      $40.3 \pm 0.9$ &     40.2 &  30 &  5 &                0.99 \\
BGL$_{221}$ &      $37.3 \pm 1.2$ &     39.5 &  30 &  5 &                0.99 \\
BGL$_{222}$ &      $38.4 \pm 1.9$ &     38.8 &  29 &  6 &                1.00 \\
BGL$_{231}$ &      $38.2 \pm 1.5$ &     40.6 &  29 &  6 &                0.96 \\
BGL$_{232}$ &      $39.0 \pm 1.5$ &     38.1 &  28 &  7 &                0.98 \\
BGL$_{311}$ &      $40.0 \pm 0.9$ &     43.3 &  30 &  5 &                0.97 \\
BGL$_{312}$ &      $39.9 \pm 1.0$ &     37.7 &  29 &  6 &                0.97 \\
BGL$_{321}$ &      $37.4 \pm 1.2$ &     39.2 &  29 &  6 &                0.96 \\
BGL$_{322}$ &      $39.4 \pm 1.8$ &     37.6 &  28 &  7 &                0.98 \\
BGL$_{331}$ &      $38.0 \pm 1.5$ &     38.4 &  28 &  7 &                0.99 \\
BGL$_{332}$ &      $38.8 \pm 2.1$ &     38.1 &  27 &  8 &                0.99 \\
\hline
\hline
\end{tabular}

\end{table}

\begin{table}
    \centering
    \caption{The result of the NHT when enforcing the unitarity bound given in Eq.~\ref{eq:unitarity_bound} on the coefficients $a_n$, $b_n$, and $c_n$. The $\rho_\mathrm{max}$ columns is the largest off-diagonal correlation coefficients and used to reject hypothesis if $\rho_\mathrm{max} \geq 95$. Highlighted in bold is the expansion order used in the main text.}
    \label{tab:nhtu}
    \begin{tabular}{llllll}
\hline
\hline
{} & $|V_{\mathrm{cb}}|$ & $\chi^2$ & dof &  N & $\rho_\mathrm{max}$ \\
\hline
BGL$_{111}$ &      $40.3 \pm 0.8$ &     45.7 &  32 &  3 &                0.71 \\
BGL$_{112}$ &      $40.8 \pm 0.8$ &     42.6 &  31 &  4 &                0.97 \\
\textbf{BGL$_{121}$} &      \textbf{$40.6 \pm 0.9$} &     \textbf{45.3} &  \textbf{31} &  \textbf{4} &                \textbf{0.62} \\
BGL$_{122}$ &      $41.4 \pm 1.0$ &     41.5 &  30 &  5 &                0.97 \\
BGL$_{131}$ &      $39.9 \pm 0.9$ &     42.4 &  30 &  5 &                0.61 \\
BGL$_{132}$ &      $40.7 \pm 1.0$ &     39.3 &  29 &  6 &                0.98 \\
BGL$_{211}$ &      $39.8 \pm 0.9$ &     42.1 &  31 &  4 &                0.99 \\
BGL$_{212}$ &      $40.4 \pm 0.9$ &     37.5 &  30 &  5 &                0.99 \\
BGL$_{221}$ &      $40.9 \pm 1.0$ &     45.1 &  30 &  5 &                0.93 \\
BGL$_{222}$ &      $39.2 \pm 1.0$ &     36.5 &  29 &  6 &                0.96 \\
BGL$_{231}$ &      $40.3 \pm 1.0$ &     41.8 &  29 &  6 &                0.94 \\
BGL$_{232}$ &      $41.0 \pm 1.0$ &     39.0 &  28 &  7 &                0.97 \\
BGL$_{311}$ &      $39.8 \pm 0.9$ &     42.1 &  30 &  5 &                0.99 \\
BGL$_{312}$ &      $40.4 \pm 0.9$ &     37.4 &  29 &  6 &                0.99 \\
BGL$_{321}$ &      $38.5 \pm 0.9$ &     39.4 &  29 &  6 &                0.65 \\
BGL$_{322}$ &      $39.2 \pm 1.0$ &     36.4 &  28 &  7 &                0.96 \\
BGL$_{331}$ &      $38.3 \pm 0.9$ &     38.1 &  28 &  7 &                0.86 \\
BGL$_{332}$ &      $38.7 \pm 1.5$ &     36.0 &  27 &  8 &                0.99 \\
\hline
\hline
\end{tabular}

\end{table}

\clearpage

\section{BGL$_{332}$ Beyond Zero-Recoil Fits}
\label{app:bgl332}
\begin{figure}
    \centering
    \includegraphics[width=\linewidth]{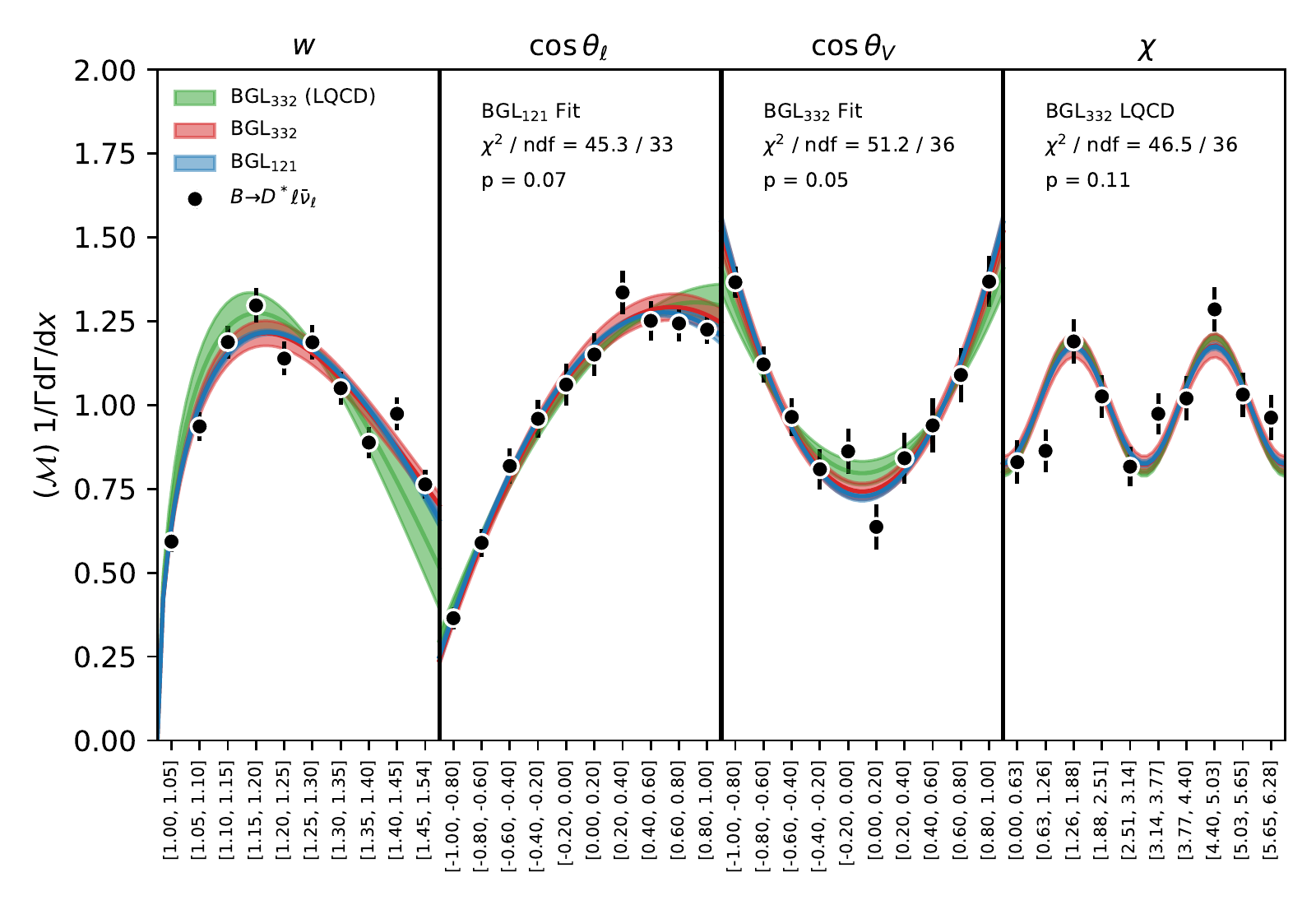}
\caption{The fitted shapes for our nominal BGL$_{121}$ (blue) and CLN (orange) scenarios from the main text using the zero-recoil point only. The result of the BGL$_{332}$ fit with the constraints from Ref.~\citep{FermilabLattice:2021cdg} on the BGL coefficients is shown in red.
    }
    \label{fig:BGL121vsBGL332}
\end{figure}

\begin{figure}
\centering
 \includegraphics[width=\linewidth]{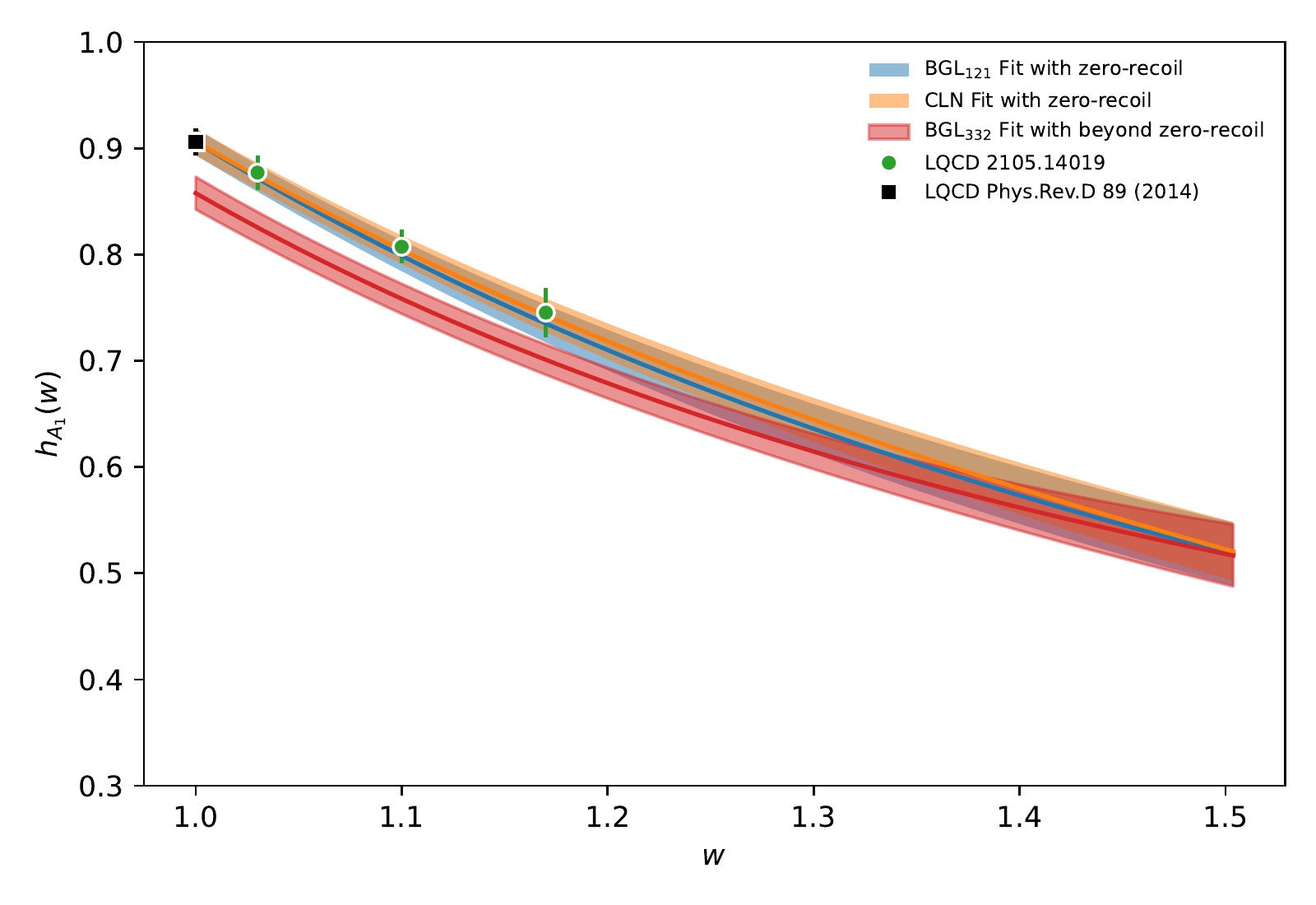}
 \caption{The fitted $h_{A_1}$ form factor for our nominal BGL$_{121}$ (blue) and CLN (orange) scenarios from the main text using the zero-recoil point only. The result of the BGL$_{332}$ fit with the constraints from Ref.~\citep{FermilabLattice:2021cdg} on the BGL coefficients is shown in red. The black data point is the zero-recoil data point from Ref.~\cite{FermilabLattice:2014ysv}, the green data points are the beyond zero-recoil data points from Ref.~\cite{FermilabLattice:2021cdg}.}
 \label{fig:hA1}
\end{figure}

\begin{figure}
\centering
 \includegraphics[width=\linewidth]{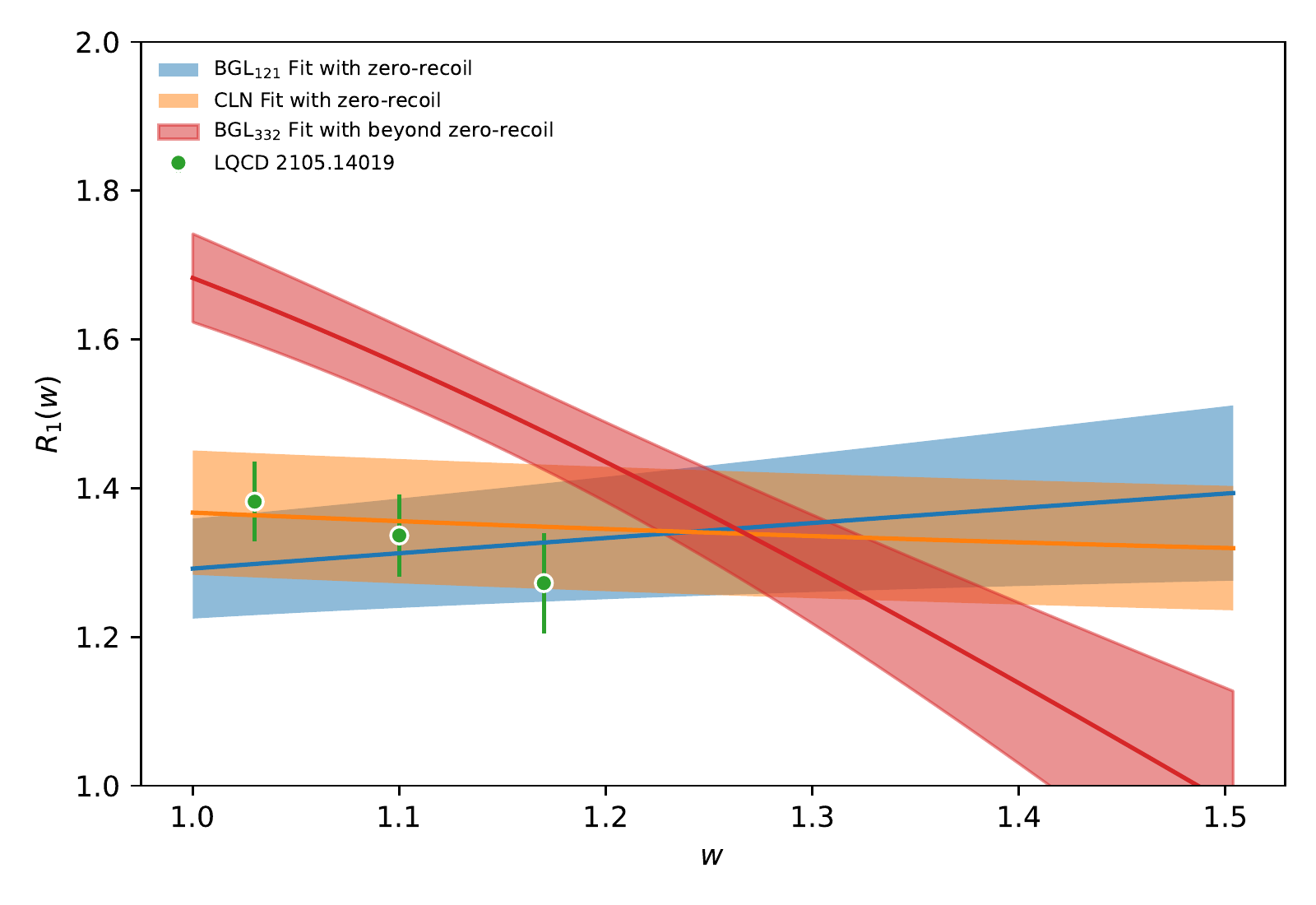}
 \caption{The fitted $R_1$ form factor for our nominal BGL$_{121}$ (blue) and CLN (orange) scenarios from the main text using the zero-recoil point only. The result of the BGL$_{332}$ fit with the constraints from Ref.~\citep{FermilabLattice:2021cdg} on the BGL coefficients is shown in red. The green data points are the beyond zero-recoil data points from Ref.~\cite{FermilabLattice:2021cdg}.}
 \label{fig:R1}
\end{figure}

\begin{figure}
\centering
 \includegraphics[width=\linewidth]{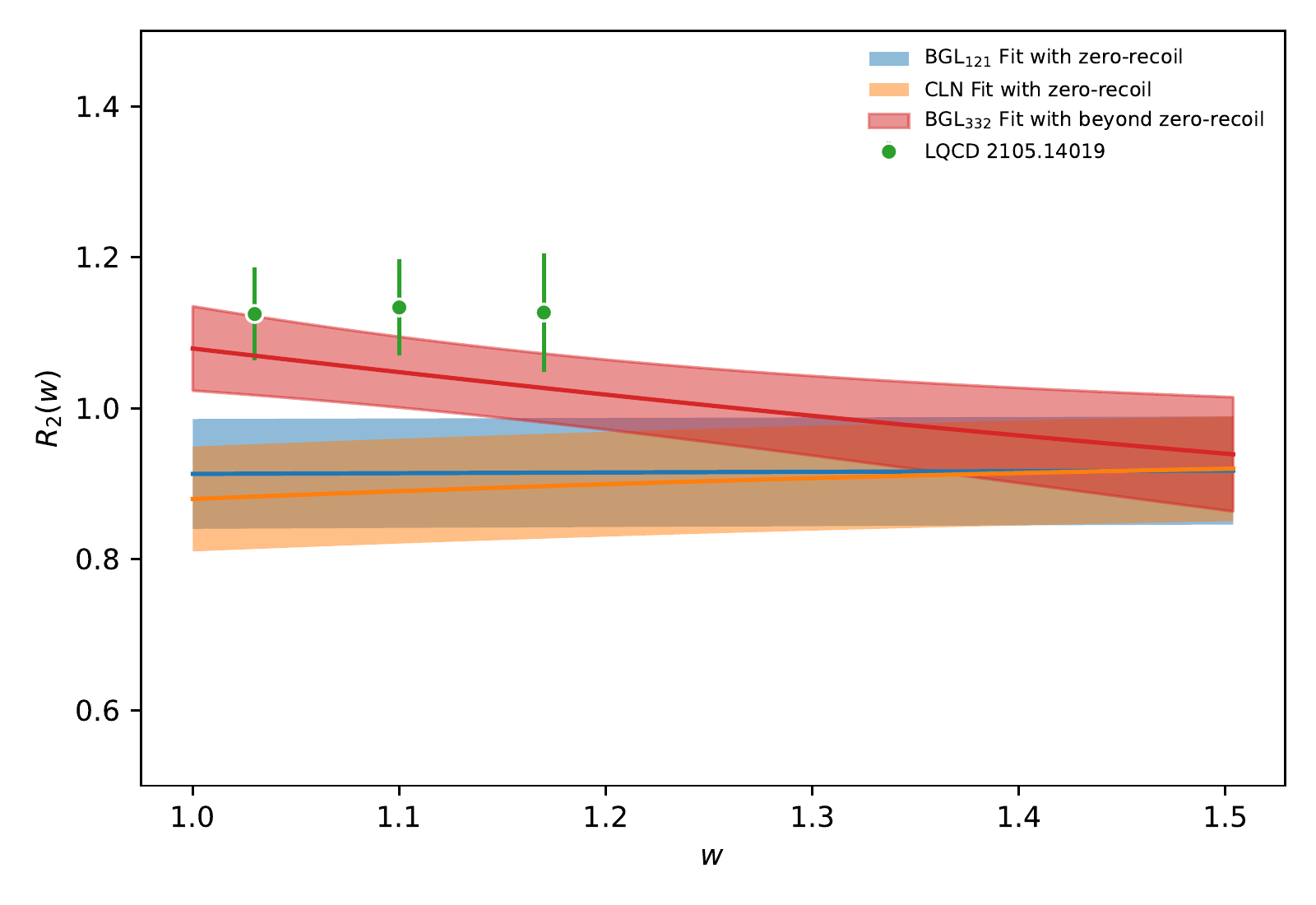}
 \caption{The fitted $R_2$ form factor for our nominal BGL$_{121}$ (blue) and CLN (orange) scenarios from the main text using the zero-recoil point only. The result of the BGL$_{332}$ fit with the constraints from Ref.~\citep{FermilabLattice:2021cdg} on the BGL coefficients is shown in red. The green data points are the beyond zero-recoil data points from Ref.~\cite{FermilabLattice:2021cdg}.}
 \label{fig:R2}
\end{figure}

In this fit we directly incorporate the lattice data in the form of constraints on the coefficients in the BGL$_{332}$ form factor expansion as provided by Ref.~\cite{FermilabLattice:2021cdg}. We use the values provided solely based on the lattice calculation, without any further experimental inputs. The corresponding term in the $\chi^2$ functions is modified to
\begin{equation}
    \chi^2_\mathrm{LCQD} = (\vec{x}_\mathrm{LQCD} - \vec{x}) C_\mathrm{LQCD}^{-1} (\vec{x}_\mathrm{LQCD} - \vec{x})
\end{equation}
with $\vec{x} = (a_0, a_1, a_2, b_0, b_1, b_2, c_1, c_2)$ and the corresponding covariance matrix $C_\mathrm{LQCD}$.
The fitted coefficients are listed in Table~\ref{tab:BGL332}. The fitted differential shape, together with our fits from the main text, are compared in Fig.~\ref{fig:BGL121vsBGL332}. We also compare the fitted form factors $h_{A_1}$, $R_1$, and $R_2$ in Fig.~\ref{fig:hA1}, Fig.~\ref{fig:R1}, and Fig.~\ref{fig:R2}, respectively.

\begin{table*}
    \centering
    \caption{Fitted BGL$_{332}$ coefficients and correlations with the constraint on the coefficients from Ref.~\citep{FermilabLattice:2021cdg}.}
    \label{tab:BGL332}
    \begin{tabular}{lrrrrrrrrrr}
\hline
\hline
{} &             Value & \multicolumn{9}{l}{Correlation} \\
\hline
$a_0 \times 10^3$      & $     32.03\pm1.11 $ &       $ 1.00$ &  $-0.43$ &  $-0.08$ &  $ 0.22$ &  $-0.06$ &  $ 0.02$ &  $ 0.07$ &  $-0.01$ &  $-0.25$ \\
$a_1 \times 10^3$      & $  -202.43\pm46.19 $ &       $-0.43$ &  $ 1.00$ &  $-0.48$ &  $ 0.04$ &  $ 0.23$ &  $-0.17$ &  $ 0.18$ &  $-0.13$ &  $-0.17$ \\
$a_2 \times 10^3$      & $ -824.26\pm952.62 $ &       $-0.08$ &  $-0.48$ &  $ 1.00$ &  $ 0.06$ &  $-0.01$ &  $-0.08$ &  $ 0.03$ &  $-0.06$ &  $-0.02$ \\
$b_0 \times 10^3$      & $     12.41\pm0.22 $ &       $ 0.22$ &  $ 0.04$ &  $ 0.06$ &  $ 1.00$ &  $-0.16$ &  $ 0.10$ &  $-0.05$ &  $ 0.01$ &  $-0.71$ \\
$b_1 \times 10^3$      & $    -7.00\pm11.21 $ &       $-0.06$ &  $ 0.23$ &  $-0.01$ &  $-0.16$ &  $ 1.00$ &  $-0.77$ &  $ 0.74$ &  $-0.62$ &  $-0.24$ \\
$b_2 \times 10^3$      & $  195.79\pm325.00 $ &       $ 0.02$ &  $-0.17$ &  $-0.08$ &  $ 0.10$ &  $-0.77$ &  $ 1.00$ &  $-0.61$ &  $ 0.54$ &  $ 0.07$ \\
$c_1 \times 10^3$      & $     -5.26\pm2.15 $ &       $ 0.07$ &  $ 0.18$ &  $ 0.03$ &  $-0.05$ &  $ 0.74$ &  $-0.61$ &  $ 1.00$ &  $-0.91$ &  $-0.33$ \\
$c_2 \times 10^3$      & $   101.53\pm40.08 $ &       $-0.01$ &  $-0.13$ &  $-0.06$ &  $ 0.01$ &  $-0.62$ &  $ 0.54$ &  $-0.91$ &  $ 1.00$ &  $ 0.21$ \\
$|V_{cb}| \times 10^3$ & $     42.67\pm0.98 $ &       $-0.25$ &  $-0.17$ &  $-0.02$ &  $-0.71$ &  $-0.24$ &  $ 0.07$ &  $-0.33$ &  $ 0.21$ &  $ 1.00$ \\
\hline
\hline
\end{tabular}

\end{table*}

\vspace{4ex}

\end{document}